\documentclass[fleqn,usenatbib]{mnras}

\usepackage[T1]{fontenc}
\usepackage{ae,aecompl}

\usepackage{graphicx}    
\usepackage{amsmath}    
\usepackage{amssymb}    
\usepackage[]{units}
\usepackage[dvipsnames]{xcolor}
\usepackage{newtxtext,newtxmath}

\title[Great Balls of FIRE II]{Great Balls of FIRE II: The evolution and destruction of star clusters across cosmic time in a Milky Way-mass galaxy}

\author[Rodriguez et al.]{
Carl L.~Rodriguez$^{1,2}$\thanks{carl.rodriguez@unc.edu}, 
Zachary Hafen$^{3}$,
Michael Y. Grudi\'{c}$^{4}$\thanks{NASA Hubble Fellow},
Astrid Lamberts$^{5,6}$,\newauthor
~Kuldeep Sharma$^{1}$,
Claude-Andr\'e Faucher-Gigu\`ere$^{7,8}$,
Andrew Wetzel$^{9}$
\\
$^{1}${McWilliams Center for Cosmology and Department of Physics, Carnegie Mellon University, Pittsburgh, PA, 15213}\\
$^{2}$ Department of Physics and Astronomy,
    University of North Carolina at Chapel Hill,
    120 E. Cameron Ave, Chapel Hill, NC, 27599, USA\\
$^{3}$ {Department of Physics and Astronomy, University of California Irvine, CA 92697, USA} \\
$^{4}$ {Carnegie Observatories, 813 Santa Barbara St, Pasadena, CA 91101, USA}\\
$^{5}${Laboratoire Lagrange, Universit\'e C\^ote d'Azur, Observatoire de la C\^ote d'Azur, CNRS,  France}\\
$^{6}${Laboratoire Artemis, Universit\'e C\^ote d'Azur, Observatoire de la C\^ote d'Azur, CNRS,  France}\\
$^{7}${ Department of Physics \& Astronomy, Northwestern University, Evanston, IL 60208, USA}\\
$^{8}${Center for Interdisciplinary Exploration \& Research in Astrophysics (CIERA), Northwestern University, Evanston, IL 60208, USA}\\
$^{9}${Department of Physics \& Astronomy, University of California, Davis, CA 95616}\\
}
\date{Accepted XXX. Received YYY; in original form ZZZ}
\defcitealias{GBOF}{Paper I}

\pubyear{2022}

\begin{document}
\label{firstpage}
\pagerange{\pageref{firstpage}--\pageref{lastpage}}
\maketitle
\begin{abstract}
The current generation of galaxy simulations can resolve individual giant molecular clouds, the progenitors of dense star clusters.  But the evolutionary fate of these young massive clusters, and whether they can become the old globular clusters (GCs) observed in many galaxies, is determined by a complex interplay of internal dynamical processes and external galactic effects.  We present the first star-by-star $N$-body models of massive ($N\sim10^5-10^7$)  star clusters formed in a FIRE-2 MHD simulation of a Milky Way-mass galaxy, with the relevant initial conditions and tidal forces extracted from the cosmological simulation.  We select 895 ($\sim 30\%$) of the YMCs with $ > 6\times10^4M_{\odot}$ from Grudi\'c et al.~2022 and integrate them to $z=0$ using the Cluster Monte Carlo Code, \texttt{CMC}.  This procedure predicts a MW-like system with 148 GCs, predominantly formed during the early, bursty mode of star formation.   Our GCs are younger, less massive, and more core-collapsed than clusters in the Milky Way or M31.  This results from the assembly history and age-metallicity relationship of the host galaxy: younger clusters are preferentially born in stronger tidal fields and initially retain fewer stellar-mass black holes, causing them to lose mass faster and reach core collapse sooner than older GCs.   Our results suggest that the masses and core/half-light radii of GCs are shaped not only by internal dynamical processes, but also by the specific evolutionary history of their host galaxies.  These results emphasize that $N$-body studies with realistic stellar physics are crucial to understanding the evolution and present-day properties of GC systems.

\end{abstract}

\begin{keywords}
globular clusters: general -- galaxies: star clusters: general -- stars: black holes -- galaxies: star formation -- Galaxy: evolution
\end{keywords}


\section{Introduction}
\label{sec:intro}

The formation, evolution, and destruction of globular clusters (GCs) has been a subject of intense study for nearly a century.  While the first GCs were observed well before the first galaxies beyond the Milky Way (MW), we now know that most galaxies with luminosities $\gtrsim 3\times10^6 L_{\odot}$ or halo masses $\gtrsim 10^9M_{\odot}$ contain GCs \citep[e.g.,][]{Harris2013,Harris2017}.  And although the exact formation scenarios for GCs are still a topic of debate, it is now thought that a significant number of GCs are simply the byproduct of normal star formation in the early Universe, where the high gas pressures in young, spheroidal, and merging galaxies allows for the efficient conversion of stars into bound clusters \citep[e.g.,][]{Kruijssen2015}.  This suggests that the formation of GCs and other old star clusters at high $z$ and the formation of young massive clusters (YMCs) in the local universe are driven by the same physical processes occurring in different galactic environments.  Furthermore, the typical masses ($\sim10^4M_{\odot}$ - $10^6M_{\odot}$), luminosities, and compact radii of GCs allow them to be resolved in both local and distant galaxies, providing a wealth of observational information about clusters across different galaxy types and morphologies \citep{Brodie2006}.   Because of this, GCs and other star clusters are an ideal probe of galaxy formation and assembly.  And while still strongly dependent on the model for cluster formation and (in many cases) the specific implementation of subgrid physics, it is now possible to model the initial conditions of star clusters in both realistic galaxies and cosmological simulations as a function of their formation environments \citep[e.g.,][]{Li2017,Pfeffer2018}.  

In concert with these advancements in our understanding of cluster formation, our ability to create realistic, fully  collisional $N$-body models of clusters has also improved by leaps and bounds.  The last decade has seen the first direct summation $N$-body simulation of old GCs with $10^6$ stars \citep{Wang2016}, while the latest generation of Monte Carlo $N$-body codes \citep[e.g.,][]{Giersz2013,Pattabiraman2013,Rodriguez2022} have had great success creating collisional models of clusters with fully-realized binary dynamics and detailed prescriptions for stellar and binary evolution.  It is now routine to create entire grids of $N$-body star cluster models with $> 10^6$ stars and binaries covering a realistic range of initial conditions that can reproduce GCs and other massive star clusters in the local universe.  But while this approach has had great success creating models of individual star clusters in the MW \citep[e.g.,][]{VanderMarel1997,Hurley2001,Baumgardt2003a,Heggie2014d,Heggie2014b,Wang2016a,Kremer2018,Ye2021}, near every study of Galactic and extra-galactic GC systems has started from idealized grids of initial conditions designed only to reproduce MW GCs.   While this method allows us to create one-to-one mappings between individual MW clusters and $N$-body models \citep[e.g.,][]{Baumgardt2018,Weatherford2020,Rui2021}, it neglects the wealth of information that cosmological models of star cluster formation can provide, such as the cluster initial mass function (CIMF), initial radii, metallicities, ages, galactic tidal fields, and more.  While these quantities are critical components of most semi-analytic models of GC evolution, they have never been self-consistently adopted by the GC modeling community.  

We have recently developed a new framework for modeling cluster formation in cosmological simulations of galaxy formation and assembly \citep[][hereafter Paper I]{GBOF}.  Using a suite of models of cloud collapse and cluster formation \citep[with resolution of $\sim 0.1$ pc and a detailed treatment of stellar feedback,][]{Grudic2021}, we created a procedure to directly link the properties of self-gravitating and collapsing giant molecular clouds (GMCs) identified in cosmological simulations \citep[using the results of][]{guszejnov_GMC_cosmic_evol} to the masses, concentrations, and radii of the clusters they eventually form.  In \citetalias{GBOF}, this framework was applied to a magnetohydrodynamical (MHD) simulation of a galaxy and its cosmological environment created as part of the Feedback In Realistic Environments (FIRE) project\footnote{See \href{http://fire.northwestern.edu}{http://fire.northwestern.edu}} \citep[][]{hopkins2014,Hopkins2017}.  \citetalias{GBOF} was focused on the properties of the YMCs, and in this paper we seek to understand what this cluster population looks like at $z=0$.  This requires not only the initial conditions of the cluster population, but a detailed treatment of the internal dynamics and evolution of collisional star clusters and their interaction with their galactic environment after formation.  

To that end, we have created the first \emph{evolved representative population} of GCs using fully-collisional, star-by-star $N$-body models.   The initial masses, radii, metallicities, and birth times for these clusters are taken directly from collapsing GMCs across cosmic time in the \texttt{m12i} MHD simulation  \citep[with a gas resolution as fine as 1 pc][]{Wetzel2016,Hopkins2020}, with unique $N$-body initial conditions generated for each YMC, while the subsequent dynamical evolution is informed by the time-dependent galactic coordinates and applied tidal forces of associated stars in the cosmological simulation.    We then integrate these clusters forward in time to $z=0$ with our Cluster Monte Carlo code \citep[\texttt{CMC},][]{Joshi1999,Pattabiraman2013,Rodriguez2022}.  The H\'enon method upon which \texttt{CMC} is based allows us to follow the evolution of dense, spherical star clusters with more than $10^7$ stars and binaries (far beyond the capabilities of direct-summation $N$-body codes), enabling star-by-star simulations of the largest clusters in the \texttt{m12i} galaxy.  Furthermore, by tracing the time-dependent tidal forces extracted from the galactic potential, we calculate the tidal boundary of each cluster along its trajectory in the galaxy, allowing us to study the relationship between the clusters' initial conditions, internal evolution, and the galactic environment they reside in. This means that we can accurately simulate the most relevant physical processes for the long-term evolution and survival of YMCs as they mature into GCs.

In \S \ref{sec:soss}, we describe the details of our cluster $N$-body simulations, how we generate initial stellar profiles from the results of the \citetalias{GBOF} catalog, and how the influence of the galactic environment (e.g.~tidal forces and dynamical friction) is modeled in \texttt{CMC}.  As a result of the cosmological environment, a significant fraction of our clusters are shown to overflow their tidal boundaries at formation, which we analyze in detail \S \ref{sec:initial}.  In \S \ref{sec:presentday}, we compare the properties of the clusters that survive to the present day ($z=0$) to the masses, metallicities, ages, and radii of GCs in the MW and M31.  Defining GCs as clusters older than 6 Gyr, we find that we can largely reproduce the correct number of MW GCs; however, our GCs are typically younger, less massive, and more core collapsed than those in the MW.  In \S \ref{sec:evolution}, we argue that this discrepancy is due to a complex interplay between the typical tidal fields experienced by clusters formed at different times, and the accelerated core collapse experienced by higher-metallicitiy (younger) clusters.  We also show that our model produces GCs that inhabit roughly the same mass-radius space as other models of long-term cluster survival in the MW \cite[e.g.][]{Gnedin1997}, and compare our results to other studies of star cluster evolution in cosmological simulations, namely the E-MOSAICS simulations of \citet{Pfeffer2018} and the ART simulations of \citet{Li2017}.  

\section{Cluster Initial Conditions and Cosmological Evolution}
\label{sec:soss}

The grids of initial conditions to attempt to reproduce the wide range of GCs observed in the MW 
\cite[e.g.][]{Morscher2015,2016MNRAS.462.2950B,Rodriguez2018c,kremerModelingDenseStar2020,2021arXiv211109223M} typically cover a wide range of initial 
star cluster masses and virial radii (defined as $r_v = -\frac{G M^2}{4 E}$, where $E$ is the total kinetic and potential energy of the cluster).  Star clusters characteristic of MW 
GCs are challenging to evolve --- even a GC with a present-day mass of $2\times10^5M_{\odot}$, near the median of the MW GC mass function 
\citep[GCMF,][]{Harris2010} must be initialized with $N\sim 8\times10^5$ particles, and an virial radius of $1-2$\,pc, beyond what direct $N$-body 
integrators can accomplish in a reasonable time frame, to say nothing of the compact radii required to model core-collapsed GCs \citep{Kremer2019} or GCs with present-day masses of $\gtrsim 10^6M_{\odot}$ 
\citep[e.g.~47 Tuc,][]{Giersz2011,Ye2021}.  Instead, studies of these largest clusters have relied upon approximate techniques, such as the Monte Carlo approach 
introduced by \citet{Henon1971a,Henon1971}.  The two most recent Monte Carlo parameter sweeps \citep{kremerModelingDenseStar2020,2021arXiv211109223M} have used 
grids of initial conditions covering a range of masses, virial radii, metallicities, and fixed tidal fields.  The stellar positions and velocities are sampled from a 
\citet{King1966} profile and evolved dynamically for approximately one Hubble time.  Both these studies have shown that they can largely cover the observed 
range of GCs and other massive star clusters observed in the MW and beyond.

But while these grid-based studies have demonstrated much success over the years, following an entire population of YMCs from formation in their galactic environments would offer a better understanding of the present-day GC population, act as a powerful probe of the process of GC formation itself (particularly given the unique tidal forces experienced by clusters during different epochs of star formation in galaxies), and may even place constraints on galaxy formation models based on the observed properties of present-day GCs.  To that end, we make several modifications to the traditional grid-based initial conditions used by Monte Carlo $N$-body studies: first, we use a subset of the catalog from \citetalias{GBOF} as our initial conditions.  These initial conditions cover a wide range of initial masses, radii, metallicities, and formation times.  We evolve each cluster forward from its birth time in the FIRE-2 \texttt{m12i} galaxy to either the present day, or its dynamical destruction.  Second, our initial cluster models are generated using Elson, Fall and Freeman density profiles \citep[EFF,][]{Elson1987} rather that the traditional King or Plummer profiles (which assume the clusters to already be in tidal equilibrium with their surrounding environments).  Third, the tidal boundary of each cluster is set by the galactic potential of the \texttt{m12i} galaxy, allowing us to resolve the effects of a realistic galactic evolutionary history on a population of YMCs.  We now discuss the details of our Monte Carlo $N$-body approach, and how each of these new physical processes are incorporated into the CMC-FIRE GC systems and their respective assumptions and limitations.

  \begin{figure*}
\centering
\includegraphics[]{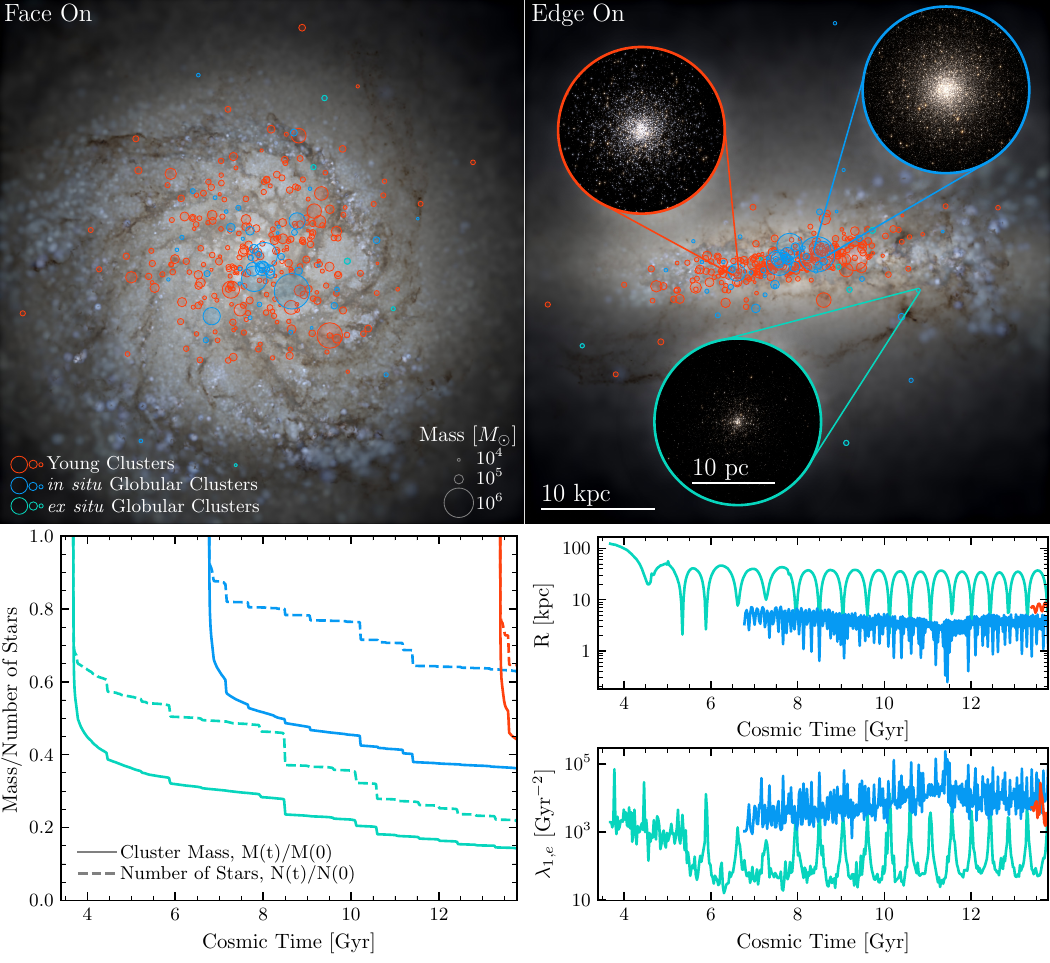}
\caption{The galactic environment and dynamical evolution of three typical clusters -- one formed \textit{in situ}, one formed \textit{ex situ}, and one formed recently -- with properties similar to GCs in the MW.  We follow each cluster from their births to the present day.  The \textbf{top} two panels show face-on and edge-on mock HST images \citep[generated using the FIRE Studio,][]{Gurvich:2021} of an MHD FIRE simulation of a MW-mass galaxy (\texttt{m12i}), with the surviving GC population in blue (for clusters born \textit{in situ} in the main galaxy) and green (for clusters born \textit{ex situ} in dwarf galaxies accreted by the main galaxy) while younger clusters are shown in red. Circle size corresponds to present-day cluster mass.  We also show three mock HST images typical clusters at $z=0$ \citep[generated with the \texttt{Fresco} package,][]{steven_rieder_2019_3553805}.  The \textbf{bottom left} panel shows the mass and the total number of particles in each cluster as a function of galactic cosmic time.  The \textbf{middle right panel} shows the galactocentric radius of the clusters over their orbits, while the \textbf{bottom right} panel shows the effective tidal strengths they experiences over their lifetimes (see \S \ref{ss:tides}).  The sharp decreases in cluster mass are caused by peaks in $\lambda_{1,e}$, as the outer parts of the clusters are stripped away by changing tidal fields as the clusters pass through the disk and near the galactic bulge.}
\label{fig:typical}
\end{figure*}

\subsection{The Cluster Monte Carlo code, \texttt{CMC}}
\label{ss:cmc}

The $N$-body models presented here were generated with \texttt{CMC}, a H\'enon-style Monte Carlo approach to collisional stellar dynamics \citep{Joshi1999,Pattabiraman2013,Rodriguez2022}. Unlike traditional $N$-body integrators, where the accelerations are calculated by directly summing pairwise gravitational forces, \texttt{CMC} assumes that for sufficiently large clusters, the cumulative effect of many two-body encounters can be understood as a statistical process.  Here, ``sufficiently large'' refers to the Fokker-Planck regime, where the relaxation timescale of any particle (that is, the time for its velocity to change by order of itself) is much longer than the orbital timescale of the particle in the cluster.  In this regime, the diffusion of energy and angular momentum between particles can be modeled via \emph{effective encounters} between neighboring particles, where the  deflection angle of the encounter is chosen to reproduce the cumulative effect of many distant two-body encounters.  This approach has been shown many times to reproduce the pre- and post-core collapse evolution of dense spherical star clusters when compared to both direct $N$-body simulations and theoretical calculations \citep[e.g.,][]{Aarseth1974,Joshi1999,Giersz2013,Rodriguez2016,Rodriguez2022}.  

\texttt{CMC} relies on pairwise interactions between neighboring stars and binaries to model the effect of many weak two-body encounters.  But this scheme allows us to model strong interactions--close encounters where additional stellar and binary physics comes into play--just as easily.  These include chaotic encounters between single and binary stars, integrated directly with the \texttt{fewbody} small-$N$ body code \citep{Fregeau2007}, direct physical collisions, and binary formation through either the Newtonian interaction of three unbound stars \citep{Morscher2012}, or two-body captures facilitated by tidal dissipation \citep{Ye2021} or gravitational-wave emission \citep{Rodriguez2018,Samsing2019}.  Each dynamical timestep allows energy and angular momentum to be exchanged between neighboring particles, pushing the stars and binaries onto new orbits within the cluster potential.  For an isolated cluster, some fraction of these orbits will naturally diffuse to positive total energies, representing the classical \emph{evaporation} of stars from the cluster.  For clusters embedded in a host galaxy, the rate of stellar loss is enhanced by the \emph{tidal stripping} of stars by the galactic potential, similar to the overflow of the Roche Lobe in binary stars \cite[e.g.,][]{spitzer_dynamical_1987,fukushige2000,Renaud2011}.  \texttt{CMC} treats both processes, removing unbound stars and binaries after every timestep (see \S \ref{ss:tides} and Appendix \ref{apx:shocking} for a description of our implementation of tidal forces in a changing galactic potential).

In addition to the relevant gravitational dynamics, \texttt{CMC} includes metallicity-dependent prescriptions for the evolution of stars and binaries using the Binary Stellar Evolution (\texttt{BSE}) package of \citet{Hurley2000,Hurley2002}.  \texttt{CMC} uses the version of \texttt{BSE} that has been upgraded as part of the \texttt{COSMIC} population synthesis code \citep{Breivik2020}.  The version of \texttt{COSMIC} employed here (v3.3) includes new prescriptions for compact-object formation and supernova \citep{Kiel2009,Fryer2012,Rodriguez2016a}, stellar winds \citep{Vink2001,Belczynski2010}, stable mass transfer \citep{Claeys2014} and more.  See \citet{Breivik2020} for details.  Because every star and binary in the cluster has a time-dependent mass, radius, and luminosity in every snapshot, it is also possible to calculate black-body spectra for both individual stars and the cluster as a whole, which can then be combined into a mock observations of the cluster through any number of appropriate filters.  See Section \S \ref{sec:presentday}.

\subsection{Initial Dynamical Profiles of Clusters}
\label{ss:init}

For old clusters which have dynamically relaxed, one can assume that the cluster has reached a sufficiently steady state that it can be described by an isothermal energy distribution.  Of course, in realistic clusters, the presence of a tidal boundary means that the energy distribution must go to zero at some boundary, as is the case with the often employed \citet{King1966} distribution function.  But YMCs, which have not had sufficient time to come into equilibrium with their surrounding environments, are neither expected nor observed to follow such trends.  In fact, observations suggest that YMCs typically follow extended power-law profiles with no discernible tidal boundary \citep{Elson1987,Ryon2015,grudic2018,brown:2021.cluster.mass.radius}.  Following these observations, our clusters are initialized using an EFF profile, which was originally used to fit surface brightness profiles of YMCs in the Magellanic Clouds \citep{Elson1987}, with a 3D density profile given by:

\begin{equation}
\rho(r) = \rho_0 \left(1 + \frac{r^2}{a^2}\right)^{-\frac{\gamma+1}{2}}~,
\label{eqn:elsonrho}
\end{equation}

\noindent where $\rho_0$ is the central cluster density, $a$ is a scale radius, and $\gamma$ corresponds to the power-law index of the outer regions of the surface brightness profile.  Note that while $\gamma=4$ corresponds to the well-known \citet{Plummer1911} profile, observations of YMCs are typically better fit with shallower profiles, where $\gamma\sim 2.2-3.2$ \citep[e.g.,][]{Mackey2003a, Mackey2003b, PortegiesZwart2010, Ryon2015}.

Each cluster in the catalog from \citetalias{GBOF} was assigned an initial EFF profile according to the cluster formation model, producing a list of $\gamma$ parameters and effective radii.  To initialize our $N$-body models, we generate the cumulative mass distribution, $M(r)$, by integrating Equation \eqref{eqn:elsonrho} for a given $\gamma$ and scale parameter $a$ (chosen to give the correct effective radius), and proceed to randomly sample stellar positions from $M(r)/M(\infty)$.  For each star, we also draw a velocity from the local velocity dispersion given by one of the 1D Jeans equations for spherical systems \citep[following][]{2008LNP...760..181K}:

\begin{equation}
\sigma^2(r) = \frac{1}{\rho(r)}\int^{\infty}_r \rho(r')\frac{GM(r')}{r'^2}dr'~~.
\end{equation}

\noindent For each star at a radius $r$, we compute $\sigma^2(r)$, and draw the individual components of the velocities from a Gaussian distribution with width $\sigma$.  This is similar to other codes used to generate $N$-body initial conditions \citep{2011MNRAS.417.2300K}, though we note that this approach does not correctly account for particles in the tail of the velocity distribution with $v > v_{\rm esc}$, which would be absent in a collisionless equilibrium. However, such particles are expected to be removed from the cluster on an orbital timescale.  

Finally, our cluster initial conditions are generated with distributions of star and binary masses and other properties typical for $N$-body simulations of star clusters.  After the particle positions and velocities are sampled, we draw random stellar masses from a \cite{Kroupa2001} initial mass function sampled between $0.08$ and 150 $M_{\odot}$.  Of these stars, 10\% are randomly selected to become binaries.  The mass ratios of the binaries are drawn from a uniform distribution between 0.1 and 1.  The semi-major axes are drawn from a flat-in-log distribution \citep{1991A&A...248..485D} with a minimum equal to the point of stellar contact for the two stars and a maximum equal to the hard-soft boundary for a binary with those masses at that point in the cluster (out to a maximum of $10^5$ AU).    Eccentricities are drawn from a thermal distribution \citep[$p(e)\propto e$,][]{Ambartsu1937}.

\subsection{YMC Catalog from Paper I}
\label{ss:weights}

The complete set of properties that determine our cluster initial conditions (as described in the prior section) are taken directly from the YMC catalog in \citetalias{GBOF}. In brief, to generate the catalog we mapped the properties of high-resolution, small-scale simulations of individual GMC collapse onto a full cosmological simulation. The metallicity, mass, surface gas density, and radius of a GMC are used as input to create distributions of star cluster masses and radii \citepalias[Equations 1-5 and 6, respectively, in][\footnote{See also \citet[][Sections 3.4-3.6]{Grudic2021} where these distributions were originally developed using specialized high-resolution MHD simulations of collapsing GMCs.}]{GBOF}. To create our population, we draw random YMC masses until all of the predicted gas mass in a specific GMC that is to be turned into gravitationally-bound stars has been converted into YMCs.  Each cluster is then assigned a half-mass radius following Equation 6 in \citetalias{GBOF} and an Elson $\gamma$ parameter from the universal relation identified in \citet[][Equation 20]{Grudic2021}, while the cluster's stellar metallicity is inherited directly from the gas metallicitiy of the GMC.  The birth times of a cluster is drawn from a uniform distribution over the time interval of the \texttt{m12i} snapshot where their parent GMC was identified.  Since most GMCs typically disperse within $\sim$ 3-10 Myr, and most snapshots were spaced 22 Myr apart, it is likely that many of the GMCs formed in the \texttt{m12i} galaxy were formed in between snapshots.  To account for this, we resample the GMC population following the procedure outlined in \citetalias[][Section 2.3.1]{GBOF}.

This initial cluster catalog contains 73,461 entries with masses from $10^4M_{\odot}$ to $10^7 M_{\odot}$ and covers a wide range of metallicities, ages, galactic positions, sizes, and stellar densities.  From this catalog, we restrict ourselves to clusters with initial masses greater than $6\times10^4M_{\odot}$, corresponding roughly to initial particle numbers of $10^5$ or greater.  This choice was motivated both by our interest in the most massive clusters in the galaxy (the progenitors of GCs), and to ensure that our clusters are sufficiently large for the H\'enon method to be reliable.\footnote{As stated in \S \ref{ss:cmc}, the H\'enon method formally requires a sufficiently large number of particles to ensure that the relaxation time of the cluster is significantly longer than the typical dynamical  time (i.e.~$T_{\rm dyn} \ll T_{\rm relax}$).  Previous work \citep{Freitag2008} has suggested that this criterion is satisfied when $N \gtrsim 3000 m_{\rm max}/\left<m\right>$, which for an average mass of $\left<m\right> \approx 0.6M_{\odot}$ and a maximum mass of $m_{\rm max} \sim 40M_{\odot}$ (the maximum BH mass after the first few Myr of stellar evolution), gives a minimum reliable initial particle number of $\sim 10^5$, corresponding to a minimum mass of $\sim6\times10^4M_{\odot}$.}  This initial cut leaves us with 3165 initial clusters which, because of limited computational resources, we randomly select 895 for integration\footnote{Even then, our cluster catalog required $\gtrsim 2$ million CPU hours to evolve, nearly twice as many as the MHD \texttt{m12i} galaxy itself!}.  

For computational tractability, we make several modifications to the initial properties of the 895 clusters we present throughout this paper.  First, for the sake of computational speed, we truncate the lower limits of our initial virial radii and cluster profile slopes to $0.8$ pc and $\gamma = 2.5$ respectively.  This was done after testing showed that clusters with very compact radii and very flat mass distributions produce unreasonably high central densities.  As an example, our most massive cluster at $5\times10^6 M_{\odot}$ would naively yield an initial central density of $\rho_0 \sim 10^{10}M_{\odot}/\rm{pc}^3$ with $r_v=0.8$ pc and $\gamma =2.01$ (versus $\rho_0 \sim 10^{8}M_{\odot}/\rm{pc}^3$ when $\gamma =2.5$).  Since even the densest nuclear star clusters observed have inferred central densities of $\sim 10^7M_{\odot}/\rm{pc}^3$ \citep[e.g.,][]{1998AJ....116.2263L,2018A&A...609A..27S}, we elect to truncate our catalog such that all clusters with initial $\gamma$ values between 2.01 and 2.5 are generated at exactly $\gamma=2.5$.   While the virial radius truncation only affects 17\% of clusters, 57\% of clusters initially had $\gamma<2.5$.  This artificial truncation means that our evolved cluster population may significantly under-predict the number of stellar mergers and runaway collisions that occur at early times \cite[e.g.][]{2004Natur.428..724P}, thereby underestimating the number of massive black holes formed \citep[e.g.,][]{2020ApJ...903...45K,2021ApJ...908L..29G}.  However, because our cluster population also has higher stellar metallicities than the aforementioned studies, any massive stellar merger products are greatly reduced by stellar winds; for example, the aforementioned massive cluster undergoes a runaway stellar merger, producing a $\sim 1000M_{\odot}$ star.  But the strong stellar winds driven by higher stellar metallicities reduce the star's mass by nearly 90\%, yielding a black hole with mass $< 100 M_{\odot}$.  We also truncate the upper metallicity of the clusters to 1.5$Z_{\odot}$ (effecting 15\% of the catalog), since our stellar evolution prescriptions are not valid above this metallicity in the original models presented in \cite{Hurley2000}.  Second, an error was discovered between our initial development of the catalog in \citetalias{GBOF} and the current published version in \citet{Grudic2021} which caused incorrect metallicities to be used when determining the distribution of initial radii for the clusters we evolve here.  While our distribution of half-mass radii only depends very weakly on metallicitiy \citepalias[$r_h\propto\left(\frac{Z_{\rm GMC}}{Z_{\odot}}\right)^{1/10}$, see Equation 6 of][]{GBOF}, this error causes our initial half-mass radii to be at most $\sim 1.5$ times smaller (in the worst case) than the actual catalog presented in \citetalias{GBOF} (where the error was corrected).  Due to computational requirements, it is prohibitive to rerun the entire catalog, so instead we assign to each cluster a weight defined by the ratio of the probability of a given cluster radius in the correct distribution to the probability of that radius in distribution it was drawn from (i.e., $w = p_{\rm correct}(r_v) / p_{\rm original}(r_v)$).  These weights serve to essentially resample our results presented here according to the correct catalog distributions.   The distribution of the weights themselves (i.e.~how much correction is required) has a median of 0.92, with 90\% of weights lying between 0.44 and 1.94, suggesting that our initial cluster population is not substantially effected by this discrepancy.  Finally, because we have only selected 895 of the total 3165 clusters with $>6\times10^4M_{\odot}$, we also multiply each of these weights an additional factor of 3.54; these weights are then used to compute all of the distributions and fractions we quote here, but note that this weighting is \emph{not} applied to scatter plots.

\subsection{Mass Loss from Time-Varying Galactic Tidal Fields}
\label{ss:tides}

  \begin{figure}
\centering
\includegraphics[scale=1.]{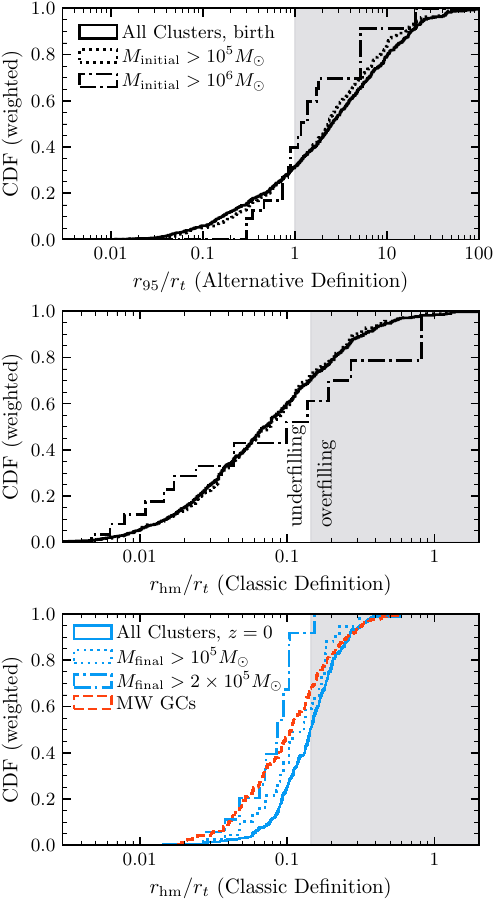}
\caption{The tidal filling of the cluster population both at birth and at $z=0$ in the \texttt{m12i} galaxy.   In the \textbf{top} panel, we show the cumulative distribution of tidally filling clusters using our alternative definition (where a cluster is overfilling if its 95\% Lagrange radius is larger than its initial tidal radius) at birth across all redshifts, across all clusters (solid line) and for clusters with masses greater than $10^5M_{\odot}$ and $10^6M_{\odot}$ in dotted and dashed-dotted, respectively.  Note that this definition (which is more appropriate for the EFF profiles our clusters are initially sampled from) suggests that nearly 70\% of clusters in our catalog are initially overflowing their tidal boundaries.  In the \textbf{middle} panel, we show the same distribution using the classic definition of tidal overfilling, $r_{\rm hm}/r_t>0.145$ (following \citealt{Henon1961}), where $r_{\rm hm}$ and $r_t$ are the half-mass and tidal radii of the clusters, respectively.  In the \textbf{bottom} panel, we show the same quantity, but for those clusters that survive to $z=0$.  We also show the distribution of $r_{\rm hm}/r_t$ for MW GCs from the catalog assembled in \citet{Baumgardt2017,Baumgardt2018,Vasiliev2021}.  The cumulative distributions are weighted following the discussion in \S \ref{ss:weights}.  The gray shaded regions in each plot indicate where clusters are overfilling, according to given definition.}
\label{fig:tides}
\end{figure}

As previously described, all star clusters will slowly shed their stars, either through natural evaporation at the high energy tail of the stellar distribution function, or through tidal stripping by the tidal fields of the cluster's host galaxy.   While these are often presented as separate processes, the mechanism is largely the same: stars still diffuse towards positive energies due to two-body relaxation, but in the presence of a galactic tidal field, the threshold for positive energy is decreased from $0$ to some $E_t$ which defines the Jacobi surface.  The only difference is that in the case of the tidal field, this zero energy surface is not spherically symmetric about the cluster, making it easier for a particle to escape in certain directions (e.g.~the L1 and L2 Lagrange points).  In general, the location of this boundary can be identified as the point where the acceleration from the cluster potential cancels that of the galactic tidal field:

\begin{equation}
    \frac{d^2 \mathbf{r'}}{dt^2} = - \nabla \phi_c(\mathbf{r'}) + \textbf{T}(\mathbf{r'})\cdot \mathbf{r'}~~,
    \label{eqn:tiderest}
\end{equation}

\noindent where $\phi_c$ is the gravitational potential of the cluster and $\mathbf{T}$ is the tidal tensor of the galactic potential $\phi_G$ at the point $\mathbf{r'}$, defined as

\begin{equation}
    \mathbf{T}^{ij}  \equiv - \left(\frac{\partial^2 \phi_G}{\partial x^{i} \partial x^{j}}\right)_\mathbf{r'}~~.
    \label{eqn:tidetensor}
\end{equation}

Equation \eqref{eqn:tiderest} describes the forces experienced by a star in the cluster in the rest frame of the galaxy.  However, what we are interested in is the force in the rest frame of the cluster.  This transformation to a rotating reference frame naturally introduces additional pseudo-forces (Coriolis, centrifugal, etc.) that depend on both the orbit of the cluster in the galaxy and the 3D structure of the cluster with respect to the galactic orbit; see, e.g., \citet{Renaud2011}.  Because \texttt{CMC} assumes clusters to be spherical anyway, we only need the spherical average of the cluster tidal boundary.  We use the prescription from Appendix C of \citet{Pfeffer2018}, and define the \emph{effective} tidal strength as $\lambda_{1,e} \equiv \lambda_1 - 0.5(\lambda_2+\lambda_3)$, where $\lambda_{1,2,3}$ are the Eigenvalues of the tidal tensor (Equation \ref{eqn:tidetensor}) sorted from largest to smallest.  Diagionalizing the tidal tensor transforms it into the rotating frame of the cluster, where $\lambda_1$ is the gravitational force along the vector pointing from the cluster center to the galactic center, and $0.5(\lambda_2+\lambda_3) = \mathbf{\Omega}^2$ is the centrifugal force of a circular orbit in a spherical potential \citep[in a true spherical potential $\lambda_1 = \lambda_2 = \mathbf{\Omega}^2$, see ][Appendix C]{Pfeffer2018}.  The instantaneous tidal radius of the cluster is then given by \citep[e.g.,][]{Renaud2011}:

\begin{equation}
r_t = \left(\frac{G M_c}{\lambda_{1,e}}\right)^{1/3}
\label{eqn:rt}
\end{equation}

\noindent where $M_c$ is the cluster mass.  With Equation \eqref{eqn:rt}, we can apply a time-dependent tidal boundary to our \texttt{CMC} clusters, so long as we can calculate $\lambda_{1,e}$.

For each cluster from the \citetalias{GBOF} catalog, we identify a single star particle in the cosmological simulation that was associated with its parent GMC in the \texttt{m12i} galaxy and use it as a tracer particle for the cluster's trajectory from birth until either $z=0$ or cluster destruction. With the exception of dynamical friction (discussed momentarily), the trajectory of the clusters is fully-resolved by the cosmological simulation --- the tracer particles have a mass resolution of $\approx 7 \times 10^3\,M_\odot$ compared to the cluster mass range of $10^4-10^7\, M_\odot$.  We then extract from every \texttt{m12i} snapshot\footnote{With the exception of the \emph{first} \texttt{m12i} snapshot, which still contains the tidal contribution of the cluster's birth GMC.  In practice, we use the tidal tensor from the second snapshot for both initial } the second derivatives of the local galactic potential (Equation \ref{eqn:tidetensor}; the potential itself is stored in the simulation output) about that star particle, as well as the local position, velocity, enclosed galactic mass, and local velocity dispersion at that point in the \texttt{m12i} galaxy (used in computing the dynamical friction timescale described below).  This procedure is similar to that we used for \textit{Behemoth}, the largest cluster from this catalog \citep[previously described in][]{Rodriguez2020}.  We show an example of the position and tidal forces experienced by typical clusters (that survives to $z=0$) in Fig.\ \ref{fig:typical}.  Once we have extracted the tidal tensor, it is passed as input to \texttt{CMC}, which then diagionalizes the tidal tensor and linearly extrapolates in time the value of $\lambda_{1,e}$ between the snapshots of the \texttt{m12i} model.  The instantaneous tidal boundary is computed using Equation \eqref{eqn:rt}.  Each timestep, \texttt{CMC} strips any star whose orbital apocenter is greater than $r_t$.  Note that this is different from previous \texttt{CMC} models and other Monte Carlo codes \citep[e.g.][]{giersz2008}, which used a stripping criterion based on the potential energy of the cluster at $r_t$; however, we have found that our choice better replicates the mass-loss rates for star clusters on eccentric orbits  (see Appendix \ref{apx:stripping}).  Because the galactic potential (and therefore the tidal tensor) is calculated on the same scale as the typical inter-particle separation \citep[$\sim 6.5$~pc for this simulation; see][\S 4.2]{Hopkins2018}, we are able to resolve nearly all relevant physical structures that can significantly influence the tidal field of our GCs.

While our approach allows us to model the effect of arbitrary tidal fields on our GC models, what it does \emph{not} capture is the affect on the cluster's orbit in the galaxy due to dynamical friction.  Of course, these two effects are not independent: as dynamical friction shrinks the orbit, the tidal fields tend to become stronger toward the denser galactic center.  In turn, stronger tidal fields strip more stars, causing the cluster to lose mass and dynamical friction to become less efficient.   These effects are particularly difficult to model within gravitationally-softened cosmological simulations, where the star particles we use to trace the clusters' orbits have similar masses to other particles \citep[see e.g.,][for similar issues relating to supermassive BHs]{2015MNRAS.451.1868T,2021MNRAS.508.1973M}.

While we cannot self-consistently calculate new cluster orbits in the cosmological simulation during a \texttt{CMC} integration, we can calculate the time it would take for dynamical friction to drive our clusters into the galactic center \cite[again following][]{Pfeffer2018}.  We use the dynamical friction timescale from \citet{1993MNRAS.262..627L}:

\begin{equation}
T_{\rm df} = \frac{\epsilon^{0.78}}{2 B\left(v_c / \sqrt{2}\sigma\right)} \frac{\sqrt{2} \sigma r_{\rm circ}^2}{G M_c \log \Lambda}~~, 
\label{eqn:tdf}
\end{equation}
  
 \noindent where $r_{\rm circ}$ is the radius of a circular orbit with the same energy E as the actual test particle, $v_c$ is the circular velocity at that radius, $M_c$ is the cluster mass,  $B(x) = \rm{erf}(x) - 2x\exp(-x^2)/\sqrt{\pi}$ is the standard velocity term for dynamical friction \citep[][]{Binney2011}, $\epsilon$ is the ratio of the angular momentum of the real orbit to that of the circular orbit at $r_c$ \citep[to correct for the effects of eccentricity,][]{1993MNRAS.262..627L}, and $\log \Lambda$ is the Coulomb logarithm, defined here as $\Lambda = 1+M_c/M_{\rm enc}$ for an enclosed mass $M_{\rm enc}$.  We pass as input to \texttt{CMC} the cluster's position, velocity, enclosed galactic mass, and local velocity dispersion, and calculate Equation \eqref{eqn:tdf} every timestep.  Because $T_{\rm df}$ can change dramatically over a single cluster orbit, we integrate each cluster until a single $T_{\rm df}$ has elapsed:
  
 \begin{equation}
 \int \frac{dt}{T_{\rm df}} > 1~,
 \label{eqn:intdf}
 \end{equation}
 
 \noindent using the same orbits extracted in the previous section.  Once \eqref{eqn:intdf} is satisfied, we assume the cluster has been destroyed. Note that this is different than the approach used in \cite{Pfeffer2018}, where a cluster is assumed to have spiraled into the galactic center when its age is greater than $T_{\rm df}$.  We explore the implications of Equation \eqref{eqn:intdf} in \S \ref{ss:comp}.
  
 Finally, we do not include any prescription for the work done by the time-dependent tidal field on cluster itself.  This periodic injection of energy, known as \emph{tidal shocking}, has been argued to have significantly shaped the evolution of the cluster mass function in the MW and other galaxies, particularly for clusters with lower masses and larger radii \citep[e.g.,][]{Spitzer1958,Ostriker1972,spitzer_dynamical_1987}.  This process is particularly difficult to implement successfully in a Monte Carlo code such as \texttt{CMC}, where the assumptions of spherical symmetry, virial equilibrium, and a timestep that is a fraction of the cluster's relaxation time explicitly preclude including processes that occur on a dynamical timescale or require the 3D positions and velocities of the stars \citep[though see][]{2014MNRAS.443.3513S}.  However, we can estimate the effect that tidal shocking would have had on the mass-loss rate of our clusters using a similar technique to \citet{Pfeffer2018}.  We do so in Appendix \ref{apx:shocking}, and find that $>99\%$ of our clusters we estimate to be destroyed by tidal shocking are already destroyed by time-dependent mass loss from tidal stripping.

\section{Initial Cluster Population}
\label{sec:initial}

It is interesting to compare the tidal truncation of our cluster catalog to the tidal radii and truncation of clusters in the MW.  The cluster formation model from \citetalias{GBOF} contains no explicit information about the local tidal field where the clusters form external to the progenitor cloud, since the cloud collapse simulations from \citet[][]{Grudic2021} were performed in isolation. As a result, star clusters in our model can occasionally be tidally overfilling immediately after formation, which does represent a significant departure from previous $N$-body studies of star clusters, where clusters are assumed to be (sometimes significantly) tidally underfilling at birth.  This is largely based on the argument that any tidally limited clusters we see today will have expanded over time to fill their Jacobi radii, and for clusters born filling their tidal boundaries, this early phase of expansion would likely lead to rapid destruction by the galactic tidal field \citep[][]{2011MNRAS.413.2509G}.  This argument has also been used to argue \emph{against} clusters being born with a significant degree of primordial mass segregation \citep[][]{2008ApJ...685..247B,2009ApJ...698..615V}, since such segregation would only increase the cluster expansion during this early phase.


In Fig.\ \ref{fig:tides}, we show the cumulative fraction of clusters that are tidally filling at birth and at $z=0$.  Following \citet{Henon1961} we define tidally 
overfilling clusters as those where the ratio of the half-mass to tidal radii, $r_{\rm hm} / r_t$, is greater than 0.145, with $r_t$ defined as the location of the outermost star in the cluster.  Using this definition, approximately 25\% of our initial cluster population is tidally overfilling at 
birth.  There is also a weak trend of the most massive clusters being more overfilling, with $\sim 40 \%$ of clusters with initial masses $> 10^6M_{\odot}$ 
overfilling their tidal boundaries initially.  However, we note that the value of 0.145 is taken from the equal-mass homological model 
presented in \citet{Henon1961}, and while we have used it here to make it easier to compare our results to the pre-existing literature, there is no reason that 
this value should apply to the non-equilibrium EFF profiles with realistic stellar mass distributions.  A more straightforward statistic would be the ratio of 
an outer Lagrange radius enclosing some large fraction of the total mass to the tidal boundary.  We show the cumulative distribution of $r_{95}/r_t$, where $r_{95}$ is the radius enclosing 95\% of the mass, in the top-right panel of Fig.\ \ref{fig:tides}.  Here, the situation is reversed: nearly 70\% of our clusters have $r_{95} > r_t$ initially, 
suggesting that the EFF profile generates more tidally overfilling clusters than predicted using the statistic taken from equal-mass homologeous models.  Furthermore, 
the more massive clusters tend to be \emph{less} overfilling under this definition than their lower-mass counterparts.

Observations of YMCs in the 31 galaxies of the Legacy Extragalactic UV Survey suggest a typical mass-radius relationship of the form $r_{\rm hm} \propto M_{\rm cl}^{1/4}$ \citep{brown:2021.cluster.mass.radius}, though with significant variation in the slopes between different galaxies (G. Brown, private comm.).  But the cluster 
tidal radii scale as $M_{\rm cl}^{1/3}$ (Equation \ref{eqn:rt}), which would suggest that, all other things being equal, the cluster filling fraction 
should very weakly ($r_{\rm hm}/r_{t}\propto M_{\rm cl}^{-1/12}$) depend on mass, with more massive clusters being very slightly more likely to overfill their tidal boundaries at birth.  This is seen in the top two panels of Fig.\ \ref{fig:tides}, where that the dependency of $r_{\rm hm} / r_t$ on mass is weak, but not consistent between the classic and alternative definitions of tidal filling.  This is also consistent with the initial cluster catalog from \citetalias{GBOF}, which exhibits a  $r_{\rm hm} \propto M_{\rm cl}^{1/4}$ scaling \citep[Fig.~10;][Section 3.3]{Grudic2021} globally, but with significant variation when binned by to the local gas surface density where each cluster was formed.  When considering clusters born in environments with similar gas densities, the mass-radius relations follow the $M_{\rm cl}^{1/3}$ scaling expected for clusters forming at constant density (albeit with different multiplicative coefficients), leaving $r_{\rm hm} / r_t$ with no dependence on cluster mass. It is only after stacking these bins together that the global mass-radius relation follows the aforementioned $M_{\rm cl}^{1/4}$ scaling.  From this, we conclude that the initial fraction of clusters that are overfilling is largely the result of the local galactic environment at formation, rather than any global trend in cluster formation.

In the bottom panel of Fig.\ \ref{fig:tides}, we show the distribution of surviving clusters that are tidally filling at the $z=0$ snapshot of the 
larger simulation (limiting ourselves to the classic $r_{\rm hm}/r_t$ definition for easier comparison to observations).  Here, we find that the distribution of 
clusters is exactly centered at $0.145$, as one would expect for a population of clusters that is both dynamically evolved (i.e.~closer to the original 
homologeous model of H\'enon) and in equilibrium with their surrounding environments.  We also compare our results to the MW GCs, using the tidal boundaries and half-mass radii from the catalog assembled in \citet{Baumgardt2018,Vasiliev2021}, which rely on a scaled grid of direct $N$-body models described in \citet{Baumgardt2017} and define $r_t$ using Equation 8 of \citet{Webb2013}.  Somewhat surprisingly, the distribution of MW GCs appears to be more tidally underfiling than the catalog clusters at $z=0$ (with only $\sim 1/3$ of the MW clusters overfilling their tidal boundaries).  However, we note that measurements of cluster tidal boundaries are extremely difficult; even the $N$-body models used by \citet{Baumgardt2017} to compare to the MW GCs were performed in isolation, making it difficult to uniquely identify a tidal boundary by comparison to $N$-body models.  Furthermore, while the overall population of cluster models at  $z=0$ seems to diverge from the MW distribution, we note that the cluster models with final masses $> 10^5M_{\odot}$ appear more consistent with the MW distribution, while clusters with final masses $> 2\times10^5M_{\odot}$ appear to match very well (for low $r_{\rm hm} / r_t$ values) with the MW population, suggesting that any disagreement may actually be the result of our lower-mass GCMF.

\section{Present-day Cluster Properties}
\label{sec:presentday}

In addition to masses and other global properties, our star-by-star $N$-body models allow us to compare the present-day structural parameters of our clusters to 
those in the MW and M31 as well.  In order to create as honest a comparison as possible, we generate for every cluster that survives to the present day an 
orbit-smoothed synthetic surface brightness profile (SBP) using the \texttt{cmctoolkit} \citep{2021zndo...4579950R,Rui2021}.  Each SBP is generated by binning the cluster into 80 
logarithmic bins of projected radii between 0.01 pc and the outermost star in the (projected) cluster profile, then applying a generic Johnson V-band filter to the cumulative light 
emitted in each bin (assuming each star to have a blackbody spectra).  GC and YMC observations typically apply magnitude cuts when creating SBPs, to ensure that the profiles are robust tracers of the underlying cluster structure, and to remove any saturated stars from the observation \citep[this is particularly important for younger clusters, e.g.,][]{1989ApJ...347L..69E,Mackey2003a}.  We apply a similar cut in luminosity by removing any star from our SBPs that contributes more than $0.5\%$ of the total luminosity of the cluster.  For most clusters, this typically cuts much less than 1\% of the stars (the exception being very small clusters which are near dissolution, for which SBPs would be extremely noisy anyway).  Throughout this paper, we distinguish between the half-light radius, $r_{\rm hl}$, which is the projected 2D radius enclosing half the visual luminosity, and the half-mass radius $r_{\rm hm}$, which is the 3D radius enclosing half the mass.  To calculate other structural parameters, such as the core radius, $r_c$, and the concentration parameter (defined as $c=\log_{10}(r_t/r_c)$, where $r_t$ is the tidal boundary), we fit each SBP to a projected \cite{King1966} profile.  The $r_c$ and $r_t$ parameters are taken directly from these fits (including the aforementioned luminosity cuts).

We also compile a series of observational catalogs to compare our results to.  For the masses of the MW GC systems, we use the V-band magnitudes from the 
\citet[][2010 Edition]{Harris1996} catalog and convert them to masses assuming a mass-to-light ratio of 2 for old stellar systems \citep{Bell2003}.  The 
metallicities, core radii, half-light radii, and King concentrations $c$ are also taken from the Harris catalog.  For the ages of the MW clusters, we use the 96
cluster ages compiled by \cite{Kruijssen2019} from the surveys presented in \citet{Forbes2010,Dotter2010,Dotter2011,Vandenberg2013}.  For the M31 GCs, we take 
the masses, ages, and metallicities from the \citet{Caldwell2011} catalog, while the King concentrations, core, and half-light radii are taken from the 
\citet{Peacock2010} catalog.  Note that the core radii are not present in the latter, but we calculate them using the provided tidal radii and King 
concentration parameters.  

Finally we must address what we actually define as a ``globular'' cluster.   While our cluster catalog is generated from initial conditions for massive spherical star clusters, they are not uniquely the old, low-metallicity clusters that we typically call ``globular'' clusters.  As stated in \citetalias{GBOF}, this galaxy does not produce massive star clusters sufficiently early ($z\gtrsim 3$) to reproduce the classical GCs of the MW.  Observed in another galaxy, many of our clusters would instead be identified as open clusters or super star clusters.  By limiting our comparison to the old clusters of the MW and M31, we are largely ignoring the contribution of lower-mass clusters (for which complete catalogs are harder to assemble).  Following previous studies \citep[e.g.~the E-MOSAICS simulation][]{Pfeffer2018}, and using the youngest GC age in M31 as a guide, we define any cluster formed more than 6 Gyr ago as a GC, while referring to clusters formed less than 6 Gyr ago as young clusters.  Finally, we restrict ourselves to clusters that lie within the main MW-like galaxy of the FIRE-2 simulation, which we define to be those within 100\,kpc of the galactic center at $z=0$.

\subsection{Cluster Mass Functions}
\label{ss:masses}

  \begin{figure}
\centering
\includegraphics[]{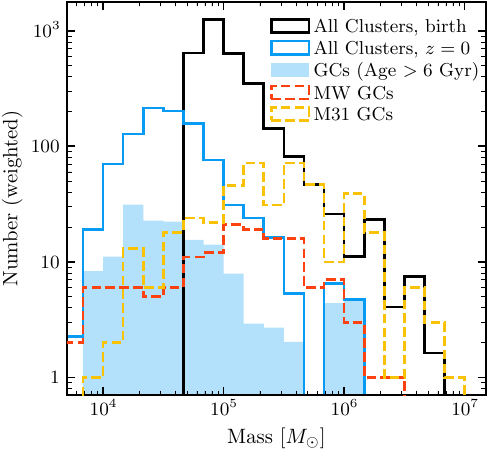}
\caption{The globular cluster mass function of our CMC-FIRE cluster system.  In black, we show the distribution of birth masses of our 895 clusters (across all 
formation redshifts), while in blue we show the population of clusters that survive to the present day.  Both distributions are weighted according to the 
description in \S \ref{ss:weights}.  We also show the masses of clusters older than 6 Gyr, which we term classical GCs, in solid blue.  For comparison, we show the masses of 
GCs in the MW \citep[][2010 Edition]{Harris1996} and M31 \citep{Caldwell2011} in dashed red and yellow respectively.}
\label{fig:gcmf}
\end{figure}

  \begin{figure}
\centering
\includegraphics[]{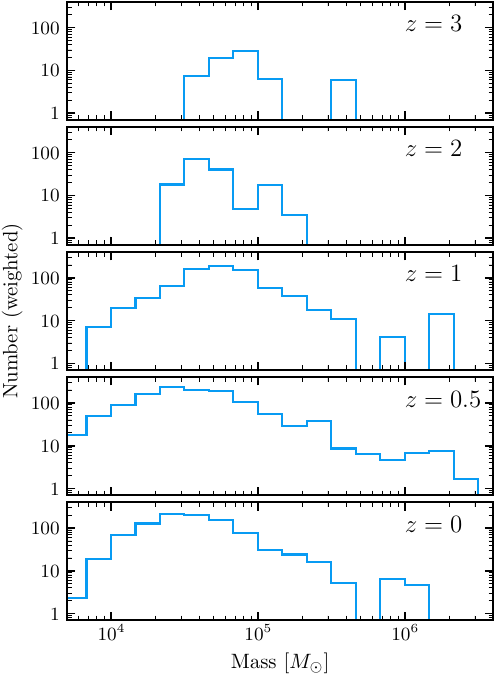}
\caption{The CMF of our CMC-FIRE cluster system in the main \texttt{m12i} galaxy at various redshifts.  Note that this includes all clusters, not just classic GCs.}
\label{fig:gcmfz}

\end{figure}

 \begin{figure*}
\centering
\includegraphics[scale=1.]{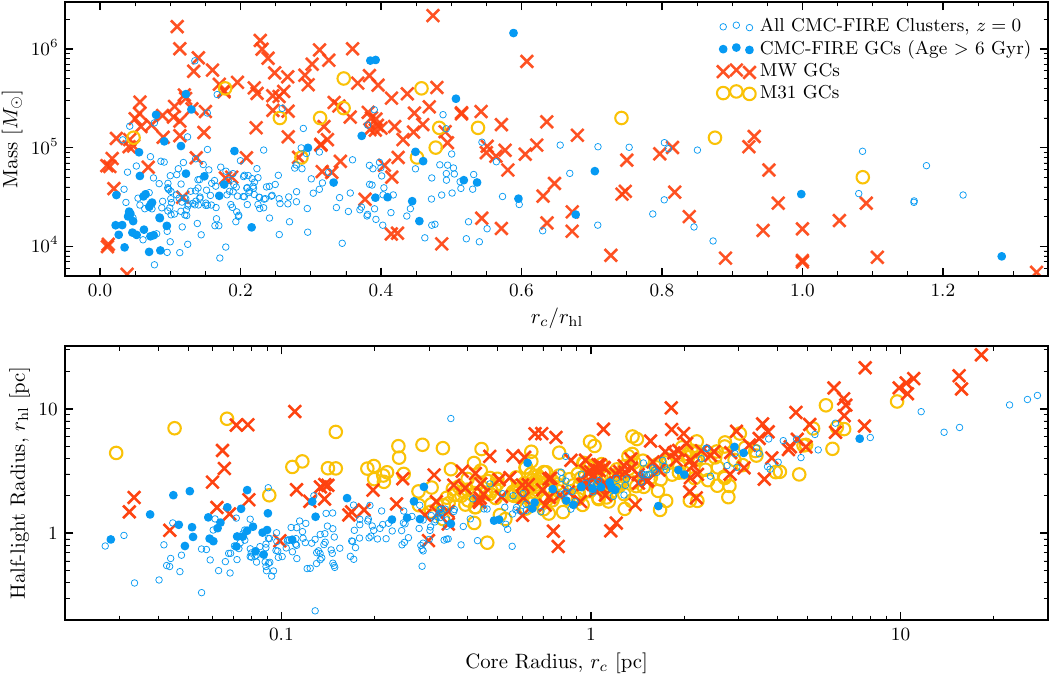}
\caption{The masses and radii of our surviving 296 (unweighted) cluster models compared to the clusters in the MW and M31.  The CMC-FIRE GCs (those older than 6 Gyr) are indicated with blue dots, while younger clusters are indicated with open blue circles.  The MW GCs are shown in red x's \citep[from the][catalog]{Harris1996} and the M31 clusters in yellow circles \citep[from the][catalog]{Peacock2010}; note that the M31 catalog does not contain masses, so we cross reference the entries there with the entries from the \citet{Caldwell2011} catalog (only a handful of clusters appear in both).   On the \textbf{top}, we show the ratio of the observational core radius $r_c$ to the half-light radius $r_{\rm hl}$ versus the cluster mass.  On the \ textbf{bottom}, we show $r_c$ and $r_{\rm hl}$ separately for all three cluster populations.  While in both cases, the CMC-FIRE catalog overpredicts the number of low-mass, compact GCs compared to both the MW and M31 (for the reasons described in Section \ref{ss:masses}), the catalog is largely able to cover the entire range of observed cluster structural parameters.}
\label{fig:mvrcrh}
\end{figure*}

  \begin{figure*}
\centering
\includegraphics[scale=1.]{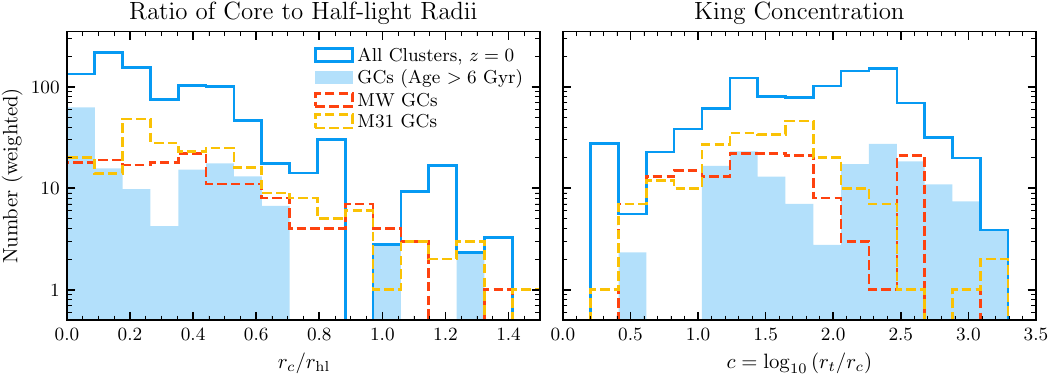}
\caption{The projected distribution of concentrations for our surviving cluster models (in blue), with the population of old ($>$6 Gyr) GCs in solid blue, compared to the concentrations of clusters in the MW (in dashed red).  On the \textbf{left}, we show the distribution of $r_c/r_{\rm hl}$ for both populations (the same as the horizontal axis of the top panel of Fig.\ \ref{fig:mvrcrh}).  On the \textbf{right}, we show the histogram of the King concentration parameter, defined as the logarithm of $r_t/r_c$, where $r_t$ is the tidal radius.  The distribution for the MW $r_c/r_{\rm hl}$ and King concentrations are taken from \citet[][2010 edition]{Harris1996}, while for M31 the distributions for taken from the \citet[][]{Peacock2010} catalog (note that the latter does not contain $r_c$, so we reconstruct it from the provided values of the King concentration and tidal radii).  Finally, note that the peak at $c=2.5$ in the \citet[][2010 edition]{Harris1996} catalog comes from \citep{1995AJ....109..218T}, where all core collapse clusters are assigned concentrations of exactly 2.5.}
\label{fig:conc}
\end{figure*}

There are multiple physical effects, both internal physical processes and external galactic influences, that determine the evolution of the CMF over cosmic time.  Immediately after formation, clusters lose a significant amount of mass through the evolution of massive stars.  For our choice of IMF and median cluster metallicity ($\sim 0.6Z_{\odot}$), nearly 20\% of the stellar mass is lost to winds and supernova within the first 50 Myr of evolution.  This mass loss causes the cluster to expand as it becomes less gravitationally bound.  For tidally filling clusters, this means that the outer regions of the cluster can suddenly find themselves beyond the tidal boundary, where they are stripped away by the galactic tidal field.  For the typical GCs shown in Fig.\ \ref{fig:typical}, the combination of mass loss and stripping conspire to decrease the cluster mass by nearly a factor of two within the first 50 Myr!  

After this initial phase of rapid evolution is complete, a typical star cluster begins a slow evolutionary phase, where mass loss is primarily driven by two-body relaxation in the forms of evaporation and tidal stripping.  In Fig.\ \ref{fig:gcmf}, we show the evolution of the CMF from birth to the present day, where the physical processes described in \S \ref{ss:tides} conspire to shape our $z=0$ mass function.  Limiting ourselves to those clusters older than 6 Gyr, it is immediately obvious that our $z=0$ CMF does not match that of either the MW or M31 GCs in the low-mass range, as our catalog contains significantly more clusters with $M \lesssim 10^5M_{\odot}$.  The median of our GCMF is at $\sim 3\times10^4 M_{\odot}$, significantly lower than the $\sim 10^5 M_{\odot}$ median of the MW GCs.  At the same time, the CMC-FIRE catalog shows good agreement for the numbers of the most massive ($\sim 10^6M_{\odot}$) GCs, as well as the number of clusters around the median of the MW GC mass function.  Furthermore, the \emph{overall number} of GCs in our (weighted) CMC-FIRE catalog is 148, astonishingly close to the actual number of GCs in the MW \citep[157,][catalog, 2010 edition]{Harris1996}.  Including the young clusters (those younger than 6 Gyr), the model predicts 941 massive clusters present in the \texttt{m12i} galaxy at the present day.

There are likely two reasons for the discrepancy in our GCMF, the first being the formation time of clusters (even those defined as GCs) within the \texttt{m12i} galaxy.  The median age of the surviving GCs in our CMC-FIRE catalog is 7.9 Gyr (versus 2 Gyr for the young clusters), in stark contrast to the $\sim12$ Gyr age of GCs in the MW \citep{Kruijssen2019}.  This is shown in Fig.\ \ref{fig:gcmfz}, where we show the CMF of our cluster population at various redshifts.  Although there is some early cluster formation at high redshifts, the majority of clusters are not present until $z\sim1$.  The number of low-mass clusters significantly increases from $z=1$ to $z=0.5$, as the massive clusters formed at $z=1$ lose mass and fill out the lower regions of the mass function.   Somewhat counter-intuitively, we will argue that these clusters lose mass at a rate that is, on average, faster than clusters formed earlier in the galaxy (c.f.~Fig.\ \ref{fig:l1e}, \S \ref{ss:metal} and \ref{ss:comp}).  This is because, in agreement with previous studies \citep[e.g.][]{Li2019,Meng2022}, we find that clusters formed earlier experience weaker tidal fields on average, since they are characteristically formed on wider orbits that are less likely to be aligned with (or lying within) the galactic disk. As we will argue, this is because most GCs are either formed during the early bursty phase of star formation in the main galaxy, when the gas has a quasi-isotropic velocity distribution  \citep[e.g.,][and \S \ref{ss:metal}]{2022arXiv220304321G}, or formed \textit{ex situ} in dwarf galaxies that are accreted by the main galaxy.  Because of this, the majority of our old, massive GCs lose more mass and are more likely to be destroyed than had they formed at an earlier epoch.  

The second reason for this discrepancy is likely the our prescription for following the effects of tidal shocking on the GC population.  As described in Appendix \ref{apx:shocking}, we find that \emph{none} of the clusters in our sample of 895 models have been significantly affected by tidal shocking over their evolution.  This is largely because the tidal tensors are calculated from the snapshots of a cosmological FIRE simulation, typically spaced every $\sim 20$ Myr in cosmic time.  However, the effectiveness of a tidal shock depends on the amount of work the tidal field is able to do on the cluster's dynamical timescale, since otherwise the cluster can undergo slow, adiabatic changes that do not significantly increase the escape rate of stars \citep[though see][for  cases where the injection of energy can be significantly more destructive]{1994AJ....108.1398W,weinberg2,weinberg3}.  Because the median dynamical time of the clusters in our catalog is on the order of $\sim 1$ Myr, well below the time resolution of tidal forces extracted from the FIRE simulation, none of our cluster would be significantly affected by tidal shocks, even if the physics had been incorporated in \texttt{CMC}.  This timestep resolution may be particularly problematic for resolving encounters with GMCs \citep[e.g.,][]{Gieles2006,Linden2021} or transits through the galactic disk \citep[e.g.,][]{Ostriker1972}, which may cause significant tidal shocks within a few Myr; see \S \ref{ss:morph}.  This means that we cannot resolve the injection of energy and subsequent expansion of the lowest mass clusters, leaving us with an excess of low-mass GCs compared to the MW or M31.  Combined, these two facts conspire to produce a lower median value for our GCMF, despite the agreement between the number of GCs in the MW and \texttt{m12i}.

As for the young clusters, the model prediction of 793 young massive clusters is in stark contrast to the observational picture in the MW, where only $\mathcal{O}(10)$ such young clusters are known \cite[e.g.,][Table 2]{PortegiesZwart2010}. However, a large part of this disagreement is definitional: restricting ourselves to clusters with ages $\lesssim 20$~Myr and $\geq 10^4M_{\odot}$ (as in the previous citation) yields no young clusters at $z=0$ in the \texttt{m12i} galaxy, while adopting a more loose age cutoff of $\lesssim 100$~Myr yields approximately 6 clusters.  What this suggests is that our cluster evolution model overpredicts the presence of \textit{intermediate}-aged massive clusters (between a few 100 Myr and a few Gyr), compared to the MW.  However, there are observations of young and intermediate-aged clusters in other galaxies, both within and beyond the Local Group \citep[e.g.,][]{Larsen2010,Richtler2012}, including other spiral galaxies \citep{Larsen1999,Larsen2004}.  As we describe in \S \ref{ss:metal} and \ref{ss:morph}, many of the discrepancies between our \texttt{m12i} galaxy and the MW may arise from the unique assembly history of both systems.

\begin{figure*}
\centering
\includegraphics[scale=1.]{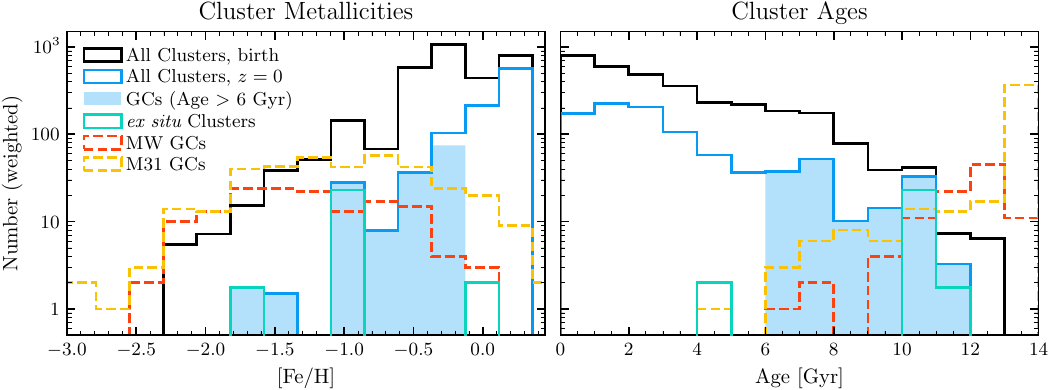}
\caption{The distribution of metallicities and ages for our initial and final population of clusters.  On the \textbf{left}, we show the cluster metallicities.  
Unlike both the MW and M31, our clusters are heavily weighted towards higher metallicities (though our GCs are still lower metallicity than the cluster population as a whole).  This is especially true for \textit{ex situ} clusters accreted from infalling dwarf galaxies (which we show in pale green). On the \textbf{right}, we show the distribution of cluster ages.  
As with our distribution of metallicities, the clusters in our catalog are significantly younger than the old GCs in both the MW and M31.}
\label{fig:metal}
\end{figure*}

\subsection{Cluster Sizes}

In Fig.\ \ref{fig:mvrcrh}, we show the observed radii of our clusters compared to those in the MW and M31.  On the top panel, we show the distribution of 
clusters in mass vs $r_c/r_{\rm hl}$ space (the latter being a proxy for the overall concentration of the cluster).  Our CMC-FIRE clusters largely cover the space of 
MW and M31 clusters, suggesting that our distribution of initial conditions is sufficiently broad to reproduce the range of observed GCs in the local universe.  
The same is true on the bottom panel of Fig.\ \ref{fig:mvrcrh}, where we show separately the values of the half-light and core radii.  The bulk of clusters in 
the MW, M31, and \texttt{m12i} galaxies lie along the same relation in $r_c$/$r_{\rm hl}$ space.

As with the mass distribution, both plots in Fig.\ \ref{fig:mvrcrh} show an excess of compact, low-mass clusters in the \texttt{m12i} galaxy as compared to the MW or M31.  This likely arises from the aforementioned coarse-grained treatment of tidal shocking and the fact that, counter-intuitively, our younger clusters undergo core collapse faster because of their high stellar metallicities (\S \ref{ss:bhs}). Of course, Fig.\ \ref{fig:mvrcrh} only shows the results of our catalog directly, without accounting for our weighting scheme or the fact that we only modeled $\sim$ 28\% of the clusters from the catalog.  In Fig.\ \ref{fig:conc}, we show the (weighted) distributions of both $r_c/r_{\rm hl}$ and the King concentration parameters.  Both the numbers and general shape of the distributions are largely correct for both the $r_c/r_{\rm hl}$ concentration distributions.  When considering all cluster ages, the CMC-FIRE system slightly over-predicts the number of clusters with large $r_c/r_{\rm hl}$  ratios. On the other hand, if we restrict ourselves to old, classic GCs, we see significantly fewer clusters with large $r_c/r_{\rm hl}$ or small $c$ values.  As we will show in \S \ref{ss:bhs}, this is largely because our clusters have higher stellar metallicities than those in the MW or M31, which in turns drives them to core collapse earlier than their low-metallicitiy counterparts.


  \begin{figure}
\centering
\includegraphics[]{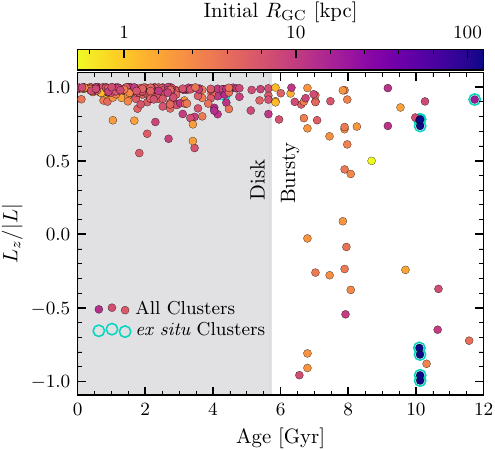}
\caption{The relationship between the ``co-rotation'' of the surviving clusters' orbits at $z=0$ and their age.  Here, we define co-rotation as the fraction of the cluster's angular momentum about the galaxy that is aligned with the rotation of the galactic disk.  Consistent with clusters in our own Galaxy, the youngest clusters preferentially rotate with the disk, while older clusters tend to be isotropically distributed.  Finally, we indicate those clusters that were originally formed outside of the main galaxy with blue-green circles.  These clusters are typically older, more metal poor, and isotropically distributed, consistent with their accretion history onto the main galaxy.  We also show in shaded regions the transition from bursty star formation to disk star formation \citep[which occurs as 6.15\,Gyr lookback time in the non-MHD \texttt{m12i} galaxy studied in][or approximately 5.7\,Gyr in the MHD \texttt{m12i} galaxy studied here]{Yu2021}.}
\label{fig:lz}
\end{figure}


\subsection{Ages, Metallicities, and Galactic Positions}
\label{ss:metal}

Arguably the largest discrepancies between our cluster population and observations is that the majority of our model clusters are significantly younger (and correspondingly higher metallicity) than the old GCs in the MW and M31.  This discrepancy is also readily apparently in the ages and metallicities of the clusters that survive to the present day, which we show in Fig.\ \ref{fig:metal}.  While there exist a handful of GCs with low metallicities ([Fe$/$H]$\sim -1.6$), similar to the classical ``blue'' GCs in the MW, the median GC metallicitiy for the CMC-FIRE catalog is -0.4, noticeably higher the median of -1.3 in the MW or -0.8 in M31.  

This behavior is a direct result of the initial conditions described in \citetalias{GBOF}, where the later formation of GCs (compared to those in the MW) is described in detail.  What is not immediately obvious is how much of this discrepancy lies in the star-formation history of the \texttt{m12i} galaxy (i.e.~whether there is some systematic trend against forming GCs early enough), and how much is simply a byproduct of the inherent galaxy-to-galaxy variation in merger history and cluster formation.  It has been shown \citep[e.g.,][Section 4.5]{Hafen2022} that the star-formation rate of FIRE-2 MW-mass galaxies can be $\sim 2$ times larger at $3-10M_{\odot}/\rm{yr}$ than observationally-based estimates \citep[$0.7-6M_{\odot}/\rm{yr}$,][]{Behroozi2013}.  Given the on average constant cluster-formation efficiency per stellar mass \citepalias[][Figure2]{GBOF}, this can certainly explain the young and intermediate age cluster described in \S \ref{ss:masses}.  At the same time, the large burst of massive cluster formation at $z\approx0.8$ is largely driven by a late galaxy merger at that redshift, in contrast to the assumed merger history of the MW \cite[where the last major merger was assumed to occur at $z\gtrsim1.5$,][]{2020MNRAS.494.3880B} (though we caution that not most starbursts occuring during the galaxy history are not associated with a merger).  In addition, it has been shown that MW-mass galaxies in Local Group environments form stars earlier, particularly at $z\gtrsim2$ \citep{santistevan:2020.fire.m12.sfh}, than isolated MW-mass galaxies.     Efforts to explore the formation of GCs in Local Group-like structures are currently underway.

Of course, one key defining feature of GCs in the MW is that they are largely found on extended orbits in the halo \citep[as opposed to younger open clusters that are preferentially found in the disk][]{PortegiesZwart2010}.  While our GCs are younger and higher metallicity than those in the MW, we do find a clear distinction between the disk-inhabiting young clusters and the more isotropically distributed GCs.  In Fig.\ \ref{fig:lz}, we show the relationship between the $z$ component of the clusters' orbital angular momentum about the galaxy (where the $z$ direction is defined as perpendicular to the galactic disk) and the age of the clusters.  While young (and high metallicity) clusters are largely co-rotating with the disk, the older clusters, formed during minor mergers and the earlier phases of galaxy assembly, appear to be on largely isotropic orbits, similar to the observed ``blue’’, metal-poor GCs in the MW.  While this galaxy may not be representative of the MW or M31 in terms of its star formation or assembly history, it is clear that the orbits of the surviving older/metal poor GCs in our model show a similar pattern \citep[e.g.,][]{Zinn1985}.

This pattern arises from two sources: first, we have limited our definition of GCs to those lying within 100\,kpc of the galactic center at $z=0$.  But this definition also includes many clusters that were formed in distant dwarf galaxies, only to later migrate inward as their hosts were accreted by the main \texttt{m12i} galaxy.  In the FIRE-2 simulations, every star particle is assigned a galaxy ID based on whichever galaxy the particle was closest to at its formation time.  We define any GC that has an initial galaxy ID different than the main galaxy and a galactocentric distance at $z=0$ of $<100$\,kpc as an \textit{ex situ} GC.  With our weighting scheme, this corresponds to 27 clusters, 25 of which are old GCs, or about 17\% of the GC population at $z=0$.  This is compatible with the low end of estimates for the fraction of accreted GCs in the MW \cite[$\sim$ 18-32\%][]{Forbes2010}.  Our \textit{ex situ} clusters are older ($\sim 10$ Gyr) and more metal poor ([Fe/H]$\sim$-1) than our other GCs, as is apparent in Fig.\ \ref{fig:metal}. This is in excellent agreement with the ages and metallicities of the accreted MW GCs ; however, in our case these clusters are older than the typical \texttt{m21i} GC, while in the MW these clusters occupy the younger of the two populations identified in Fig.\ 1 of \cite{Forbes2010}.

Second, Fig.\ \ref{fig:lz} shows that the \textit{ex situ} GCs are not the only ones on isotropic or even retrograde orbits about the galaxy.  Even the GCs formed at low initial $R_{\rm GC}$ in the main galaxy are on isotropic orbits.  This is because while the present day orbits of \textit{ex situ} clusters are set by the orbits of their infalling hosts, the orbits of the \textit{in situ} CMC-FIRE clusters are largely set by the morphology and dynamical state of the galaxy at the time that the clusters formed.  Approximately 6 Gyr ago, the MHD \texttt{m12i} galaxy transitioned from a bursty, chaotic phase of star formation to a smooth phase where most star formation occurred in the newly-formed thick disk \citep{Ma2017, Stern2021, Yu2021,2022arXiv220304321G}\footnote{Note that while these references analyzed to the non-MHD version of the \texttt{m12i} galaxy, the MHD version we employ here also undergoes a similar transition \cite[at $5.7$\,Gyr lookback time, instead of the 6.15\,Gyr identified in][]{Yu2021}.  See Fig.\ 1 of \citetalias{GBOF}.}.  Because we have defined our GCs as those clusters older than 6 Gyr, they are preferentially formed on isotropically distributed orbits \citep[following the velocity distribution of the gas, ][]{2022arXiv220304321G}, while the younger clusters are preferentially formed within the disk \citep[though we emphasize that our choice of 6 Gyr was actually based upon the youngest GCs in the MW, and for consistency with the literature, e.g.,][]{Pfeffer2018}.  Thus, we can conclude that the main differences between our GC population and that of the MW arise from the specific assembly history of the \texttt{m12i} galaxy itself.  Had the transition occurred at an earlier time in \texttt{m12i}, our GC population would likely be closer in age and metallicitiy to those in the MW.  This suggests that there may be a connection between the orbits of metal-rich and metal-poor clusters in the MW and the formation of the most massive GCs during the transition from bursty to disky star formation; we will explore this potential connection in a future work.

  \begin{figure}
\centering
\includegraphics[]{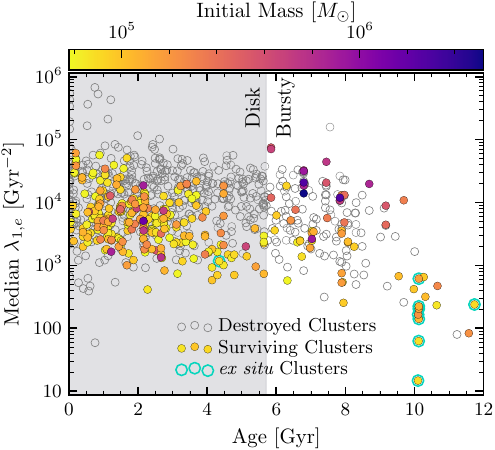}
\caption{The relationship between the formation time of all clusters in our catalog to the median effective tidal field they experience over their lifetime (see \S \ref{ss:tides}).  The open circles indicate clusters destroyed before $z=0$, while the filled colors indicate those that survive to the present day.  The color bar indicates the mass of the cluster {at formation}, with clusters formed \textit{ex situ} outlined in blue-green. }
\label{fig:l1e}
\end{figure}

  \begin{figure}
\centering
\includegraphics[]{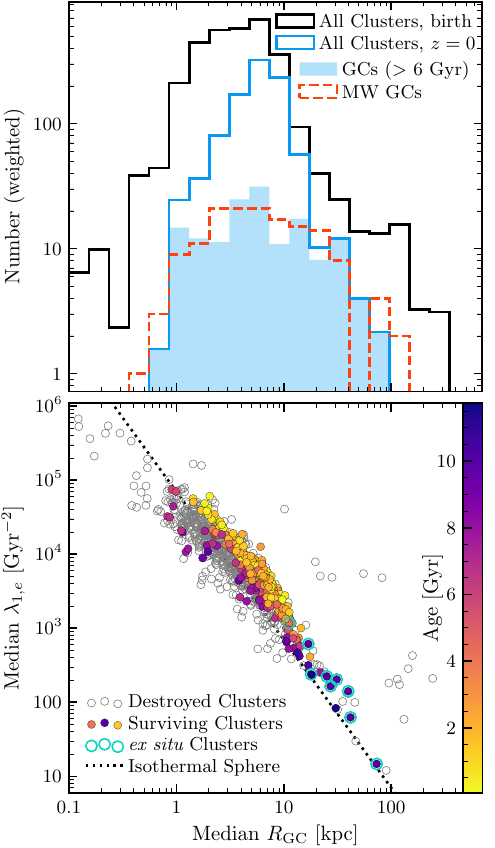}
\caption{The galactocentric radii of the CMC-FIRE clusters and the tidal fields they experience.  On the \textbf{top}, we show the distribution of median galactocentric distances for all clusters over their lifetimes (in black), as well as those that survive to $z=0$ (blue line) and those surviving clusters we define as GCs (solid blue).  We also show the galactocentric distances of the MW GCs from the \citet[][2010 edition]{Harris1996} catalog.  While these are not the same quantity, the median $R_{\rm GC}$ of an orbit over time represents a fair draw from the distributions of that orbit, making it a reasonable comparison to the MW GCs.  On the \textbf{bottom}, we show the relationship between the median radii to the median effective tidal field they experience over their lifetime (see \S \ref{ss:tides}).  The open circles indicate clusters destroyed before $z=0$, while the filled colors indicate those that survive to the present day.  The color bar indicates the age of the surviving clusters.  We also show in dotted black the relationship between tidal strength and radius for clusters in an isothermal sphere with the same effective velocity dispersion as the \texttt{m12i} galaxy.}
\label{fig:l1e_rgc}
\end{figure}

  \begin{figure}
\centering
\includegraphics[scale=1.]{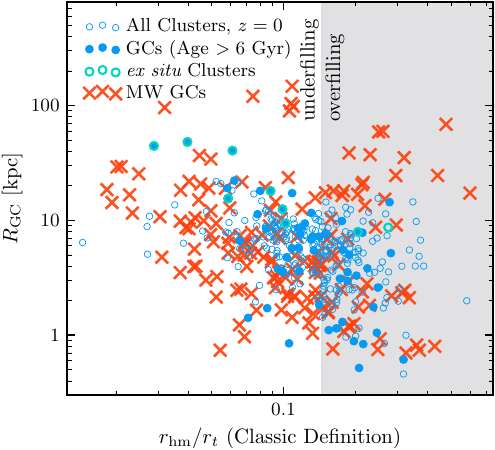}
\caption{The relation between clusters that are tidally overfilling and their galactocentric distance, for both our model clusters (in open blue circles all clusters and filled blue circles for old GCs) and the MW GCs (red crosses). We also outline clusters formed \textit{ex situ} in blue-green.  The shaded region indicates clusters that are tidally overfilling (see discussion c.f.~Fig.\ \ref{fig:tides}).  Our GCs largely span the space of clusters in the MW, though we do not reproduce the tidally filling population at large galactocentric radii that \citet{2010MNRAS.401.1832B} suggested are in the process of disrupting.}
\label{fig:rhmrt_rgc}
\end{figure}

  \begin{figure*}
\centering
\includegraphics[scale=1.]{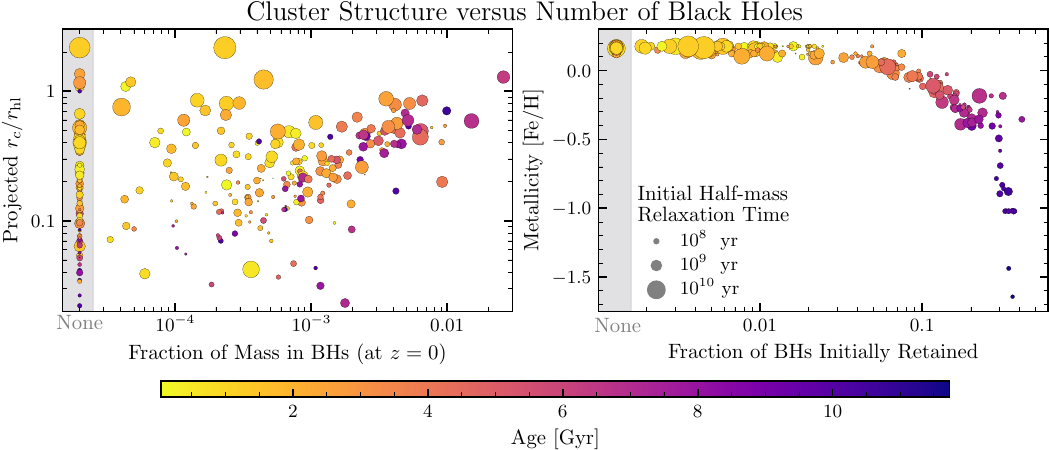}
\caption{The structural parameters of the clusters that survive to the present day and their dependence on the fraction of black holes that are initially retained. We define the BH retention fraction as the number of black holes that remain bound to the cluster immediately after the collapse/supernova of their progenitor stars, divided by the total number of black holes formed.  Note that we exclude BHs formed though mergers involving other BHs.  The size of each point indicates the cluster's initial half-mass relaxation time (Equation \ref{eqn:rlx}), while the color indicates the age of the cluster in Gyr.    On the \textbf{left} panel, we show the relationship between $r_c/r_{\rm hl}$ and the fraction of BHs that have been retained up to the present day.  Older clusters follow a roughly linear correlation, with clusters with significant mass in BHs have larger core radii compared to those with fewer BHs.  However, most younger clusters have a mass fraction in BHs $\lesssim 10^{-3}$ and larger core radii.  On the \textbf{right}, we show why this is the case: clusters that are younger have higher metallicities and retain fewer of their BHs at birth because of the larger supernova natal kicks associated with less-massive BHs. }
\label{fig:bhfrac}
\end{figure*}

  \begin{figure}
\centering
\includegraphics[scale=1.]{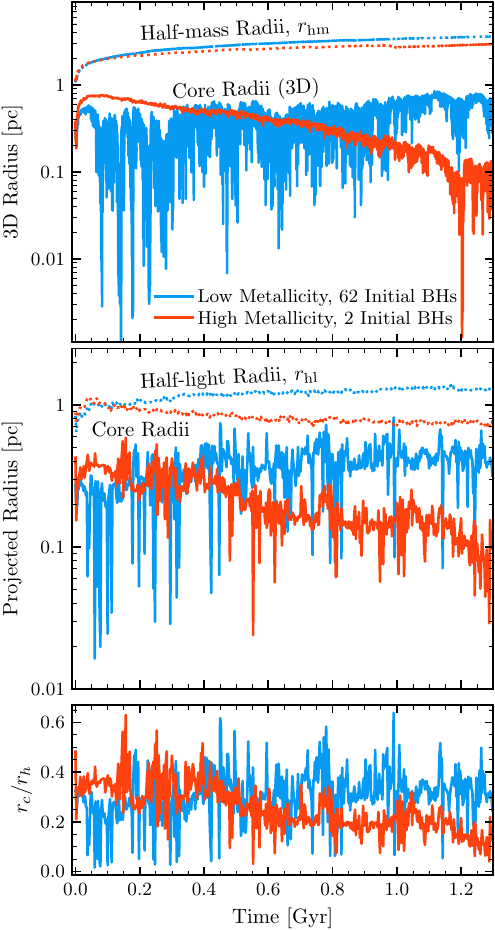}
\caption{The core and half-mass/light radii for two typical clusters ($N\approx 10^5$,  virial radius $\approx 2$ pc, $\gamma \approx 3$), whose primary difference is their birth redshifts ($z=0.9$ for blue and $z=0.1$ for red) and the number of BHs they initially retain. On the \textbf{top}, we show the theoretical 3D core and half-mass radii, while in the \textbf{middle} we show the projected 2D core and half-light radii.  Here, the model with 62 BHs collapses quickly,but then rapidly re-expands due to heating from BH burning.  The model with 2 BHs, on the other hand, collapses much more slowly, but does not re-expand.  As a result, the cluster appears core collapsed at a much younger age than its metal-poor counterpart, which does not appear core collapsed until it is $\sim 4.5$ Gyr old (not shown).  Finally, on the \textbf{bottom}, we show the ratio of the core- to half-light radii.}
\label{fig:rcrh_single}
\end{figure}

 \section{Cluster Evolution and Destruction}
 \label{sec:evolution}
 
 \subsection{Galactic Morphology Controls Cluster Survival}
 \label{ss:morph}
 
 Since cluster formation in this model traces the normal star formation of the \texttt{m12i} galaxy, it is hardly surprising that cluster orbits should be set 
 by the morphology and dynamical state of the galaxy when they were formed.  But even after the clusters are formed, the galaxy morphology has a significant 
 effect on the long-term survival of GCs: clusters on elliptical or inclined  orbits can experience significant tidal mass loss during close encounters with 
 GMCs \citep[e.g.,][]{1987MNRAS.224..193T,Gieles2006,Linden2021}, transits through the galactic disk \citep[e.g.,][]{Ostriker1972}, encounters with galactic spiral arms \citep[e.g.,][]{gieles2007}, and close pericenter passages about the galactic bulge \citep[e.g.,][]{spitzer_dynamical_1987}.  Overall, we expect clusters in the disk to experience stronger tidal fields, ensuring that they dissolve faster than their halo counterparts.  Given our 6 Gyr dividing line between GCs and young clusters, this largely implies that our GCs should live, on average, longer than the massive young and open clusters formed once the galaxy transitions to a disk morphology.  
 
 In Fig.\ \ref{fig:l1e}, we show the median tidal strength experienced by all of our clusters throughout their lifetime as a function of cluster age.  The median tidal strength experienced by clusters is lowest during the earliest phases of galaxy formation, but as the galactic disk begins to assemble, the tidal strength steadily increases as \texttt{m12i} transitions from its bursty mode of star formation into a thick disk galaxy.  Once clusters begin forming in the disk, the tidal field is largely constant, with a wide scatter determined by the cluster's initial orbital position and eccentricity in the disk.  This carves out a roughly triangular region in Fig.\ \ref{fig:l1e} where older clusters are more preferentially destroyed at higher median $\lambda_{1,e}$.  This has important implications for the long-term survival (or lack thereof) of disk clusters, particularly the open clusters observed in the MW.  A better understanding of this process will require both a higher-resolution tracking of the potential experienced by clusters in the disk and a real-time treatment of tidal shocking during the cluster integrations; see \S \ref{ss:tides} and Appendix \ref{apx:shocking}.   

The fact that clusters formed earlier in the process of galaxy assembly are more likely to survive creates an obvious mechanism for promoting GC survival over younger massive star clusters.  Furthermore, clusters that were formed \textit{ex situ} are also more likely to survive for longer periods, owing to their significantly wider halo orbits than clusters formed \textit{in situ} (Fig.\ \ref{fig:l1e_rgc}).   Ironically, these clusters actually experience stronger tidal fields while still living in their birth galaxies; it is only after they are accreted by the main \texttt{m12i} galaxy that they experience sufficiently weak tidal fields to survive to the present day.  This suggests that some \textit{ex situ} clusters have \emph{only} survived because of their accretion into a larger galaxy.  This characteristic decrease in tidal field strength is shown in the example \textit{ex situ} cluster in Fig.\ \ref{fig:typical}, and has been observed in previous studies of cluster formation in galaxy simulations \citep[][and \S \ref{ss:comp}]{Li2019,Meng2022}.  This result suggests that accreted clusters in the MW and other galaxies may actually be a better tracer of old star formation in dwarf galaxies than clusters in present-day dwarf galaxies!

We note in Fig.\ \ref{fig:l1e_rgc} the excellent agreement between the median radial distribution of our CMC-FIRE clusters and the present-day radial distribution of GCs in the MW, and the strong correlation between the median $\lambda_{1,e}$ of the clusters and their median galactocentric distance.   This correlation follows a roughly $1/R_{\rm GC}$ power law, similar to the predicted tidal field experienced by clusters on circular orbits in an isothermal sphere, where $\lambda_{1,e} = 2 \sigma^2 / R_{\rm GC}^2$ (and $\sigma\approx180\rm{km}/\rm{s}$ is the effective velocity dispersion of the potential for the \texttt{m12i} galaxy).  Taken together, this suggests that the tidal fields experienced by our CMC-FIRE population are representative of those experienced by the MW GCs.  

Finally, it has been suggested \citep{2010MNRAS.401.1832B} that there exist two distinct populations of clusters beyond 8 kpc when viewed in the plane of galactocentric distance versus tidal filling: a compact population with $r_{\rm hm}/r_t$ less than 0.05, and a second population of tidally filling clusters with $0.1 < r_{\rm hm}/r_t < 0.3$.  We show this in Fig.\ \ref{fig:rhmrt_rgc}.  While we are able to reproduce the lower population in galactocentric radius, we do not seem to reproduce the population of tidally-filling, large distant clusters.  \citep{2010MNRAS.401.1832B} suggested that these clusters in the MW were in the process of being disrupted.  Given that the closest clusters from our population are all \textit{ex situ} clusters, this population may also simply be the remnants of a dwarf galaxy accredited by the MW (for which \texttt{m12i} has no counterpart).

 \subsection{Cluster Structure as Determined by Black Holes}
 \label{ss:bhs}

Over the past decade, a growing consensus has been developing that the overall size of a GC’s core, and the ratio between the core and half-mass (or half-light) radii, is determined by the number of the black holes (BHs) that have been retained by the cluster up to the present day.  A cluster only reaches ``core collapse’’ (defined observationally as having a SBP that increases continuously towards the cluster center) once it has ejected nearly all of its initial BHs \citep{2004ApJ...608L..25M,Mackey2008,Breen2013,Kremer2018}.  Both analytic arguments \citep{Breen2011} and numerical simulations \citep{Breen2013,ArcaSedda2018,Kremer2019} have shown that the ejection rate of BHs is largely controlled by the relaxation time of the cluster at the half-mass radius, given by \cite[e.g.,][]{spitzer_dynamical_1987}:  

\begin{equation}
T_{\rm rlx} = 0.138 \frac{N}{\log\Lambda}\left(\frac{r_{\rm hm}^3}{GM}\right)^{1/2}~,
\label{eqn:rlx}
\end{equation}

\noindent at the cluster’s half-mass radius\footnote{Note that $\log\Lambda$ here is the Coulomb logarithm for the internal cluster evolution, typically assumed to be $\log(\gamma N)$ with $\gamma = 0.01$.}.  Because clusters require BH ``burning’’ to supply the energy to hold the cluster against continued gravothermal collapse (producing the flat core profiles characteristic of the King or Plummer models), the clusters with the largest cores are expected to be those with longer relaxation timescales, that have been unable to eject all of their BHs by the present day.   In other words, GCs with smaller cores are thought to be dynamically older, having had many relaxation times over which to eject their BH subsystems.

But then how are we to understand distribution of $r_c/r_{\rm hl}$ and concentration parameters presented in Fig.\ \ref{fig:conc}, where our GCs are more compact than those in the MW, despite being younger overall?  In Fig.\ \ref{fig:bhfrac}, we show the relationship between fraction of mass in BHs for our $z=0$ cluster population and the ratio of the core to half-mass radii.  For the dynamically old clusters, the relationship is nearly linear, in good agreement with previous results that have studied the presence of BH subsystems in GCs and their observational properties \citep{Morscher2015,Kremer2019,Weatherford2020}.  However, we note that there also exists a population of younger clusters that do not appear to lie on this linear trend.  

This feature actually arises from a complex interplay between the age-metallicity relationship of the \texttt{m12i} galaxy and the initial retention of stellar-mass BHs.  Younger clusters have higher stellar metallicities, which in turn drive stronger winds for the massive stars in the clusters at birth.  However, this means that the BHs that are formed from collapsing stars are typically lower mass ($\lesssim 15M_{\odot}$) than those formed in low-metallicity clusters (which are thought to produce many of the ``heavy’’ 30-40 $M_{\odot}$ BHs observed by LIGO/Virgo).  The supernova that form these lower-mass BHs produce large amounts of ejecta \citep[according to the standard prescriptions available in COSMIC, from ][]{Fryer2012}, which in turn impart larger natal kicks to the BHs at birth \citep[as high as the $\sim$ 400 km/s kicks experienced by neutron stars, ][]{Hobbs2005}.  This is in contrast to the 30 $M_{\odot}$ BHs formed in low-metallicity environments, which are largely thought to form via direct collapse \citep{Fryer2001}. As a result, higher-metallicity GCs and young clusters retain fewer BHs after $\sim20$\,Myr, and can dynamically eject their remaining BH population faster.

This relationship is shown explicitly in the right panel of Fig.\ \ref{fig:bhfrac}. Clusters with metallicities above solar (roughly those 4 Gyr or younger) retain at most 10\% of their BHs initially.  This correlation has a far stronger influence on the initial retention of BHs than either the cluster mass or initial radius (which directly effect both the initial half-mass relaxation time and the cluster escape speed).  Only a handful of clusters are so massive that they deviate from the trend \citep[e.g. the \textit{Behemoth} cluster previously described in][]{Rodriguez2020}.  It is immediately clear that the younger clusters will never follow the linear relationship between $M_{\rm BH}$ and $r_c/r_{\rm hl}$, because they do not have a significant number of BHs to begin with!  Our GCs--with a median age of $\sim 8$ Gyr--are able to retain more (10\% - 20\%) of their initial BHs.  But this is still a factor of $\sim$2 below the BH retention for the truly old, metal poor GCs (like those observed in the MW).  

What this means is that, despite being younger, our GCs appear dynamically older and more core collapsed than those in the MW.   Indeed, for clusters that begin with very few BHs, the picture of core collapse resembles the more classical picture of GC evolution \citep[e.g.,][]{spitzer_dynamical_1987}, where, for a Plummer sphere with equal masses, the core-collapse time is approximately $16 t_{\rm rlx}$ \citep[e.g.,][and references therein]{Freitag2001}. To illustrate this, we pick two clusters, one old and one young, with nearly identical initial conditions--$\sim 10^5$ stars, $2$ pc virial radii, and Elson $\gamma$ parameters of $\sim$3--and plot the evolution of their radii over time in Fig.\ \ref{fig:rcrh_single}.  Of course, because of their different cosmic birth times ($z=0.9$ versus $z=0.1$), the old cluster was born with a metallicity of $0.5 Z_{\odot}$ and retains 62 BHs initially, while the young cluster was born with a metallicity of $2 Z_{\odot}$ (which we truncate to $1.5Z_{\odot}$, see \S \ref{ss:weights}) and retains only 2 BHs.  In the old cluster, the \emph{initial} core collapse occurs within $\sim75$ Myr.  This first collapse is driven by the mass segregation of the BHs, which scales as  

\begin{equation}
T_{\rm cc} \approx \frac{\left<m\right>}{m_{\rm BH}} T_{\rm rlx}
\end{equation}

\noindent where $m_{\rm BH}$ is the mass of the BHs (or the most massive objects in the cluster) and $\left < m \right>$ is the average stellar mass.  However, immediately after core collapse, the core suddenly re-expands as the energy generated by the formation of BH binaries is injected into the cluster core.  While the initial collapse is visible in the 3D core radius, it is barely detectable (beyond standard fluctuations) in the observed core radius shown in the bottom panel if Fig.\ \ref{fig:rcrh_single}.  This cluster will not undergo a true, permanent core collapse \citep[sometimes called ``second core collapse'',][]{Breen2013,Heggie2014b} until much later (4.5 Gyr after its formation, not shown here).

On the other hand, the younger cluster, being deprived of any BH subsystem by natal kicks, only experiences a single core collapse event.  In Fig.\ \ref{fig:rcrh_single}, this occurs at approximately 1.2 Gyr (or 6 $T_{\rm rlx}$, where $T_{\rm rlx}=200$ Myr for this cluster) after formation.  This is closer to the 16 $T_{\rm rlx}$ core collapse time for equal-mass clusters \citep[with the remaining difference arising from the non-equal mass stellar IMF, e.g.,][]{1971ApJ...164..399S,1985ApJ...292..339I,2010MNRAS.408L..16G}, but still significantly faster than the 4.5 Gyr collapse time for the older low-metallicity cluster.  Despite its age, the younger cluster appears dynamically older than its older sibling.  This counter-intuitive results suggests that one must be cautious when using GCs as tracers of galactic star formation and evolution, since any measurement of their dynamical age strongly depends upon the initial retention of BHs and the age-metallicity relation of their birth galaxies in addition to their initial relaxation time as determined by their birth masses and radii.  

Finally, it should be noted that our analysis has utilized almost exclusively a relatively new understanding of GC evolution, where a GC's core radius is largely determined by the size of its BH subsystem \citep[e.g.,][]{Mackey2008,Breen2013,Heggie2014b,Kremer2019}, which disperses over a timescale set by the cluster's half-mass relaxation time, c.f.~Equation \eqref{eqn:rlx}.  Of course, this new framework relies on the initial presence of a large BH population; in the absence of these BHs (as occurs in our high-metallicity clusters), the size of a cluster's core is instead determined by the classic ``binary burning'' of primordial stellar binaries in the central regions of the core \citep[e.g.,][]{Gao1991,Wilkinson2003,Chatterjee2010}, which in turn would suggest a dependence on the initial binary fraction of our simulations (10\%, see \S \ref{ss:init}).  However, recent observational evidence \citep{Moe2019} has shown that binary fraction (for close solar-type stars) is largely anti-correlated wtih metallicity, decreasing to $f_b \lesssim 25$\% above [Fe/H] of -0.5 (where our cluster models in Figure \ref{fig:bhfrac} retain less than 10\% of their initial BHs) cluster.  Secondly, detailed $N$-body simulations \citep[][Fig.~17 and 18]{Heggie2006}, have shown that the primordial binary fraction does not significantly change either $r_c/r_{\rm hm}$ or the core collapse time (unless the primordial binary fraction is zero), at least not before the clusters can be tidally disrupted \citep{Trenti2007}.  As such, we conclude that our primordial binary fraction does not significantly influence the results presented here.

 \subsection{Comparison of GC Survival to Other Studies}
 \label{ss:comp}
   
   \begin{figure*}
\centering
\includegraphics[]{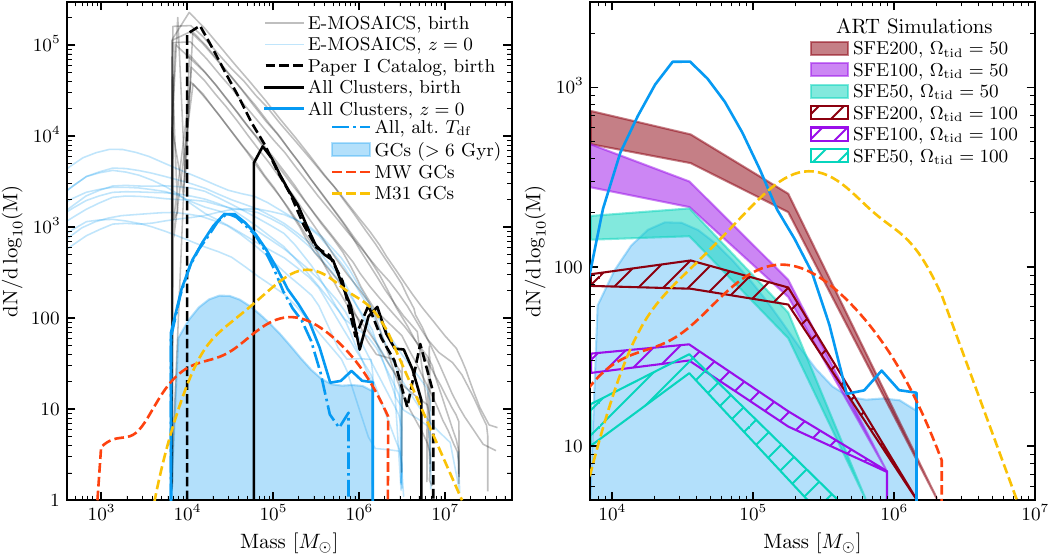}
\caption{Comparisons between the mass functions of our CMC-FIRE clusters and the E-MOSAICS and ART simulations.  On the \textbf{left}, we show the ICMF and GCMF from 10 MW-mass galaxy cluster systems in the E-MOSAICS models \citep[Fig.\ 16 from][]{Pfeffer2018}.  We show two ICMFs: the full ICMF from \citetalias{GBOF} (dashed black), and the reduced ICMF of 895 clusters that were employed here (solid black, and multiplied by the weights described in \S \ref{ss:weights}).  Note that the GCMF from E-MOSAICS only includes clusters older than 6 Gyr.  To better compare, we show our full population of surviving clusters in solid blue, and our population of surviving clusters older than 6 Gyr in dotted blue.  We also show the mass functions from the MW \citep[][2010 edition]{Harris1996} and from M31 \citep{Caldwell2011}.  On the \textbf{right}, we show the results of comparisons between our models and the GCMF distributions from the ART simulations \citep[][Fig.\ 4]{Li2019}, including the mass functions for several different galaxy formation simulations and assumed tidal strengths.}
\label{fig:emosaics}
\end{figure*}
 
   \begin{figure*}
\centering
\includegraphics[]{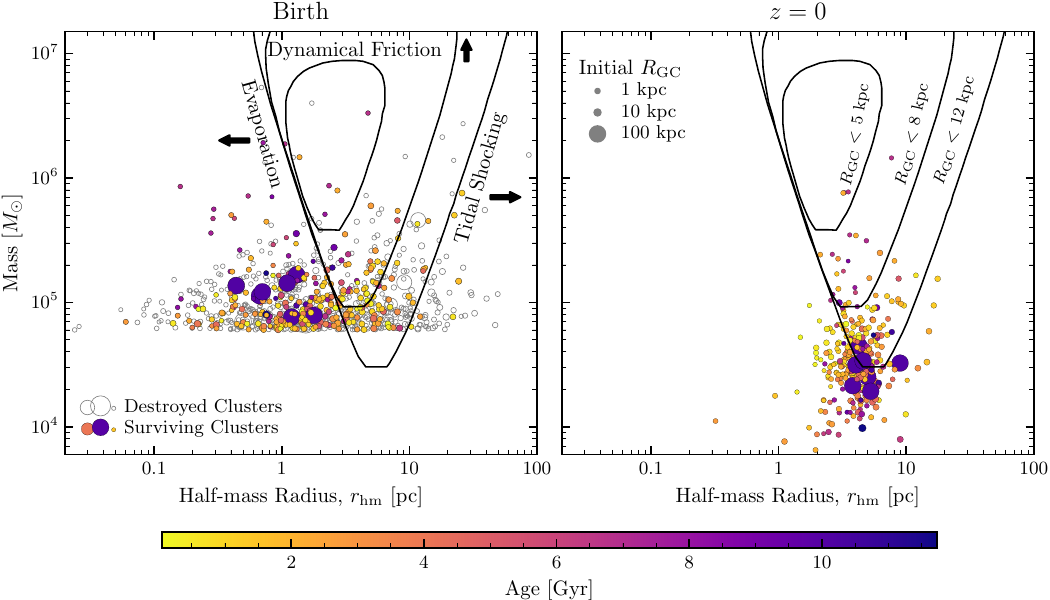}
\caption{The distribution of our cluster populations in mass-radius space (note we use the real 3D half-mass radius of each cluster).  On the \textbf{left}, we show the initial distribution of clusters (with disrupted clusters shown with gray circles), while the \textbf{right} shows the population that survives to $z=0$.  The filled color of the surviving clusters indicates their age, while the size indicates the initial galactocentric radius at which each cluster was formed.  Finally, the black lines show the original vital diagrams from \citet{Gnedin1997} based on the galactic model of \citet{Ostriker1983}, assuming isotropic velocities.  Each of the three sides of the triangles show the region beyond which the process indicated with arrows would cause a cluster to be destroyed within 10 Gyr (e.g.~clusters that lie to the right of the triangles will succumb to tidal shocking within a Hubble time, and so on).  }
\label{fig:vital}
\end{figure*}

 Many previous studies have explored the long-term evolution and survival of the MW GC system, both using observations of our own Galaxy \cite[using the  best-available knowledge of it's galactic potential,  e.g.,][]{Gnedin1997,LongOstriker1992,Chernoff1990,Prieto2008,Vesperini1997,Baumgardt1998}, and cosmological simulations of MW-like galaxies  \cite[e.g.][]{Rieder2013,Pfeffer2018,Li2017}. The initial results here occupy a unique space in this pre-existing literature: on the one hand, our models of star cluster evolution are the highest resolution of any presented so far, including realistic prescriptions for stellar evolution, binary formation and interactions, and realistic surface brightness profiles to compare directly to observations.  Of course, this precision comes at a cost, as we have only simulated a single galaxy's cluster system.  As such, it is important to compare our results to other cosmological studies of cluster formation in MW-mass galaxies, as well as studies of the MW GC system itself.
 
 The first obvious comparison is between our results and those of the E-MOSAICS simulations \citep{Pfeffer2018}.  In this series of simulations, the authors explored the populations of GCs formed in several MW-mass galaxies in the EAGLE simulations.  Starting from those initial star clusters, they then follow the evolution and destruction of clusters from birth to $z=0$ using semi-analytic prescriptions for the cluster's mass and radius and its subsequent evolution in a time-varying galactic potential \citep[based on the models of][]{kruijssen2011}.  While $N$-body cluster evolution models are more sophisticated in their internal physics, our method of computing the tidal fields experienced by the clusters and their dynamical friction timescales are both taken directly from \citet{Pfeffer2018}.  See \S \ref{ss:tides} and Appendix \ref{apx:shocking}, though note that our modeling of tidal shocking is done in post-processing, and does not follow the instantaneous increase in cluster radius.
 
 In the left panel of Fig.\ \ref{fig:emosaics}, we compare our initial and final CMFs to those from the 10 MW-like cluster systems of the E-MOSAICS simulations.  Both the initial masses of our full catalog and the reduced ICMF of the 895 clusters we evolved here are consistent with the masses of  clusters identified by E-MOSAICS (see \citetalias{GBOF} for a more in-depth discussion).  When comparing our $z=0$ mass function, we find consistency between our results, despite the different choices we employed in calculating the dynamical friction timescales (see the discussion in \S \ref{ss:tides}) and our  limited treatment of tidal shocking.  Like the present study, the GCMFs presented in \cite{Pfeffer2018} only considers clusters older than 6 Gyr to be true GCs.  When comparing only GCs, we find find a decrease in lower-mass clusters compared to E-MOSAICS, particularly around the peak ($\sim 4\times10^5M_{\odot}$) of our GCMF.  
 
 
 On the one hand, our GCMF is arguably closer in number to the number of GCs in the MW (148 vs 157) than the E-MOSAICS simulations.  However, this could easily be a byproduct of our $6\times10^4M_{\odot}$ cutoff in our simulated initial CMF (i.e.~had our ICMF extended down to $10^4M_{\odot}$, we may have produced more low-mass GCs).  But at the same time, the number density of CMC-FIRE clusters at $\sim10^5M_{\odot}$ is closer to that of the MW than the E-MOSAICS GCMFs (while the number density of M31 GCs at that mass lies roughly between both models).  While it is difficult to disentangle the effects of galaxy-to-galaxy variation here, it is obvious that we produce fewer GCs across all masses than the E-MOSAICS simulations. However, we note that both the E-MOSAICS GCMFs and our CMC-FIRE GCMF peak significantly below the GCMFs of the MW or M31, indicating that both approaches either form too many low mass clusters or are insufficiently efficient at destroying them.  
 
 This is also suggested by comparisons to the ART simulations \citep{Li2017,Li2018,Li2019}, which use high-resolution hydro simulations of MW-mass galaxies to study the formation and evolution of star clusters formed before $z=1.5$.  In the right panel of Fig.\ \ref{fig:emosaics}, we show the $z=0$ GCMFs from Fig.\ 4 of \citet{Li2019} and compare it to our own CMF and GCMF.  Here, the different GCMFs represent star cluster systems from simulations with different star-formation efficiencies, while the solid and dashed bands represent different assumptions for the strength of the galactic tidal field \citep[see][for details]{Li2019}.  Limiting ourselves only to old GCs, we find that our GCMF is equivalent to one found in either a galaxy with high star-formation efficiency and strong tides (i.e.~a high destruction rate) or low star-formation efficiency and weak tides, both of which produce a decrease in the number of GCs surviving to the present day.  We note that the galaxies do not show a late transition from bursty to disk-based star formation like the \texttt{m12i} galaxy, possibly because the galaxy simulations themselves did not proceed beyond $z=1.5$.  But at the same time, we do find the same trend for the tidal fields experienced by individual clusters to decrease over time (particularly for those clusters accreted from infalling dwarf galaxies; see \S \ref{ss:morph} and Fig.\ \ref{fig:typical}).  This is consistent with the mechanism proposed in \cite{Kruijssen2014,Forbes2018}, where GCs rapidly migrate from gas-rich environments (where they are formed) following galaxy mergers to regions with lower tidal field strength, thereby promoting their long-term survival.
 
 However, we do not find the same trend of decreasing tidal field for the \textit{in situ} clusters identified by \cite{Meng2022}. While this may be because of our treatment of the tidal field in the initial \texttt{m12i} snapshot (see footnote in \S \ref{ss:tides}), the present study's measurement of the tidal tensors (with typical separations of $\sim6.5$ pc) is significantly higher resolution than the $\mathcal{O}(100)$ pc resolution of either the E-MOSAICS or ART simulations, and would allow us to resolve the early tidal effects of star-forming regions.  More likely, we believe this is because our \textit{in situ} clusters do not show the same trend of outward migration identified in their study. Since both studies (as well as E-MOSAICS) use tracer particles from the cosmological simulation to extract the tidal fields, these differences likely arise from the different evolution of the cosmological simulations themselves.  It should also be noted that the ART simulations use a different approximation for the tidal field experienced by the clusters (using the maximum absolute value of the Eigenvalues of the tidal tensor, instead of the $\lambda_{1,e}$ that we and E-MOSAICS use), which can be a factor of $\sim$2 lower than our approach \citep[See][Appendix B]{Meng2022}.  While our approximation has been shown to accurately reproduce the tidal fields from a spherical Plummer potential \citep[See Figure C1 from][]{Pfeffer2018}, we note that both approaches are necessary simplifications from the true tidal field.  Future work (di Carlo et al., in prep.) will explore the validity of these prescriptions when compared to $N$-body integrations in a realistic galactic tidal potential.
 
While previous studies have suggested that tidal fields, and particular tidal shocking, play a critical role in sculpting the $z=0$ GC populations, our results are substantially less definitive.  As stated in \S \ref{ss:tides} and Appendix \ref{apx:shocking}, our post-processing analysis suggests that only one of our 895 clusters would have been destroyed from the injection of energy by tidal shocking, and even then, that cluster was destroyed by our tidal stripping prescription at the same time that our shocking analysis predicted its destruction.  Of course, it is unsurprising that these two effects are linked: both are computed directly from the tidal tensor  (Equation \ref{eqn:tidetensor}), and any significant change in the tidal tensor that would inject energy into the cluster would \emph{also} be accompanied by a reduction in the tidal boundary, instantaneously removing a significant fraction of the cluster mass from our simulations. 
 
 In previous studies, the cumulative effects of evaporation, tidal shocking, and dynamical friction on the survival of the MW GC system have been explored by the use of ``vital diagrams'', where the typical destruction rate of clusters is given by the sum of the destruction rate for the three processes.  Assuming  that the rate of cluster destruction by each mechanism can be expressed as the inverse of the typical timescale for destruction by that mechanism, the vital 
 diagram for clusters can be computed as the level sets of the function  
 
 \begin{equation}
 \frac{1}{T_{\rm dest}} = \frac{1}{T_{\rm rlx}} + \frac{1}{T_{\rm ts}} + \frac{1}{T_{\rm df}}
 \label{eqn:vital}
 \end{equation}
 
 \noindent where, to study the old clusters in our own Galaxy, $T_{\rm dest}$ is typically set to a Hubble time \cite[or 10 Gyr in][]{Gnedin1997}.
 
 In \citet{Gnedin1997}, these diagrams were generated using fits to the orbits and masses of MW GCs, along with a model for the MW gravitational potential.  The  level sets of Equation \eqref{eqn:vital} were then computed for different initial galactocentric radii (adjusting both the tidal field strength in the given  potential and the rate and strength of tidal shocking).  In Fig.\ \ref{fig:vital}, we show the vital diagrams from \citet{Gnedin1997} in (3D) mass and radius space and compare them to our own CMC-FIRE population.  The initial population spans several orders of magnitude in radius when compared to the predictions,  but our final population agrees reasonably well with the vital diagrams for the MW.  In particular, if we restrict ourselves to old GCs (close to the $\sim$ 10  Gyr lifetime considered in the original study), our results agree extremely well with the original predictions of \citet{Gnedin1997}, arguably better than the actual MW!
 
 However, this result is somewhat shocking, as we have not included any run-time prescription for tidal shocking in the models.   Instead, our models replace  both the evaporation and tidal shocking timescales with the time-varying change in the tidal boundary of the cluster.  This process appears to work surprisingly well, destroying many of the clusters that were born in the region where tidal shocking should have operated. However, because of the difference in ages between our GC population and the MW, as well as the approximations made in our implementation of the external tidal potential within \texttt{CMC}, it is not immediately obvious if this truly means that tidal shocking does not operate on the GC population, or if our initial conditions largely occupy a region of $M-r_{\rm hm}$ space that is unaffected by tidal shocking.  By the same token, however, it is also possible that the MW clusters themselves were born in a region largely impervious to tidal shocking, and that while the process may be important in some regions, it is largely absent in the formation of our own Galaxy.   Work exploring the effects of tidal shocking using direct $N$-body simulations is currently underway.
 
 Lastly, we can use these comparisons as an opportunity to explore the implications of our prescription for dynamical friction.  As previously stated (\S \ref{ss:tides}), while our computation of $T_{\rm df}$ is identical to that employed in \citet{Pfeffer2018}, our criterion for termination is less strict, since we require the cluster to have an age older than the time-integrated value of $T_{\rm df}$ before it is considered destroyed (whereas the E-MOSAICS simulation assumes any cluster where the instantaneous values of $T_{\rm df}$ is greater than the cluster age is destroyed).  In Fig.\ \ref{fig:emosaics}, we show the effect that adopting the E-MOSAICS prescription would have on our CMF\footnote{Though note that our implementation is actually \emph{more} strict that the E-MOSAICS assumptions, since here we calculate $T_{\rm df}$ assuming the initial mass of the cluster is fixed}.  This increases the number of clusters that are destroyed by dynamical friction from 4 to 45, though we note that in neither case is dynamical friction the dominant destruction mechanism for star clusters.  In reality, of course, neither prescription is accurately capturing the change in the cluster's orbit due to dynamical friction.  These choices will have important implications for the survival of the most massive clusters, as well as the creation of nuclear star clusters found in the centers of many galaxies \citep[many of which are thought to have formed from the infall of massive GCs, eg.,][]{1975ApJ...196..407T,2020A&ARv..28....4N}.

 \section{Conclusion}
 
 We have presented the first end-to-end simulation of massive star clusters that combines realistic star-by-star $N$-body simulations with a GMC-resolving cosmological simulation of galaxy evolution to model cluster evolution and destruction.  Starting from the MHD \texttt{m12i} Latte simulation from the FIRE-2 suite of galaxy models, in \citetalias{GBOF} we identified all of the collapsing GMCs and, using a catalog of high-resolution cloud-collapse simulations as a guide, developed an initial population of 73,461 YMCs formed throughout the galaxy across cosmic time.  In this paper, we evolved 895 of the most massive clusters ($\sim 30\%$ of those more massive than $6\times10^4M_{\odot}$) to the present day using the masses, radii, concentrations, metallicities, and birth times as initial conditions for our \texttt{CMC} integrations.  To account for the continued influence of the galaxy on the star clusters, we included prescriptions for time-dependent tidal mass loss into our $N$-body code, where the tidal fields were calculated from the gravitational potential experienced along the orbit of a test particle that comprised the parent GMC.
 
 Our cluster model predicts a total of 941 clusters in the galaxy by the present day, 148 of which we define as classical GCs (older than 6 Gyr).  These GCs were all formed during the early phases of galaxy assembly, where star formation occurred in a bursty fashion, or in dwarf galaxies which were later accreted by the host galaxy.  As a result, these clusters are largely on orbits that are isotropically distributed about the galactic disk.  The younger clusters, meanwhile, are largely formed once the galaxy transitions to a disk profile and largely lie within the disk (where they experience tidal fields that are, on average, stronger than those on isotropic halo orbits).    
 
 Our $z=0$ GC population still shows several key differences from GCs observed in the MW or M31.  The CMC-FIRE GCs that formed \textit{in situ} in the main galaxy are younger on average (with a median age of 7.9 Gyr) than those in the MW or M31, while the clusters created, \textit{ex situ} (those accreted from dwarf galaxies) are largely consistent in number and with similar ages and metallicities to those in the MW.  The most massive clusters in our catalog are formed \textit{in situ} at $z\sim0.8$ during a minor merger between a dwarf galaxy and the main \texttt{m12i} disk galaxy.  Because of their age, the bulk of our GCs tend to be less massive than those in the MW, since clusters formed later in the process of galaxy assembly experience stronger tidal fields, on average, than those formed during early bursty phases of star formation (or GCs accreted from infalling dwarf galaxies).  Furthermore, the later formation times of our GCs means they also form with higher stellar metallicities ([Fe/H] $\sim-0.4$) than many GCs in the MW.  These higher metallicities lead massive stars in our clusters to lose more mass to stellar winds, producing BHs with lower masses than those expected to form in low-metallicity GCs.  This, in turn, leads to higher BH natal kicks, reducing the number of retained BHs and accelerating core collapse in these clusters (despite their younger ages).  
 
 While these differences mean our GC population is not a perfect comparison to the GCs in the MW, the CMC-FIRE clusters offer an illuminating link between the ages, masses, metallicities, and radii of GCs in galaxies and the star formation and assembly histories of those galaxies.   Orbits of present-day \textit{in situ} clusters are largely set by the mode of star formation the galaxy was in when the clusters were born.  In the \texttt{m12i} galaxy, all our GCs were formed during the bursty phase of star formation and retain their isotropically-distributed halo orbits up to the present day.  In other galaxies, such as the MW, the transition from bursty to disk-based star formation may have occurred while GCs were still forming, creating a correlation between the age/metallicity of GCs and their orbits.  The age/metallicity relationship is also encoded in the radii of the surviving clusters: since higher-metallicity clusters retain fewer stellar-mass BHs at birth, and undergo core collapse more rapidly, suggesting that cluster radii, in addition to masses and numbers, are a key tracer of star formation and galaxy assembly at high redshifts.    This, in particular, is uniquely possible because of our $N$-body approach to modeling GCs (with its detailed treatment of single and binary stellar evolution).

Several previous studies have explored the evolution and destruction of star clusters in cosmological simulations using semi-analytic treatments of the cluster evolution \citep[e.g.,][]{Li2017,Pfeffer2018,Li2018,Li2019}.  This study represents the first attempt to attack this problem from the realm of collisional $N$-body modeling of star clusters.  In doing so, we can not only answer questions about the formation, evolution, and survival of the MW and other GC systems, but we can begin to perform systematic studies of the internal properties of GCs and their potential contribution to many transients observed in the local universe.  In a followup study (Lamberts et al., in prep.), we will explore the implications of these realistic initial conditions and our treatment of cluster evolution and destruction on the production of gravitational-wave sources.
 
 \section*{Acknowledgments}

The authors are grateful to Mike Boylan-Kolchin, Jeremy Webb, Sterl Phinney, Kyle Kremer, Fred Rasio, and Katie Breivik useful discussions.  CR was supported by NSF Grant AST-2009916 and a New Investigator Research Grant from the Charles E.~Kaufman Foundation. ZH was supported by a Gary A. McCue postdoctoral fellowship at UC Irvine. MYG was supported by a CIERA Postdoctoral Fellowship and a NASA Hubble Fellowship (award HST-HF2-51479). AL acknowledges funding from the Observatoire de la C\^ote d'Azur and the Centre National de la Recherche Scientifique through the Programme National des Hautes Energies and the Programme National de Physique Stellaire as well as the ANR COSMERGE project, grant ANR-20-CE31-001 of the French Agence Nationale de la Recherche.  CAFG was supported by NSF through grants AST-1715216, AST-2108230,  and CAREER award AST-1652522; by NASA through grant 17-ATP17-0067; by STScI through grant HST-AR-16124.001-A; and by the Research Corporation for Science Advancement through a Cottrell Scholar Award.  AW received support from: NSF via CAREER award AST-2045928 and grant AST-2107772; NASA ATP grant 80NSSC20K0513; HST grants AR-15809, GO-15902, GO-16273 from STScI.  This work used the Extreme Science and Engineering Discovery Environment (XSEDE), which is supported by National Science Foundation grant number ACI-1548562. Specifically, it used both the Bridges-2 system, which is supported by NSF award number ACI-1928147 at the Pittsburgh Supercomputing Center (PSC), and the San Diego Supercomputing Center Comet cluster under XSEDE allocation PHY180017.  Additional computations were run on the FASRC Cannon cluster supported by the FAS Division of Science Research Computing Group and by the Black Hole Initiative (funded by JTF and GBMF grants), both at Harvard University.  This work also used computational resources provided by TACC Frontera allocations AST-20019 and AST-21002.  Images of the \texttt{m12i} galaxy were generated with FIRE studio (\url{github.com/agurvich/FIRE_studio}), an open source Python visualization package designed with the FIRE simulations in mind.

\section*{Data Availability}

The data supporting the plots within this article are available upon request to the corresponding author. A public version of the {\texttt CMC} code is available at \url{https://clustermontecarlo.github.io/}.
 
\bibliographystyle{mnras}

\begin{thebibliography}{}
\makeatletter
\relax
\def\mn@urlcharsother{\let\do\@makeother \do\$\do\&\do\#\do\^\do\_\do\%\do\~}
\def\mn@doi{\begingroup\mn@urlcharsother \@ifnextchar [ {\mn@doi@}
  {\mn@doi@[]}}
\def\mn@doi@[#1]#2{\def\@tempa{#1}\ifx\@tempa\@empty \href
  {http://dx.doi.org/#2} {doi:#2}\else \href {http://dx.doi.org/#2} {#1}\fi
  \endgroup}
\def\mn@eprint#1#2{\mn@eprint@#1:#2::\@nil}
\def\mn@eprint@arXiv#1{\href {http://arxiv.org/abs/#1} {{\tt arXiv:#1}}}
\def\mn@eprint@dblp#1{\href {http://dblp.uni-trier.de/rec/bibtex/#1.xml}
  {dblp:#1}}
\def\mn@eprint@#1:#2:#3:#4\@nil{\def\@tempa {#1}\def\@tempb {#2}\def\@tempc
  {#3}\ifx \@tempc \@empty \let \@tempc \@tempb \let \@tempb \@tempa \fi \ifx
  \@tempb \@empty \def\@tempb {arXiv}\fi \@ifundefined
  {mn@eprint@\@tempb}{\@tempb:\@tempc}{\expandafter \expandafter \csname
  mn@eprint@\@tempb\endcsname \expandafter{\@tempc}}}

\bibitem[\protect\citeauthoryear{Aarseth}{Aarseth}{2003}]{Aarseth2003}
Aarseth S.~J.,  2003, Gravitational N-Body Simulations, by Sverre J. Aarseth,
  pp. 430. ISBN 0521432723. Cambridge, UK: Cambridge University Press, November
  2003.

\bibitem[\protect\citeauthoryear{{Aarseth}, {Henon}  \& {Wielen}}{{Aarseth}
  et~al.}{1974}]{Aarseth1974}
{Aarseth} S.~J.,  {Henon} M.,   {Wielen} R.,  1974, \aap, \href
  {https://ui.adsabs.harvard.edu/abs/1974A&A....37..183A} {37, 183}

\bibitem[\protect\citeauthoryear{Ambartsumian}{Ambartsumian}{1937}]{Ambartsu1937}
Ambartsumian V.,  1937, Astronomicheskii Zhurnal, 14, 207

\bibitem[\protect\citeauthoryear{{Arca Sedda}, {Askar}  \& {Giersz}}{{Arca
  Sedda} et~al.}{2018}]{ArcaSedda2018}
{Arca Sedda} M.,  {Askar} A.,   {Giersz} M.,  2018, \mn@doi [\mnras]
  {10.1093/mnras/sty1859}, \href
  {https://ui.adsabs.harvard.edu/abs/2018MNRAS.479.4652A} {479, 4652}

\bibitem[\protect\citeauthoryear{{Baumgardt}}{{Baumgardt}}{1998}]{Baumgardt1998}
{Baumgardt} H.,  1998, \aap, \href
  {https://ui.adsabs.harvard.edu/abs/1998A&A...330..480B} {330, 480}

\bibitem[\protect\citeauthoryear{Baumgardt}{Baumgardt}{2001}]{baumgardt2001}
Baumgardt H.,  2001, \mn@doi [MNRAS] {10.1046/j.1365-8711.2001.04272.x}, 325,
  1323

\bibitem[\protect\citeauthoryear{Baumgardt}{Baumgardt}{2017}]{Baumgardt2017}
Baumgardt H.,  2017, \mn@doi [MNRAS] {10.1093/mnras/stw2488}, 464, 2174

\bibitem[\protect\citeauthoryear{{Baumgardt} \& {Hilker}}{{Baumgardt} \&
  {Hilker}}{2018}]{Baumgardt2018}
{Baumgardt} H.,  {Hilker} M.,  2018, \mn@doi [\mnras] {10.1093/mnras/sty1057},
  \href {https://ui.adsabs.harvard.edu/abs/2018MNRAS.478.1520B} {478, 1520}

\bibitem[\protect\citeauthoryear{{Baumgardt}, {Hut}, {Makino}, {McMillan}  \&
  {Portegies Zwart}}{{Baumgardt} et~al.}{2003}]{Baumgardt2003a}
{Baumgardt} H.,  {Hut} P.,  {Makino} J.,  {McMillan} S.,   {Portegies Zwart}
  S.,  2003, \mn@doi [\apjl] {10.1086/367537}, \href
  {https://ui.adsabs.harvard.edu/abs/2003ApJ...582L..21B} {582, L21}

\bibitem[\protect\citeauthoryear{{Baumgardt}, {De Marchi}  \&
  {Kroupa}}{{Baumgardt} et~al.}{2008}]{2008ApJ...685..247B}
{Baumgardt} H.,  {De Marchi} G.,   {Kroupa} P.,  2008, \mn@doi [\apj]
  {10.1086/590488}, \href
  {https://ui.adsabs.harvard.edu/abs/2008ApJ...685..247B} {685, 247}

\bibitem[\protect\citeauthoryear{{Baumgardt}, {Parmentier}, {Gieles}  \&
  {Vesperini}}{{Baumgardt} et~al.}{2010}]{2010MNRAS.401.1832B}
{Baumgardt} H.,  {Parmentier} G.,  {Gieles} M.,   {Vesperini} E.,  2010,
  \mn@doi [\mnras] {10.1111/j.1365-2966.2009.15758.x}, \href
  {https://ui.adsabs.harvard.edu/abs/2010MNRAS.401.1832B} {401, 1832}

\bibitem[\protect\citeauthoryear{Behroozi, Wechsler  \& Conroy}{Behroozi
  et~al.}{2013}]{Behroozi2013}
Behroozi P.~S.,  Wechsler R.~H.,   Conroy C.,  2013, \mn@doi [ApJ]
  {10.1088/0004-637X/770/1/57}, 770, 57

\bibitem[\protect\citeauthoryear{Belczynski, Dominik, Bulik, O’Shaughnessy,
  Fryer  \& Holz}{Belczynski et~al.}{2010}]{Belczynski2010}
Belczynski K.,  Dominik M.,  Bulik T.,  O’Shaughnessy R.,  Fryer C.,   Holz
  D.~E.,  2010, \mn@doi [ApJ] {10.1088/2041-8205/715/2/L138}, 715, L138

\bibitem[\protect\citeauthoryear{Bell, McIntosh, Katz  \& Weinberg}{Bell
  et~al.}{2003}]{Bell2003}
Bell E.~F.,  McIntosh D.~H.,  Katz N.,   Weinberg M.~D.,  2003, \mn@doi [\apjs]
  {10.1086/378847}, 149, 289

\bibitem[\protect\citeauthoryear{{Belloni}, {Giersz}, {Askar}, {Leigh}  \&
  {Hypki}}{{Belloni} et~al.}{2016}]{2016MNRAS.462.2950B}
{Belloni} D.,  {Giersz} M.,  {Askar} A.,  {Leigh} N.,   {Hypki} A.,  2016,
  \mn@doi [\mnras] {10.1093/mnras/stw1841}, \href
  {https://ui.adsabs.harvard.edu/abs/2016MNRAS.462.2950B} {462, 2950}

\bibitem[\protect\citeauthoryear{{Belokurov}, {Sanders}, {Fattahi}, {Smith},
  {Deason}, {Evans}  \& {Grand}}{{Belokurov}
  et~al.}{2020}]{2020MNRAS.494.3880B}
{Belokurov} V.,  {Sanders} J.~L.,  {Fattahi} A.,  {Smith} M.~C.,  {Deason}
  A.~J.,  {Evans} N.~W.,   {Grand} R. J.~J.,  2020, \mn@doi [\mnras]
  {10.1093/mnras/staa876}, \href
  {https://ui.adsabs.harvard.edu/abs/2020MNRAS.494.3880B} {494, 3880}

\bibitem[\protect\citeauthoryear{{Binney} \& {Tremaine}}{{Binney} \&
  {Tremaine}}{2008}]{Binney2011}
{Binney} J.,  {Tremaine} S.,  2008, {Galactic Dynamics: Second Edition}.
Princeton University Press

\bibitem[\protect\citeauthoryear{Breen \& Heggie}{Breen \&
  Heggie}{2011}]{Breen2011}
Breen P.~G.,  Heggie D.~C.,  2011, MNRAS, 420, 13

\bibitem[\protect\citeauthoryear{Breen \& Heggie}{Breen \&
  Heggie}{2013}]{Breen2013}
Breen P.~G.,  Heggie D.~C.,  2013, \mn@doi [MNRAS] {10.1093/mnras/stt628}, 432,
  2779

\bibitem[\protect\citeauthoryear{{Breivik} et~al.,}{{Breivik}
  et~al.}{2020}]{Breivik2020}
{Breivik} K.,  et~al., 2020, \mn@doi [\apj] {10.3847/1538-4357/ab9d85}, \href
  {https://ui.adsabs.harvard.edu/abs/2020ApJ...898...71B} {898, 71}

\bibitem[\protect\citeauthoryear{Brodie \& Strader}{Brodie \&
  Strader}{2006}]{Brodie2006}
Brodie J.~P.,  Strader J.,  2006, \mn@doi [ARAA]
  {10.1146/annurev.astro.44.051905.092441}, 44, 193

\bibitem[\protect\citeauthoryear{{Brown} \& {Gnedin}}{{Brown} \&
  {Gnedin}}{2021}]{brown:2021.cluster.mass.radius}
{Brown} G.,  {Gnedin} O.~Y.,  2021, arXiv e-prints, \href
  {https://ui.adsabs.harvard.edu/abs/2021arXiv210612420B} {p. arXiv:2106.12420}

\bibitem[\protect\citeauthoryear{{Caldwell}, {Schiavon}, {Morrison}, {Rose}  \&
  {Harding}}{{Caldwell} et~al.}{2011}]{Caldwell2011}
{Caldwell} N.,  {Schiavon} R.,  {Morrison} H.,  {Rose} J.~A.,   {Harding} P.,
  2011, \mn@doi [\aj] {10.1088/0004-6256/141/2/61}, \href
  {https://ui.adsabs.harvard.edu/abs/2011AJ....141...61C} {141, 61}

\bibitem[\protect\citeauthoryear{Chatterjee, Fregeau, Umbreit  \&
  Rasio}{Chatterjee et~al.}{2010}]{Chatterjee2010}
Chatterjee S.,  Fregeau J.~M.,  Umbreit S.,   Rasio F.~A.,  2010, ApJ, 719, 915

\bibitem[\protect\citeauthoryear{Chernoff \& Weinberg}{Chernoff \&
  Weinberg}{1990}]{Chernoff1990}
Chernoff D.~F.,  Weinberg M.~D.,  1990, \mn@doi [ApJ] {10.1086/168451}, 351,
  121

\bibitem[\protect\citeauthoryear{Claeys, Pols, Izzard, Vink  \& Verbunt}{Claeys
  et~al.}{2014}]{Claeys2014}
Claeys J. S.~W.,  Pols O.~R.,  Izzard R.~G.,  Vink J.,   Verbunt F. W.~M.,
  2014, \mn@doi [Astronomy \& Astrophysics] {10.1051/0004-6361/201322714}, 563,
  A83

\bibitem[\protect\citeauthoryear{{Dotter} et~al.,}{{Dotter}
  et~al.}{2010}]{Dotter2010}
{Dotter} A.,  et~al., 2010, \mn@doi [\apj] {10.1088/0004-637X/708/1/698}, \href
  {https://ui.adsabs.harvard.edu/abs/2010ApJ...708..698D} {708, 698}

\bibitem[\protect\citeauthoryear{{Dotter}, {Sarajedini}  \&
  {Anderson}}{{Dotter} et~al.}{2011}]{Dotter2011}
{Dotter} A.,  {Sarajedini} A.,   {Anderson} J.,  2011, \mn@doi [\apj]
  {10.1088/0004-637X/738/1/74}, \href
  {https://ui.adsabs.harvard.edu/abs/2011ApJ...738...74D} {738, 74}

\bibitem[\protect\citeauthoryear{{Duquennoy} \& {Mayor}}{{Duquennoy} \&
  {Mayor}}{1991}]{1991A&A...248..485D}
{Duquennoy} A.,  {Mayor} M.,  1991, \aap, \href
  {https://ui.adsabs.harvard.edu/abs/1991A&A...248..485D} {500, 337}

\bibitem[\protect\citeauthoryear{Elson, Fall  \& Freeman}{Elson
  et~al.}{1987}]{Elson1987}
Elson R. A.~W.,  Fall S.~M.,   Freeman K.~C.,  1987, \mn@doi [ApJ]
  {10.1086/165807}, 323, 54

\bibitem[\protect\citeauthoryear{{Elson}, {Freeman}  \& {Lauer}}{{Elson}
  et~al.}{1989}]{1989ApJ...347L..69E}
{Elson} R. A.~W.,  {Freeman} K.~C.,   {Lauer} T.~R.,  1989, \mn@doi [\apjl]
  {10.1086/185610}, \href
  {https://ui.adsabs.harvard.edu/abs/1989ApJ...347L..69E} {347, L69}

\bibitem[\protect\citeauthoryear{{Forbes} \& {Bridges}}{{Forbes} \&
  {Bridges}}{2010}]{Forbes2010}
{Forbes} D.~A.,  {Bridges} T.,  2010, \mn@doi [\mnras]
  {10.1111/j.1365-2966.2010.16373.x}, \href
  {https://ui.adsabs.harvard.edu/abs/2010MNRAS.404.1203F} {404, 1203}

\bibitem[\protect\citeauthoryear{{Forbes} et~al.,}{{Forbes}
  et~al.}{2018}]{Forbes2018}
{Forbes} D.~A.,  et~al., 2018, \mn@doi [Proceedings of the Royal Society of
  London Series A] {10.1098/rspa.2017.0616}, \href
  {https://ui.adsabs.harvard.edu/abs/2018RSPSA.47470616F} {474, 20170616}

\bibitem[\protect\citeauthoryear{Fregeau \& Rasio}{Fregeau \&
  Rasio}{2007}]{Fregeau2007}
Fregeau J.~M.,  Rasio F.~A.,  2007, ApJ, 658, 1047

\bibitem[\protect\citeauthoryear{Freitag}{Freitag}{2008}]{Freitag2008}
Freitag M.,  2008, The {Cambridge} {N}-{Body} {Lectures}.
 Vol. 760, Springer Netherlands, Dordrecht, \mn@doi{10.1007/978-1-4020-8431-7},
  \url {http://www.springerlink.com/index/10.1007/978-1-4020-8431-7}

\bibitem[\protect\citeauthoryear{Freitag \& Benz}{Freitag \&
  Benz}{2001}]{Freitag2001}
Freitag M.,  Benz W.,  2001, \mn@doi [A\&A] {10.1051/0004-6361:20010706}, 375,
  711

\bibitem[\protect\citeauthoryear{Fryer \& Kalogera}{Fryer \&
  Kalogera}{2001}]{Fryer2001}
Fryer C.~L.,  Kalogera V.,  2001, \mn@doi [ApJ] {10.1086/321359}, 554, 548

\bibitem[\protect\citeauthoryear{Fryer, Belczynski, Wiktorowicz, Dominik,
  Kalogera  \& Holz}{Fryer et~al.}{2012}]{Fryer2012}
Fryer C.~L.,  Belczynski K.,  Wiktorowicz G.,  Dominik M.,  Kalogera V.,   Holz
  D.~E.,  2012, \mn@doi [ApJ] {10.1088/0004-637X/749/1/91}, 749, 91

\bibitem[\protect\citeauthoryear{Fukushige \& Heggie}{Fukushige \&
  Heggie}{2000}]{fukushige2000}
Fukushige T.,  Heggie D.~C.,  2000, \mn@doi [MNRAS]
  {10.1046/j.1365-8711.2000.03811.x}, 318, 753

\bibitem[\protect\citeauthoryear{{Gao}, {Goodman}, {Cohn}  \& {Murphy}}{{Gao}
  et~al.}{1991}]{Gao1991}
{Gao} B.,  {Goodman} J.,  {Cohn} H.,   {Murphy} B.,  1991, \mn@doi [\apj]
  {10.1086/169843}, \href
  {https://ui.adsabs.harvard.edu/abs/1991ApJ...370..567G} {370, 567}

\bibitem[\protect\citeauthoryear{Gieles, Zwart, Baumgardt, Athanassoula,
  Lamers, Sipior  \& Leenaarts}{Gieles et~al.}{2006}]{Gieles2006}
Gieles M.,  Zwart S. F.~P.,  Baumgardt H.,  Athanassoula E.,  Lamers H. J. G.
  L.~M.,  Sipior M.,   Leenaarts J.,  2006, \mn@doi [MNRAS]
  {10.1111/j.1365-2966.2006.10711.x}, 371, 793

\bibitem[\protect\citeauthoryear{Gieles, Athanassoula  \&
  Portegies~Zwart}{Gieles et~al.}{2007}]{gieles2007}
Gieles M.,  Athanassoula E.,   Portegies~Zwart S.~F.,  2007, \mn@doi [MNRAS]
  {10.1111/j.1365-2966.2007.11477.x}, 376, 809

\bibitem[\protect\citeauthoryear{{Gieles}, {Baumgardt}, {Heggie}  \&
  {Lamers}}{{Gieles} et~al.}{2010}]{2010MNRAS.408L..16G}
{Gieles} M.,  {Baumgardt} H.,  {Heggie} D.~C.,   {Lamers} H. J.~G.~L.~M.,
  2010, \mn@doi [\mnras] {10.1111/j.1745-3933.2010.00919.x}, \href
  {https://ui.adsabs.harvard.edu/abs/2010MNRAS.408L..16G} {408, L16}

\bibitem[\protect\citeauthoryear{{Gieles}, {Heggie}  \& {Zhao}}{{Gieles}
  et~al.}{2011}]{2011MNRAS.413.2509G}
{Gieles} M.,  {Heggie} D.~C.,   {Zhao} H.,  2011, \mn@doi [\mnras]
  {10.1111/j.1365-2966.2011.18320.x}, \href
  {https://ui.adsabs.harvard.edu/abs/2011MNRAS.413.2509G} {413, 2509}

\bibitem[\protect\citeauthoryear{Giersz \& Heggie}{Giersz \&
  Heggie}{2011}]{Giersz2011}
Giersz M.,  Heggie D.~C.,  2011, \mn@doi [MNRAS]
  {10.1111/j.1365-2966.2010.17648.x}, 410, 2698

\bibitem[\protect\citeauthoryear{Giersz, Heggie  \& Hurley}{Giersz
  et~al.}{2008}]{giersz2008}
Giersz M.,  Heggie D.~C.,   Hurley J.~R.,  2008, \mn@doi [MNRAS]
  {10.1111/j.1365-2966.2008.13407.x}, 388, 429

\bibitem[\protect\citeauthoryear{Giersz, Heggie, Hurley  \& Hypki}{Giersz
  et~al.}{2013}]{Giersz2013}
Giersz M.,  Heggie D.~C.,  Hurley J.~R.,   Hypki A.,  2013, \mn@doi [MNRAS]
  {10.1093/mnras/stt307}, 431, 2184

\bibitem[\protect\citeauthoryear{Gnedin \& Ostriker}{Gnedin \&
  Ostriker}{1997}]{Gnedin1997}
Gnedin O.~Y.,  Ostriker J.~P.,  1997, \mn@doi [ApJ] {10.1086/303441}, 474, 223

\bibitem[\protect\citeauthoryear{Gnedin \& Ostriker}{Gnedin \&
  Ostriker}{1999}]{gnedin1999}
Gnedin O.~Y.,  Ostriker J.~P.,  1999, \mn@doi [ApJ] {10.1086/306864}, 513, 626

\bibitem[\protect\citeauthoryear{{Gonz{\'a}lez}, {Kremer}, {Chatterjee},
  {Fragione}, {Rodriguez}, {Weatherford}, {Ye}  \& {Rasio}}{{Gonz{\'a}lez}
  et~al.}{2021}]{2021ApJ...908L..29G}
{Gonz{\'a}lez} E.,  {Kremer} K.,  {Chatterjee} S.,  {Fragione} G.,  {Rodriguez}
  C.~L.,  {Weatherford} N.~C.,  {Ye} C.~S.,   {Rasio} F.~A.,  2021, \mn@doi
  [\apjl] {10.3847/2041-8213/abdf5b}, \href
  {https://ui.adsabs.harvard.edu/abs/2021ApJ...908L..29G} {908, L29}

\bibitem[\protect\citeauthoryear{{Grudi{\'c}}, {Kruijssen},
  {Faucher-Gigu{\`e}re}, {Hopkins}, {Ma}, {Quataert}  \&
  {Boylan-Kolchin}}{{Grudi{\'c}} et~al.}{2021}]{Grudic2021}
{Grudi{\'c}} M.~Y.,  {Kruijssen} J.~M.~D.,  {Faucher-Gigu{\`e}re} C.-A.,
  {Hopkins} P.~F.,  {Ma} X.,  {Quataert} E.,   {Boylan-Kolchin} M.,  2021,
  \mn@doi [\mnras] {10.1093/mnras/stab1894}, \href
  {https://ui.adsabs.harvard.edu/abs/2021MNRAS.506.3239G} {506, 3239}

\bibitem[\protect\citeauthoryear{{Grudi{\'c}}, {Hafen}, {Rodriguez},
  {Guszejnov}, {Lamberts}, {Wetzel}, {Boylan-Kolchin}  \&
  {Faucher-Gigu{\`e}re}}{{Grudi{\'c}} et~al.}{2022}]{GBOF}
{Grudi{\'c}} M.~Y.,  {Hafen} Z.,  {Rodriguez} C.~L.,  {Guszejnov} D.,
  {Lamberts} A.,  {Wetzel} A.,  {Boylan-Kolchin} M.,   {Faucher-Gigu{\`e}re}
  C.-A.,  2022, arXiv e-prints, \href
  {https://ui.adsabs.harvard.edu/abs/2022arXiv220305732G} {p. arXiv:2203.05732}

\bibitem[\protect\citeauthoryear{Grudić, Guszejnov, Hopkins, Lamberts,
  Boylan-Kolchin, Murray  \& Schmitz}{Grudić et~al.}{2018}]{grudic2018}
Grudić M.~Y.,  Guszejnov D.,  Hopkins P.~F.,  Lamberts A.,  Boylan-Kolchin M.,
   Murray N.,   Schmitz D.,  2018, \mn@doi [MNRAS] {10.1093/mnras/sty2303},
  481, 688

\bibitem[\protect\citeauthoryear{Gurvich}{Gurvich}{2021}]{Gurvich:2021}
Gurvich A.~B.,  2021, {FIRE Studio: Movie making utilities for the FIRE
  simulations}, \mn@doi{2022ascl.soft02006G}, \url
  {https://github.com/agurvich/FIRE_studio}

\bibitem[\protect\citeauthoryear{{Gurvich} et~al.,}{{Gurvich}
  et~al.}{2022}]{2022arXiv220304321G}
{Gurvich} A.~B.,  et~al., 2022, arXiv e-prints, \href
  {https://ui.adsabs.harvard.edu/abs/2022arXiv220304321G} {p. arXiv:2203.04321}

\bibitem[\protect\citeauthoryear{{Guszejnov}, {Grudi{\'c}}, {Offner},
  {Boylan-Kolchin}, {Faucher-Gig{\`e}re}, {Wetzel}, {Benincasa}  \&
  {Loebman}}{{Guszejnov} et~al.}{2020}]{guszejnov_GMC_cosmic_evol}
{Guszejnov} D.,  {Grudi{\'c}} M.~Y.,  {Offner} S. S.~R.,  {Boylan-Kolchin} M.,
  {Faucher-Gig{\`e}re} C.-A.,  {Wetzel} A.,  {Benincasa} S.~M.,   {Loebman} S.,
   2020, \mn@doi [\mnras] {10.1093/mnras/stz3527}, \href
  {https://ui.adsabs.harvard.edu/abs/2020MNRAS.492..488G} {492, 488}

\bibitem[\protect\citeauthoryear{{Hafen} et~al.,}{{Hafen}
  et~al.}{2022}]{Hafen2022}
{Hafen} Z.,  et~al., 2022, \mn@doi [\mnras] {10.1093/mnras/stac1603}, \href
  {https://ui.adsabs.harvard.edu/abs/2022MNRAS.514.5056H} {514, 5056}

\bibitem[\protect\citeauthoryear{Harris}{Harris}{1996}]{Harris1996}
Harris W.~E.,  1996, AJ, 112, 1487

\bibitem[\protect\citeauthoryear{Harris}{Harris}{2010}]{Harris2010}
Harris W.~E.,  2010, \mn@doi [Philosophical transactions. Series A,
  Mathematical, physical, and engineering sciences] {10.1098/rsta.2009.0256},
  368, 889

\bibitem[\protect\citeauthoryear{{Harris}, {Harris}  \& {Alessi}}{{Harris}
  et~al.}{2013}]{Harris2013}
{Harris} W.~E.,  {Harris} G. L.~H.,   {Alessi} M.,  2013, \mn@doi [\apj]
  {10.1088/0004-637X/772/2/82}, \href
  {https://ui.adsabs.harvard.edu/abs/2013ApJ...772...82H} {772, 82}

\bibitem[\protect\citeauthoryear{{Harris}, {Blakeslee}  \& {Harris}}{{Harris}
  et~al.}{2017}]{Harris2017}
{Harris} W.~E.,  {Blakeslee} J.~P.,   {Harris} G. L.~H.,  2017, \mn@doi [\apj]
  {10.3847/1538-4357/836/1/67}, \href
  {https://ui.adsabs.harvard.edu/abs/2017ApJ...836...67H} {836, 67}

\bibitem[\protect\citeauthoryear{Heggie}{Heggie}{2014}]{Heggie2014b}
Heggie D.~C.,  2014, \mn@doi [MNRAS] {10.1093/mnras/stu1976}, 445, 3435

\bibitem[\protect\citeauthoryear{Heggie \& Giersz}{Heggie \&
  Giersz}{2014}]{Heggie2014d}
Heggie D.~C.,  Giersz M.,  2014, MNRAS, astro-ph.G, 2459

\bibitem[\protect\citeauthoryear{{Heggie}, {Trenti}  \& {Hut}}{{Heggie}
  et~al.}{2006}]{Heggie2006}
{Heggie} D.~C.,  {Trenti} M.,   {Hut} P.,  2006, \mn@doi [\mnras]
  {10.1111/j.1365-2966.2006.10122.x}, \href
  {https://ui.adsabs.harvard.edu/abs/2006MNRAS.368..677H} {368, 677}

\bibitem[\protect\citeauthoryear{{H{\'e}non}}{{H{\'e}non}}{1961}]{Henon1961}
{H{\'e}non} M.,  1961, Annales d'Astrophysique, \href
  {https://ui.adsabs.harvard.edu/abs/1961AnAp...24..369H} {24, 369}

\bibitem[\protect\citeauthoryear{{Henon}}{{Henon}}{1969}]{Henon1969}
{Henon} M.,  1969, \aap, \href
  {https://ui.adsabs.harvard.edu/abs/1969A&A.....2..151H} {2, 151}

\bibitem[\protect\citeauthoryear{Hobbs, Lorimer, Lyne  \& Kramer}{Hobbs
  et~al.}{2005}]{Hobbs2005}
Hobbs G.,  Lorimer D.~R.,  Lyne A.~G.,   Kramer M.,  2005, \mn@doi [MNRAS]
  {10.1111/j.1365-2966.2005.09087.x}, 360, 974

\bibitem[\protect\citeauthoryear{Hopkins}{Hopkins}{2017}]{Hopkins2017}
Hopkins P.~F.,  2017, arXiv:1712.01294 [astro-ph, physics:physics]

\bibitem[\protect\citeauthoryear{Hopkins, Kereš, Oñorbe, Faucher-Giguère,
  Quataert, Murray  \& Bullock}{Hopkins et~al.}{2014}]{hopkins2014}
Hopkins P.~F.,  Kereš D.,  Oñorbe J.,  Faucher-Giguère C.-A.,  Quataert E.,
  Murray N.,   Bullock J.~S.,  2014, \mn@doi [MNRAS] {10.1093/mnras/stu1738},
  445, 581

\bibitem[\protect\citeauthoryear{{Hopkins} et~al.,}{{Hopkins}
  et~al.}{2018}]{Hopkins2018}
{Hopkins} P.~F.,  et~al., 2018, \mn@doi [\mnras] {10.1093/mnras/sty1690}, \href
  {https://ui.adsabs.harvard.edu/abs/2018MNRAS.480..800H} {480, 800}

\bibitem[\protect\citeauthoryear{{Hopkins} et~al.,}{{Hopkins}
  et~al.}{2020}]{Hopkins2020}
{Hopkins} P.~F.,  et~al., 2020, \mn@doi [\mnras] {10.1093/mnras/stz3321}, \href
  {https://ui.adsabs.harvard.edu/abs/2020MNRAS.492.3465H} {492, 3465}

\bibitem[\protect\citeauthoryear{Hurley, Pols  \& Tout}{Hurley
  et~al.}{2000}]{Hurley2000}
Hurley J.~R.,  Pols O.~R.,   Tout C.~A.,  2000, \mn@doi [MNRAS]
  {10.1046/j.1365-8711.2000.03426.x}, 315, 543

\bibitem[\protect\citeauthoryear{Hurley, Tout, Aarseth  \& Pols}{Hurley
  et~al.}{2001}]{Hurley2001}
Hurley J.~R.,  Tout C.~A.,  Aarseth S.~J.,   Pols O.~R.,  2001, \mn@doi [MNRAS]
  {10.1046/j.1365-8711.2001.04220.x}, 323, 630

\bibitem[\protect\citeauthoryear{Hurley, Tout  \& Pols}{Hurley
  et~al.}{2002}]{Hurley2002}
Hurley J.~R.,  Tout C.~A.,   Pols O.~R.,  2002, \mn@doi [MNRAS]
  {10.1046/j.1365-8711.2002.05038.x}, 329, 897

\bibitem[\protect\citeauthoryear{Hénon}{Hénon}{1971a}]{Henon1971a}
Hénon M.,  1971a, Astrophysics and Space Science, 13, 284

\bibitem[\protect\citeauthoryear{Hénon}{Hénon}{1971b}]{Henon1971}
Hénon M.,  1971b, \mn@doi [Astrophysics and Space Science]
  {10.1007/BF00649201}, 14, 151

\bibitem[\protect\citeauthoryear{{Inagaki} \& {Saslaw}}{{Inagaki} \&
  {Saslaw}}{1985}]{1985ApJ...292..339I}
{Inagaki} S.,  {Saslaw} W.~C.,  1985, \mn@doi [\apj] {10.1086/163164}, \href
  {https://ui.adsabs.harvard.edu/abs/1985ApJ...292..339I} {292, 339}

\bibitem[\protect\citeauthoryear{Joshi, Rasio, Zwart  \& Portegies~Zwart}{Joshi
  et~al.}{2000}]{Joshi1999}
Joshi K.~J.,  Rasio F.~A.,  Zwart S.~P.,   Portegies~Zwart S.,  2000, \mn@doi
  [ApJ] {10.1086/309350}, 540, 969

\bibitem[\protect\citeauthoryear{Kiel \& Hurley}{Kiel \&
  Hurley}{2009}]{Kiel2009}
Kiel P.~D.,  Hurley J.~R.,  2009, \mn@doi [MNRAS]
  {10.1111/j.1365-2966.2009.14711.x}, 395, 2326

\bibitem[\protect\citeauthoryear{{King}}{{King}}{1959}]{King1959}
{King} I.,  1959, \mn@doi [\aj] {10.1086/107954}, \href
  {https://ui.adsabs.harvard.edu/abs/1959AJ.....64..351K} {64, 351}

\bibitem[\protect\citeauthoryear{King}{King}{1966}]{King1966}
King I.~R.,  1966, \mn@doi [AJ] {10.1086/109918}, 71, 276

\bibitem[\protect\citeauthoryear{Kremer, Ye, Chatterjee, Rodriguez  \&
  Rasio}{Kremer et~al.}{2018}]{Kremer2018}
Kremer K.,  Ye C.~S.,  Chatterjee S.,  Rodriguez C.~L.,   Rasio F.~A.,  2018,
  \mn@doi [\apjl] {10.3847/2041-8213/aab26c}, 855, L15

\bibitem[\protect\citeauthoryear{Kremer, Chatterjee, Ye, Rodriguez  \&
  Rasio}{Kremer et~al.}{2019}]{Kremer2019}
Kremer K.,  Chatterjee S.,  Ye C.~S.,  Rodriguez C.~L.,   Rasio F.~A.,  2019,
  \mn@doi [ApJ] {10.3847/1538-4357/aaf646}, 871, 38

\bibitem[\protect\citeauthoryear{Kremer et~al.,}{Kremer
  et~al.}{2020a}]{kremerModelingDenseStar2020}
Kremer K.,  et~al., 2020a, \mn@doi [\apjs] {10.3847/1538-4365/ab7919}, 247, 48

\bibitem[\protect\citeauthoryear{{Kremer} et~al.,}{{Kremer}
  et~al.}{2020b}]{2020ApJ...903...45K}
{Kremer} K.,  et~al., 2020b, \mn@doi [\apj] {10.3847/1538-4357/abb945}, \href
  {https://ui.adsabs.harvard.edu/abs/2020ApJ...903...45K} {903, 45}

\bibitem[\protect\citeauthoryear{Kroupa}{Kroupa}{2001}]{Kroupa2001}
Kroupa P.,  2001

\bibitem[\protect\citeauthoryear{{Kroupa}}{{Kroupa}}{2008}]{2008LNP...760..181K}
{Kroupa} P.,  2008, {Initial Conditions for Star Clusters}.
p.~181, \mn@doi{10.1007/978-1-4020-8431-7\_8}

\bibitem[\protect\citeauthoryear{{Kruijssen}}{{Kruijssen}}{2014}]{Kruijssen2014}
{Kruijssen} J.~M.~D.,  2014, \mn@doi [Classical and Quantum Gravity]
  {10.1088/0264-9381/31/24/244006}, \href
  {https://ui.adsabs.harvard.edu/abs/2014CQGra..31x4006K} {31, 244006}

\bibitem[\protect\citeauthoryear{Kruijssen}{Kruijssen}{2015}]{Kruijssen2015}
Kruijssen J. M.~D.,  2015, \mn@doi [MNRAS] {10.1093/mnras/stv2026}, 454, 1658

\bibitem[\protect\citeauthoryear{Kruijssen, Pelupessy, Lamers, Portegies~Zwart
  \& Icke}{Kruijssen et~al.}{2011}]{kruijssen2011}
Kruijssen J. M.~D.,  Pelupessy F.~I.,  Lamers H. J. G. L.~M.,  Portegies~Zwart
  S.~F.,   Icke V.,  2011, \mn@doi [MNRAS] {10.1111/j.1365-2966.2011.18467.x},
  414, 1339

\bibitem[\protect\citeauthoryear{Kruijssen, Pfeffer, Reina-Campos, Crain  \&
  Bastian}{Kruijssen et~al.}{2019}]{Kruijssen2019}
Kruijssen J. M.~D.,  Pfeffer J.~L.,  Reina-Campos M.,  Crain R.~A.,   Bastian
  N.,  2019, \mn@doi [MNRAS] {10.1093/mnras/sty1609}, 486, 3180

\bibitem[\protect\citeauthoryear{{Kundic} \& {Ostriker}}{{Kundic} \&
  {Ostriker}}{1995}]{1995ApJ...438..702K}
{Kundic} T.,  {Ostriker} J.~P.,  1995, \mn@doi [\apj] {10.1086/175114}, \href
  {https://ui.adsabs.harvard.edu/abs/1995ApJ...438..702K} {438, 702}

\bibitem[\protect\citeauthoryear{{K{\"u}pper}, {Maschberger}, {Kroupa}  \&
  {Baumgardt}}{{K{\"u}pper} et~al.}{2011}]{2011MNRAS.417.2300K}
{K{\"u}pper} A. H.~W.,  {Maschberger} T.,  {Kroupa} P.,   {Baumgardt} H.,
  2011, \mn@doi [\mnras] {10.1111/j.1365-2966.2011.19412.x}, \href
  {https://ui.adsabs.harvard.edu/abs/2011MNRAS.417.2300K} {417, 2300}

\bibitem[\protect\citeauthoryear{{K{\"u}pper}, {Lane}  \&
  {Heggie}}{{K{\"u}pper} et~al.}{2012}]{2012MNRAS.420.2700K}
{K{\"u}pper} A. H.~W.,  {Lane} R.~R.,   {Heggie} D.~C.,  2012, \mn@doi [\mnras]
  {10.1111/j.1365-2966.2011.20242.x}, \href
  {https://ui.adsabs.harvard.edu/abs/2012MNRAS.420.2700K} {420, 2700}

\bibitem[\protect\citeauthoryear{{Lacey} \& {Cole}}{{Lacey} \&
  {Cole}}{1993}]{1993MNRAS.262..627L}
{Lacey} C.,  {Cole} S.,  1993, \mn@doi [\mnras] {10.1093/mnras/262.3.627},
  \href {https://ui.adsabs.harvard.edu/abs/1993MNRAS.262..627L} {262, 627}

\bibitem[\protect\citeauthoryear{{Larsen}}{{Larsen}}{2010}]{Larsen2010}
{Larsen} S.~S.,  2010, \mn@doi [Philosophical Transactions of the Royal Society
  of London Series A] {10.1098/rsta.2009.0255}, \href
  {https://ui.adsabs.harvard.edu/abs/2010RSPTA.368..867L} {368, 867}

\bibitem[\protect\citeauthoryear{{Larsen} \& {Richtler}}{{Larsen} \&
  {Richtler}}{1999}]{Larsen1999}
{Larsen} S.~S.,  {Richtler} T.,  1999, \aap, \href
  {https://ui.adsabs.harvard.edu/abs/1999A&A...345...59L} {345, 59}

\bibitem[\protect\citeauthoryear{{Larsen} \& {Richtler}}{{Larsen} \&
  {Richtler}}{2004}]{Larsen2004}
{Larsen} S.~S.,  {Richtler} T.,  2004, \mn@doi [\aap]
  {10.1051/0004-6361:20040547}, \href
  {https://ui.adsabs.harvard.edu/abs/2004A&A...427..495L} {427, 495}

\bibitem[\protect\citeauthoryear{{Lauer}, {Faber}, {Ajhar}, {Grillmair}  \&
  {Scowen}}{{Lauer} et~al.}{1998}]{1998AJ....116.2263L}
{Lauer} T.~R.,  {Faber} S.~M.,  {Ajhar} E.~A.,  {Grillmair} C.~J.,   {Scowen}
  P.~A.,  1998, \mn@doi [\aj] {10.1086/300617}, \href
  {https://ui.adsabs.harvard.edu/abs/1998AJ....116.2263L} {116, 2263}

\bibitem[\protect\citeauthoryear{{Li} \& {Gnedin}}{{Li} \&
  {Gnedin}}{2019}]{Li2019}
{Li} H.,  {Gnedin} O.~Y.,  2019, \mn@doi [\mnras] {10.1093/mnras/stz1114},
  \href {https://ui.adsabs.harvard.edu/abs/2019MNRAS.486.4030L} {486, 4030}

\bibitem[\protect\citeauthoryear{{Li}, {Gnedin}, {Gnedin}, {Meng}, {Semenov}
  \& {Kravtsov}}{{Li} et~al.}{2017}]{Li2017}
{Li} H.,  {Gnedin} O.~Y.,  {Gnedin} N.~Y.,  {Meng} X.,  {Semenov} V.~A.,
  {Kravtsov} A.~V.,  2017, \mn@doi [\apj] {10.3847/1538-4357/834/1/69}, \href
  {https://ui.adsabs.harvard.edu/abs/2017ApJ...834...69L} {834, 69}

\bibitem[\protect\citeauthoryear{{Li}, {Gnedin}  \& {Gnedin}}{{Li}
  et~al.}{2018}]{Li2018}
{Li} H.,  {Gnedin} O.~Y.,   {Gnedin} N.~Y.,  2018, \mn@doi [\apj]
  {10.3847/1538-4357/aac9b8}, \href
  {https://ui.adsabs.harvard.edu/abs/2018ApJ...861..107L} {861, 107}

\bibitem[\protect\citeauthoryear{{Linden} et~al.,}{{Linden}
  et~al.}{2021}]{Linden2021}
{Linden} S.~T.,  et~al., 2021, \mn@doi [\apj] {10.3847/1538-4357/ac2892}, \href
  {https://ui.adsabs.harvard.edu/abs/2021ApJ...923..278L} {923, 278}

\bibitem[\protect\citeauthoryear{{Long}, {Ostriker}  \& {Aguilar}}{{Long}
  et~al.}{1992}]{LongOstriker1992}
{Long} K.,  {Ostriker} J.~P.,   {Aguilar} L.,  1992, \mn@doi [\apj]
  {10.1086/171159}, \href
  {https://ui.adsabs.harvard.edu/abs/1992ApJ...388..362L} {388, 362}

\bibitem[\protect\citeauthoryear{{Ma}, {Hopkins}, {Wetzel}, {Kirby},
  {Angl{\'e}s-Alc{\'a}zar}, {Faucher-Gigu{\`e}re}, {Kere{\v{s}}}  \&
  {Quataert}}{{Ma} et~al.}{2017}]{Ma2017}
{Ma} X.,  {Hopkins} P.~F.,  {Wetzel} A.~R.,  {Kirby} E.~N.,
  {Angl{\'e}s-Alc{\'a}zar} D.,  {Faucher-Gigu{\`e}re} C.-A.,  {Kere{\v{s}}} D.,
    {Quataert} E.,  2017, \mn@doi [\mnras] {10.1093/mnras/stx273}, \href
  {https://ui.adsabs.harvard.edu/abs/2017MNRAS.467.2430M} {467, 2430}

\bibitem[\protect\citeauthoryear{{Ma}, {Hopkins}, {Ma},
  {Angl{\'e}s-Alc{\'a}zar}, {Faucher-Gigu{\`e}re}  \& {Kelley}}{{Ma}
  et~al.}{2021}]{2021MNRAS.508.1973M}
{Ma} L.,  {Hopkins} P.~F.,  {Ma} X.,  {Angl{\'e}s-Alc{\'a}zar} D.,
  {Faucher-Gigu{\`e}re} C.-A.,   {Kelley} L.~Z.,  2021, \mn@doi [\mnras]
  {10.1093/mnras/stab2713}, \href
  {https://ui.adsabs.harvard.edu/abs/2021MNRAS.508.1973M} {508, 1973}

\bibitem[\protect\citeauthoryear{{Mackey} \& {Gilmore}}{{Mackey} \&
  {Gilmore}}{2003a}]{Mackey2003a}
{Mackey} A.~D.,  {Gilmore} G.~F.,  2003a, \mn@doi [\mnras]
  {10.1046/j.1365-8711.2003.06021.x}, \href
  {https://ui.adsabs.harvard.edu/abs/2003MNRAS.338...85M} {338, 85}

\bibitem[\protect\citeauthoryear{{Mackey} \& {Gilmore}}{{Mackey} \&
  {Gilmore}}{2003b}]{Mackey2003b}
{Mackey} A.~D.,  {Gilmore} G.~F.,  2003b, \mn@doi [\mnras]
  {10.1046/j.1365-8711.2003.06022.x}, \href
  {https://ui.adsabs.harvard.edu/abs/2003MNRAS.338..120M} {338, 120}

\bibitem[\protect\citeauthoryear{Mackey, Wilkinson, Davies  \& Gilmore}{Mackey
  et~al.}{2008}]{Mackey2008}
Mackey A.~D.,  Wilkinson M.~I.,  Davies M.~B.,   Gilmore G.~F.,  2008, \mn@doi
  [MNRAS] {10.1111/j.1365-2966.2008.13052.x}, 386, 65

\bibitem[\protect\citeauthoryear{{Madrid}, {Leigh}, {Hurley}  \&
  {Giersz}}{{Madrid} et~al.}{2017}]{Madrid2017}
{Madrid} J.~P.,  {Leigh} N. W.~C.,  {Hurley} J.~R.,   {Giersz} M.,  2017,
  \mn@doi [\mnras] {10.1093/mnras/stx1350}, \href
  {https://ui.adsabs.harvard.edu/abs/2017MNRAS.470.1729M} {470, 1729}

\bibitem[\protect\citeauthoryear{{Maliszewski}, {Giersz},
  {Gondek-Rosi{\'n}ska}, {Askar}  \& {Hypki}}{{Maliszewski}
  et~al.}{2021}]{2021arXiv211109223M}
{Maliszewski} K.,  {Giersz} M.,  {Gondek-Rosi{\'n}ska} D.,  {Askar} A.,
  {Hypki} A.,  2021, arXiv e-prints, \href
  {https://ui.adsabs.harvard.edu/abs/2021arXiv211109223M} {p. arXiv:2111.09223}

\bibitem[\protect\citeauthoryear{{Meng} \& {Gnedin}}{{Meng} \&
  {Gnedin}}{2022}]{Meng2022}
{Meng} X.,  {Gnedin} O.~Y.,  2022, \mn@doi [\mnras] {10.1093/mnras/stac1751},
  \href {https://ui.adsabs.harvard.edu/abs/2022MNRAS.515.1065M} {515, 1065}

\bibitem[\protect\citeauthoryear{{Merritt}, {Piatek}, {Portegies Zwart}  \&
  {Hemsendorf}}{{Merritt} et~al.}{2004}]{2004ApJ...608L..25M}
{Merritt} D.,  {Piatek} S.,  {Portegies Zwart} S.,   {Hemsendorf} M.,  2004,
  \mn@doi [\apjl] {10.1086/422252}, \href
  {https://ui.adsabs.harvard.edu/abs/2004ApJ...608L..25M} {608, L25}

\bibitem[\protect\citeauthoryear{{Miyamoto} \& {Nagai}}{{Miyamoto} \&
  {Nagai}}{1975}]{1975PASJ...27..533M}
{Miyamoto} M.,  {Nagai} R.,  1975, \pasj, \href
  {https://ui.adsabs.harvard.edu/abs/1975PASJ...27..533M} {27, 533}

\bibitem[\protect\citeauthoryear{{Moe}, {Kratter}  \& {Badenes}}{{Moe}
  et~al.}{2019}]{Moe2019}
{Moe} M.,  {Kratter} K.~M.,   {Badenes} C.,  2019, \mn@doi [\apj]
  {10.3847/1538-4357/ab0d88}, \href
  {https://ui.adsabs.harvard.edu/abs/2019ApJ...875...61M} {875, 61}

\bibitem[\protect\citeauthoryear{Morscher, Umbreit, Farr  \& Rasio}{Morscher
  et~al.}{2013}]{Morscher2012}
Morscher M.,  Umbreit S.,  Farr W.~M.,   Rasio F.~A.,  2013, \mn@doi [ApJ]
  {10.1088/2041-8205/763/1/L15}, 763, L15

\bibitem[\protect\citeauthoryear{Morscher, Pattabiraman, Rodriguez, Rasio  \&
  Umbreit}{Morscher et~al.}{2015}]{Morscher2015}
Morscher M.,  Pattabiraman B.,  Rodriguez C.,  Rasio F.~A.,   Umbreit S.,
  2015, \mn@doi [ApJ] {10.1088/0004-637X/800/1/9}, 800, 9

\bibitem[\protect\citeauthoryear{{Neumayer}, {Seth}  \& {B{\"o}ker}}{{Neumayer}
  et~al.}{2020}]{2020A&ARv..28....4N}
{Neumayer} N.,  {Seth} A.,   {B{\"o}ker} T.,  2020, \mn@doi [\aapr]
  {10.1007/s00159-020-00125-0}, \href
  {https://ui.adsabs.harvard.edu/abs/2020A&ARv..28....4N} {28, 4}

\bibitem[\protect\citeauthoryear{{Ostriker} \& {Caldwell}}{{Ostriker} \&
  {Caldwell}}{1983}]{Ostriker1983}
{Ostriker} J.~P.,  {Caldwell} J.~A.~R.,  1983, in {Shuter} W.~L.~H.,  ed.,
  Astrophysics and Space Science Library Vol. 100, Kinematics, Dynamics and
  Structure of the Milky Way. pp 249--257,
  \mn@doi{10.1007/978-94-009-7060-1\_35}

\bibitem[\protect\citeauthoryear{{Ostriker}, {Spitzer}  \&
  {Chevalier}}{{Ostriker} et~al.}{1972}]{Ostriker1972}
{Ostriker} J.~P.,  {Spitzer} Lyman J.,   {Chevalier} R.~A.,  1972, \mn@doi
  [\apjl] {10.1086/181018}, \href
  {https://ui.adsabs.harvard.edu/abs/1972ApJ...176L..51O} {176, L51}

\bibitem[\protect\citeauthoryear{Pattabiraman, Umbreit, Liao, Choudhary,
  Kalogera, Memik  \& Rasio}{Pattabiraman et~al.}{2013}]{Pattabiraman2013}
Pattabiraman B.,  Umbreit S.,  Liao W.-k.,  Choudhary A.,  Kalogera V.,  Memik
  G.,   Rasio F.~A.,  2013, \mn@doi [\apjs] {10.1088/0067-0049/204/2/15}, 204,
  15

\bibitem[\protect\citeauthoryear{{Peacock}, {Maccarone}, {Knigge}, {Kundu},
  {Waters}, {Zepf}  \& {Zurek}}{{Peacock} et~al.}{2010}]{Peacock2010}
{Peacock} M.~B.,  {Maccarone} T.~J.,  {Knigge} C.,  {Kundu} A.,  {Waters}
  C.~Z.,  {Zepf} S.~E.,   {Zurek} D.~R.,  2010, \mn@doi [\mnras]
  {10.1111/j.1365-2966.2009.15952.x}, \href
  {https://ui.adsabs.harvard.edu/abs/2010MNRAS.402..803P} {402, 803}

\bibitem[\protect\citeauthoryear{{Pfeffer}, {Kruijssen}, {Crain}  \&
  {Bastian}}{{Pfeffer} et~al.}{2018}]{Pfeffer2018}
{Pfeffer} J.,  {Kruijssen} J.~M.~D.,  {Crain} R.~A.,   {Bastian} N.,  2018,
  \mn@doi [\mnras] {10.1093/mnras/stx3124}, \href
  {https://ui.adsabs.harvard.edu/abs/2018MNRAS.475.4309P} {475, 4309}

\bibitem[\protect\citeauthoryear{{Plummer}}{{Plummer}}{1911}]{Plummer1911}
{Plummer} H.~C.,  1911, \mn@doi [\mnras] {10.1093/mnras/71.5.460}, \href
  {https://ui.adsabs.harvard.edu/abs/1911MNRAS..71..460P} {71, 460}

\bibitem[\protect\citeauthoryear{{Portegies Zwart}, {Baumgardt}, {Hut},
  {Makino}  \& {McMillan}}{{Portegies Zwart}
  et~al.}{2004}]{2004Natur.428..724P}
{Portegies Zwart} S.~F.,  {Baumgardt} H.,  {Hut} P.,  {Makino} J.,   {McMillan}
  S. L.~W.,  2004, \mn@doi [\nat] {10.1038/nature02448}, \href
  {https://ui.adsabs.harvard.edu/abs/2004Natur.428..724P} {428, 724}

\bibitem[\protect\citeauthoryear{Portegies~Zwart, McMillan  \&
  Gieles}{Portegies~Zwart et~al.}{2010}]{PortegiesZwart2010}
Portegies~Zwart S.~F.,  McMillan S.~L.,   Gieles M.,  2010, \mn@doi [ARAA]
  {10.1146/annurev-astro-081309-130834}, 48, 431

\bibitem[\protect\citeauthoryear{Price-Whelan}{Price-Whelan}{2017}]{gala}
Price-Whelan A.~M.,  2017, \mn@doi [The Journal of Open Source Software]
  {10.21105/joss.00388}, 2

\bibitem[\protect\citeauthoryear{Prieto \& Gnedin}{Prieto \&
  Gnedin}{2008}]{Prieto2008}
Prieto J.~L.,  Gnedin O.~Y.,  2008, \mn@doi [ApJ] {10.1086/591777}, 689, 919

\bibitem[\protect\citeauthoryear{Renaud \& Gieles}{Renaud \&
  Gieles}{2015}]{Renaud2015}
Renaud F.,  Gieles M.,  2015, \mn@doi [MNRAS] {10.1093/mnras/stv245}, 448, 3416

\bibitem[\protect\citeauthoryear{Renaud, Gieles  \& Boily}{Renaud
  et~al.}{2011}]{Renaud2011}
Renaud F.,  Gieles M.,   Boily C.~M.,  2011, \mn@doi [MNRAS]
  {10.1111/j.1365-2966.2011.19531.x}, 418, 759

\bibitem[\protect\citeauthoryear{{Richtler}, {Bassino}, {Dirsch}  \&
  {Kumar}}{{Richtler} et~al.}{2012}]{Richtler2012}
{Richtler} T.,  {Bassino} L.~P.,  {Dirsch} B.,   {Kumar} B.,  2012, \mn@doi
  [\aap] {10.1051/0004-6361/201118589}, \href
  {https://ui.adsabs.harvard.edu/abs/2012A&A...543A.131R} {543, A131}

\bibitem[\protect\citeauthoryear{Rieder \& Pelupessy}{Rieder \&
  Pelupessy}{2019}]{steven_rieder_2019_3553805}
Rieder S.,  Pelupessy I.,  2019, rieder/Fresco: Fresco 0.6,
  \mn@doi{10.5281/zenodo.3553805}, \url
  {https://doi.org/10.5281/zenodo.3553805}

\bibitem[\protect\citeauthoryear{{Rieder}, {Ishiyama}, {Langelaan}, {Makino},
  {McMillan}  \& {Portegies Zwart}}{{Rieder} et~al.}{2013}]{Rieder2013}
{Rieder} S.,  {Ishiyama} T.,  {Langelaan} P.,  {Makino} J.,  {McMillan} S.
  L.~W.,   {Portegies Zwart} S.,  2013, \mn@doi [\mnras]
  {10.1093/mnras/stt1848}, \href
  {https://ui.adsabs.harvard.edu/abs/2013MNRAS.436.3695R} {436, 3695}

\bibitem[\protect\citeauthoryear{Rodriguez, Chatterjee  \& Rasio}{Rodriguez
  et~al.}{2016a}]{Rodriguez2016a}
Rodriguez C.~L.,  Chatterjee S.,   Rasio F.~A.,  2016a, \mn@doi [\prd]
  {10.1103/PhysRevD.93.084029}, 93, 084029

\bibitem[\protect\citeauthoryear{Rodriguez, Morscher, Wang, Chatterjee, Rasio
  \& Spurzem}{Rodriguez et~al.}{2016b}]{Rodriguez2016}
Rodriguez C.~L.,  Morscher M.,  Wang L.,  Chatterjee S.,  Rasio F.~A.,
  Spurzem R.,  2016b, \mn@doi [MNRAS] {10.1093/mnras/stw2121}, 463, 2109

\bibitem[\protect\citeauthoryear{Rodriguez, Zevin, Pankow, Kalogera  \&
  Rasio}{Rodriguez et~al.}{2016c}]{Rodriguez2016c}
Rodriguez C.~L.,  Zevin M.,  Pankow C.,  Kalogera V.,   Rasio F.~A.,  2016c,
  \mn@doi [ApJ] {10.3847/2041-8205/832/1/L2}, 832, L2

\bibitem[\protect\citeauthoryear{Rodriguez, Amaro-Seoane, Chatterjee, Kremer,
  Rasio, Samsing, Ye  \& Zevin}{Rodriguez et~al.}{2018a}]{Rodriguez2018c}
Rodriguez C.~L.,  Amaro-Seoane P.,  Chatterjee S.,  Kremer K.,  Rasio F.~A.,
  Samsing J.,  Ye C.~S.,   Zevin M.,  2018a, \mn@doi [\prd]
  {10.1103/PhysRevD.98.123005}, 98, 123005

\bibitem[\protect\citeauthoryear{Rodriguez, Amaro-Seoane, Chatterjee  \&
  Rasio}{Rodriguez et~al.}{2018b}]{Rodriguez2018}
Rodriguez C.~L.,  Amaro-Seoane P.,  Chatterjee S.,   Rasio F.~A.,  2018b,
  \mn@doi [\prl] {10.1103/PhysRevLett.120.151101}, 120, 151101

\bibitem[\protect\citeauthoryear{{Rodriguez} et~al.,}{{Rodriguez}
  et~al.}{2020}]{Rodriguez2020}
{Rodriguez} C.~L.,  et~al., 2020, \mn@doi [\apjl] {10.3847/2041-8213/ab961d},
  \href {https://ui.adsabs.harvard.edu/abs/2020ApJ...896L..10R} {896, L10}

\bibitem[\protect\citeauthoryear{{Rodriguez} et~al.,}{{Rodriguez}
  et~al.}{2022}]{Rodriguez2022}
{Rodriguez} C.~L.,  et~al., 2022, \mn@doi [\apjs] {10.3847/1538-4365/ac2edf},
  \href {https://ui.adsabs.harvard.edu/abs/2022ApJS..258...22R} {258, 22}

\bibitem[\protect\citeauthoryear{{Rui}, {Kremer}, {Weatherford}, {Chatterjee},
  {Rasio}, {Rodriguez}  \& {Ye}}{{Rui} et~al.}{2021a}]{2021zndo...4579950R}
{Rui} N.~Z.,  {Kremer} K.,  {Weatherford} N.~C.,  {Chatterjee} S.,  {Rasio}
  F.~A.,  {Rodriguez} C.~L.,   {Ye} C.~S.,  2021a, {NicholasRui/cmctoolkit:
  First release}, \mn@doi{10.5281/zenodo.4579950}

\bibitem[\protect\citeauthoryear{{Rui}, {Kremer}, {Weatherford}, {Chatterjee},
  {Rasio}, {Rodriguez}  \& {Ye}}{{Rui} et~al.}{2021b}]{Rui2021}
{Rui} N.~Z.,  {Kremer} K.,  {Weatherford} N.~C.,  {Chatterjee} S.,  {Rasio}
  F.~A.,  {Rodriguez} C.~L.,   {Ye} C.~S.,  2021b, \mn@doi [\apj]
  {10.3847/1538-4357/abed49}, \href
  {https://ui.adsabs.harvard.edu/abs/2021ApJ...912..102R} {912, 102}

\bibitem[\protect\citeauthoryear{{Ryon} et~al.,}{{Ryon}
  et~al.}{2015}]{Ryon2015}
{Ryon} J.~E.,  et~al., 2015, \mn@doi [\mnras] {10.1093/mnras/stv1282}, \href
  {https://ui.adsabs.harvard.edu/abs/2015MNRAS.452..525R} {452, 525}

\bibitem[\protect\citeauthoryear{{Samsing}, {D'Orazio}, {Kremer}, {Rodriguez}
  \& {Askar}}{{Samsing} et~al.}{2020}]{Samsing2019}
{Samsing} J.,  {D'Orazio} D.~J.,  {Kremer} K.,  {Rodriguez} C.~L.,   {Askar}
  A.,  2020, \mn@doi [\prd] {10.1103/PhysRevD.101.123010}, \href
  {https://ui.adsabs.harvard.edu/abs/2020PhRvD.101l3010S} {101, 123010}

\bibitem[\protect\citeauthoryear{{Santistevan}, {Wetzel}, {El-Badry},
  {Bland-Hawthorn}, {Boylan-Kolchin}, {Bailin}, {Faucher-Gigu{\`e}re}  \&
  {Benincasa}}{{Santistevan} et~al.}{2020}]{santistevan:2020.fire.m12.sfh}
{Santistevan} I.~B.,  {Wetzel} A.,  {El-Badry} K.,  {Bland-Hawthorn} J.,
  {Boylan-Kolchin} M.,  {Bailin} J.,  {Faucher-Gigu{\`e}re} C.-A.,
  {Benincasa} S.,  2020, \mn@doi [\mnras] {10.1093/mnras/staa1923}, \href
  {https://ui.adsabs.harvard.edu/abs/2020MNRAS.497..747S} {497, 747}

\bibitem[\protect\citeauthoryear{{Sch{\"o}del}, {Gallego-Cano}, {Dong},
  {Nogueras-Lara}, {Gallego-Calvente}, {Amaro-Seoane}  \&
  {Baumgardt}}{{Sch{\"o}del} et~al.}{2018}]{2018A&A...609A..27S}
{Sch{\"o}del} R.,  {Gallego-Cano} E.,  {Dong} H.,  {Nogueras-Lara} F.,
  {Gallego-Calvente} A.~T.,  {Amaro-Seoane} P.,   {Baumgardt} H.,  2018,
  \mn@doi [\aap] {10.1051/0004-6361/201730452}, \href
  {https://ui.adsabs.harvard.edu/abs/2018A&A...609A..27S} {609, A27}

\bibitem[\protect\citeauthoryear{{Sollima} \& {Mastrobuono Battisti}}{{Sollima}
  \& {Mastrobuono Battisti}}{2014}]{2014MNRAS.443.3513S}
{Sollima} A.,  {Mastrobuono Battisti} A.,  2014, \mn@doi [\mnras]
  {10.1093/mnras/stu1426}, \href
  {https://ui.adsabs.harvard.edu/abs/2014MNRAS.443.3513S} {443, 3513}

\bibitem[\protect\citeauthoryear{{Spitzer}}{{Spitzer}}{1958}]{Spitzer1958}
{Spitzer} Lyman J.,  1958, \mn@doi [\apj] {10.1086/146435}, \href
  {https://ui.adsabs.harvard.edu/abs/1958ApJ...127...17S} {127, 17}

\bibitem[\protect\citeauthoryear{Spitzer}{Spitzer}{1987}]{spitzer_dynamical_1987}
Spitzer L.,  1987, Princeton, NJ, Princeton University Press, 1987, 191 p.

\bibitem[\protect\citeauthoryear{{Spitzer} \& {Hart}}{{Spitzer} \&
  {Hart}}{1971}]{1971ApJ...164..399S}
{Spitzer} Lyman J.,  {Hart} M.~H.,  1971, \mn@doi [\apj] {10.1086/150855},
  \href {https://ui.adsabs.harvard.edu/abs/1971ApJ...164..399S} {164, 399}

\bibitem[\protect\citeauthoryear{{Stern} et~al.,}{{Stern}
  et~al.}{2021}]{Stern2021}
{Stern} J.,  et~al., 2021, \mn@doi [\apj] {10.3847/1538-4357/abd776}, \href
  {https://ui.adsabs.harvard.edu/abs/2021ApJ...911...88S} {911, 88}

\bibitem[\protect\citeauthoryear{{Terlevich}}{{Terlevich}}{1987}]{1987MNRAS.224..193T}
{Terlevich} E.,  1987, \mn@doi [\mnras] {10.1093/mnras/224.1.193}, \href
  {https://ui.adsabs.harvard.edu/abs/1987MNRAS.224..193T} {224, 193}

\bibitem[\protect\citeauthoryear{{Trager}, {King}  \& {Djorgovski}}{{Trager}
  et~al.}{1995}]{1995AJ....109..218T}
{Trager} S.~C.,  {King} I.~R.,   {Djorgovski} S.,  1995, \mn@doi [\aj]
  {10.1086/117268}, \href
  {https://ui.adsabs.harvard.edu/abs/1995AJ....109..218T} {109, 218}

\bibitem[\protect\citeauthoryear{{Tremaine}, {Ostriker}  \&
  {Spitzer}}{{Tremaine} et~al.}{1975}]{1975ApJ...196..407T}
{Tremaine} S.~D.,  {Ostriker} J.~P.,   {Spitzer} L. J.,  1975, \mn@doi [\apj]
  {10.1086/153422}, \href
  {https://ui.adsabs.harvard.edu/abs/1975ApJ...196..407T} {196, 407}

\bibitem[\protect\citeauthoryear{{Tremmel}, {Governato}, {Volonteri}  \&
  {Quinn}}{{Tremmel} et~al.}{2015}]{2015MNRAS.451.1868T}
{Tremmel} M.,  {Governato} F.,  {Volonteri} M.,   {Quinn} T.~R.,  2015, \mn@doi
  [\mnras] {10.1093/mnras/stv1060}, \href
  {https://ui.adsabs.harvard.edu/abs/2015MNRAS.451.1868T} {451, 1868}

\bibitem[\protect\citeauthoryear{{Trenti}, {Heggie}  \& {Hut}}{{Trenti}
  et~al.}{2007}]{Trenti2007}
{Trenti} M.,  {Heggie} D.~C.,   {Hut} P.,  2007, \mn@doi [\mnras]
  {10.1111/j.1365-2966.2006.11166.x}, \href
  {https://ui.adsabs.harvard.edu/abs/2007MNRAS.374..344T} {374, 344}

\bibitem[\protect\citeauthoryear{{VandenBerg}, {Brogaard}, {Leaman}  \&
  {Casagrande}}{{VandenBerg} et~al.}{2013}]{Vandenberg2013}
{VandenBerg} D.~A.,  {Brogaard} K.,  {Leaman} R.,   {Casagrande} L.,  2013,
  \mn@doi [\apj] {10.1088/0004-637X/775/2/134}, \href
  {https://ui.adsabs.harvard.edu/abs/2013ApJ...775..134V} {775, 134}

\bibitem[\protect\citeauthoryear{{Vasiliev} \& {Baumgardt}}{{Vasiliev} \&
  {Baumgardt}}{2021}]{Vasiliev2021}
{Vasiliev} E.,  {Baumgardt} H.,  2021, \mn@doi [\mnras]
  {10.1093/mnras/stab1475}, \href
  {https://ui.adsabs.harvard.edu/abs/2021MNRAS.505.5978V} {505, 5978}

\bibitem[\protect\citeauthoryear{{Vesperini} \& {Heggie}}{{Vesperini} \&
  {Heggie}}{1997}]{Vesperini1997}
{Vesperini} E.,  {Heggie} D.~C.,  1997, \mn@doi [\mnras]
  {10.1093/mnras/289.4.898}, \href
  {https://ui.adsabs.harvard.edu/abs/1997MNRAS.289..898V} {289, 898}

\bibitem[\protect\citeauthoryear{{Vesperini}, {McMillan}  \& {Portegies
  Zwart}}{{Vesperini} et~al.}{2009}]{2009ApJ...698..615V}
{Vesperini} E.,  {McMillan} S. L.~W.,   {Portegies Zwart} S.,  2009, \mn@doi
  [\apj] {10.1088/0004-637X/698/1/615}, \href
  {https://ui.adsabs.harvard.edu/abs/2009ApJ...698..615V} {698, 615}

\bibitem[\protect\citeauthoryear{Vink, de Koter  \& Lamers}{Vink
  et~al.}{2001}]{Vink2001}
Vink J.~S.,  de Koter A.,   Lamers H. J. G. L.~M.,  2001, \mn@doi [A\&A]
  {10.1051/0004-6361:20010127}, 369, 574

\bibitem[\protect\citeauthoryear{Wang, Jia  \& Li}{Wang
  et~al.}{2016a}]{Wang2016a}
Wang C.,  Jia K.,   Li X.-D.,  2016a, p.~10

\bibitem[\protect\citeauthoryear{Wang et~al.,}{Wang et~al.}{2016b}]{Wang2016}
Wang L.,  et~al., 2016b, \mn@doi [MNRAS] {10.1093/mnras/stw274}, 458, 1450

\bibitem[\protect\citeauthoryear{{Wang}, {Iwasawa}, {Nitadori}  \&
  {Makino}}{{Wang} et~al.}{2020}]{2020MNRAS.497..536W}
{Wang} L.,  {Iwasawa} M.,  {Nitadori} K.,   {Makino} J.,  2020, \mn@doi
  [\mnras] {10.1093/mnras/staa1915}, \href
  {https://ui.adsabs.harvard.edu/abs/2020MNRAS.497..536W} {497, 536}

\bibitem[\protect\citeauthoryear{{Weatherford}, {Chatterjee}, {Kremer}  \&
  {Rasio}}{{Weatherford} et~al.}{2020}]{Weatherford2020}
{Weatherford} N.~C.,  {Chatterjee} S.,  {Kremer} K.,   {Rasio} F.~A.,  2020,
  \mn@doi [\apj] {10.3847/1538-4357/ab9f98}, \href
  {https://ui.adsabs.harvard.edu/abs/2020ApJ...898..162W} {898, 162}

\bibitem[\protect\citeauthoryear{{Webb}, {Harris}, {Sills}  \& {Hurley}}{{Webb}
  et~al.}{2013}]{Webb2013}
{Webb} J.~J.,  {Harris} W.~E.,  {Sills} A.,   {Hurley} J.~R.,  2013, \mn@doi
  [\apj] {10.1088/0004-637X/764/2/124}, \href
  {https://ui.adsabs.harvard.edu/abs/2013ApJ...764..124W} {764, 124}

\bibitem[\protect\citeauthoryear{Webb, Sills, Harris  \& Hurley}{Webb
  et~al.}{2014}]{Webb2014}
Webb J.~J.,  Sills A.,  Harris W.~E.,   Hurley J.~R.,  2014, \mn@doi [MNRAS]
  {10.1093/mnras/stu1763}, 445, 1048

\bibitem[\protect\citeauthoryear{{Weinberg}}{{Weinberg}}{1994a}]{1994AJ....108.1398W}
{Weinberg} M.~D.,  1994a, \mn@doi [\aj] {10.1086/117161}, \href
  {https://ui.adsabs.harvard.edu/abs/1994AJ....108.1398W} {108, 1398}

\bibitem[\protect\citeauthoryear{{Weinberg}}{{Weinberg}}{1994b}]{weinberg2}
{Weinberg} M.~D.,  1994b, \mn@doi [\aj] {10.1086/117162}, \href
  {https://ui.adsabs.harvard.edu/abs/1994AJ....108.1403W} {108, 1403}

\bibitem[\protect\citeauthoryear{{Weinberg}}{{Weinberg}}{1994c}]{weinberg3}
{Weinberg} M.~D.,  1994c, \mn@doi [\aj] {10.1086/117163}, \href
  {https://ui.adsabs.harvard.edu/abs/1994AJ....108.1414W} {108, 1414}

\bibitem[\protect\citeauthoryear{Wetzel, Hopkins, Kim, {Faucher-Gigu{\`e}re},
  Kere{\v s}  \& Quataert}{Wetzel et~al.}{2016}]{Wetzel2016}
Wetzel A.~R.,  Hopkins P.~F.,  Kim J.-h.,  {Faucher-Gigu{\`e}re} C.-A.,
  Kere{\v s} D.,   Quataert E.,  2016, \mn@doi [ApJ]
  {10.3847/2041-8205/827/2/L23}, 827, L23

\bibitem[\protect\citeauthoryear{{Wilkinson}, {Hurley}, {Mackey}, {Gilmore}  \&
  {Tout}}{{Wilkinson} et~al.}{2003}]{Wilkinson2003}
{Wilkinson} M.~I.,  {Hurley} J.~R.,  {Mackey} A.~D.,  {Gilmore} G.~F.,   {Tout}
  C.~A.,  2003, \mn@doi [\mnras] {10.1046/j.1365-8711.2003.06749.x}, \href
  {https://ui.adsabs.harvard.edu/abs/2003MNRAS.343.1025W} {343, 1025}

\bibitem[\protect\citeauthoryear{{Ye}, {Kremer}, {Rodriguez}, {Rui},
  {Weatherford}, {Chatterjee}, {Fragione}  \& {Rasio}}{{Ye}
  et~al.}{2021}]{Ye2021}
{Ye} C.~S.,  {Kremer} K.,  {Rodriguez} C.~L.,  {Rui} N.~Z.,  {Weatherford}
  N.~C.,  {Chatterjee} S.,  {Fragione} G.,   {Rasio} F.~A.,  2021, arXiv
  e-prints, \href {https://ui.adsabs.harvard.edu/abs/2021arXiv211005495Y} {p.
  arXiv:2110.05495}

\bibitem[\protect\citeauthoryear{{Yu} et~al.,}{{Yu} et~al.}{2021}]{Yu2021}
{Yu} S.,  et~al., 2021, \mn@doi [\mnras] {10.1093/mnras/stab1339}, \href
  {https://ui.adsabs.harvard.edu/abs/2021MNRAS.505..889Y} {505, 889}

\bibitem[\protect\citeauthoryear{{Zinn}}{{Zinn}}{1985}]{Zinn1985}
{Zinn} R.,  1985, \mn@doi [\apj] {10.1086/163249}, \href
  {https://ui.adsabs.harvard.edu/abs/1985ApJ...293..424Z} {293, 424}

\bibitem[\protect\citeauthoryear{{van der Marel}, {Sigurdsson}  \&
  {Hernquist}}{{van der Marel} et~al.}{1997}]{VanderMarel1997}
{van der Marel} R.~P.,  {Sigurdsson} S.,   {Hernquist} L.,  1997, \mn@doi
  [\apj] {10.1086/304605}, \href
  {https://ui.adsabs.harvard.edu/abs/1997ApJ...487..153V} {487, 153}

\makeatother
\end{thebibliography}

\appendix

\section{Tidal Stripping in Spherical $N$-body Codes}
\label{apx:stripping}

\begin{figure*}
\centering
\includegraphics[scale=1.]{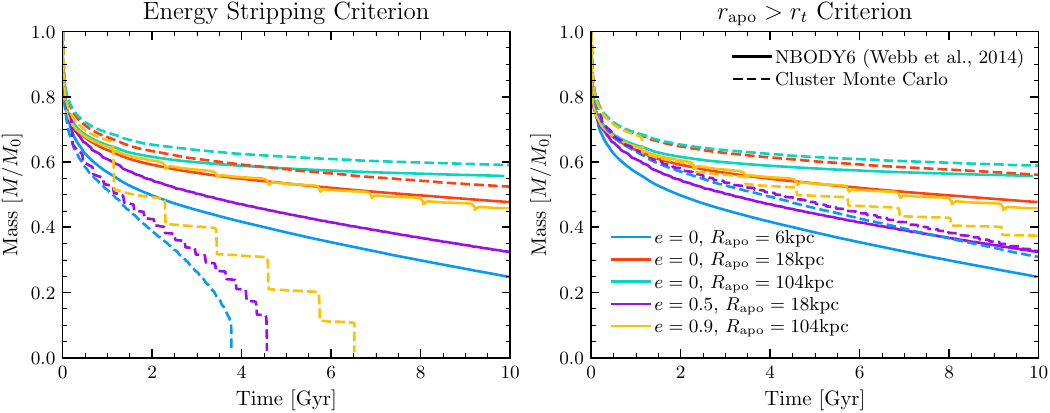}
\caption{Comparison between direct $N$-body simulations of \citet{Webb2014} and \texttt{CMC} models evolved with the same initial conditions, galactic potential, and orbit.  The different colors correspond to orbits with different eccentricities ($e$) and apogalacticons ($R_{\rm apo}$) in the galactic potential described in the text.  On the \textbf{left}, we show the normalized mass loss rates for both approaches using the energy criterion of tidal mass loss from \citet{giersz2008}, while on the \textbf{right} we show the same using the more simplistic radial criterion, where any stars with $r_{\rm apo} > r_t$ are stripped.  Surprisingly, the radial criterion matches the $N$-body mass loss rates significantly better than the default energy criterion, and as such we adopt it for this study.   Note that the $N$-body models on highly-eccentric orbits show apparent mass \emph{gain} after they pass through perigalacticon.  This reflects the recapture of stars by the cluster potential after a close perigalactic passage \citep[e.g.][]{2012MNRAS.420.2700K}, which cannot be modeled in \texttt{CMC} at present.}
\label{fig:tidal}
\end{figure*}

While \texttt{CMC} has been shown to produce extremely accurate models of spherical star clusters (when compared to direct $N$-body integrators), a large part of the speed of the method derives from the assumption of spherical symmetry.  Of course, the application of an external tidal potential is explicitly non-spherical.  For direct $N$-body integrators, this does not pose a significant issue: the external field can either be incorporated into the cluster using the instantaneous tidal tensor \citep[e.g.,][]{Renaud2011}, or directly using the galactic potential itself \citep[e.g.][]{Renaud2015,2020MNRAS.497..536W}, with the external force being added directly to the equations of motion.  However, for \texttt{CMC} (and other approaches that rely on spherical symmetry), the forces cannot be added in 3 dimensions self-consistently, which prohibits us from calculating when stars move beyond the classic Roche-Lobe shape of the energy surface.  

What Equation \eqref{eqn:rt} provides us with is the distance from the cluster center to the L1 Lagrange point of the cluster-galaxy system where the combined gravitational influence of both potentials and the centrifugal force from the cluster’s galactic orbit sum to zero (while neglecting the Coriolis and Eulerian pseudo forces that arise in the rotating reference frame of the cluster). We assess two methods of implementing tidal stripping. The most naive way to impose this tidal boundary on our clusters would be to simply remove any stars whose apocenter lies beyond that tidal boundary.  However as stated in \S \ref{ss:tides} the criterion for whether a star is bound to the cluster relies on the energy of the orbit, not its position.  As such, detailed comparisons to direct $N$-body simulations have suggested that a modified energy criterion, where any stars with energies greater than $\alpha \phi_t$ are stripped from the cluster, is often the preferred approach to tidal stripping  \citep{giersz2008,Chatterjee2010}.  Here, $\phi_t$ is the potential of the cluster at the tidal boundary $r_t$ defined in Equation \eqref{eqn:rt}, and $\alpha$ is a parameter tuned to direct $N$-body simulations, given by 

\begin{equation}
\alpha = 1.5 - 3\left(\frac{\log(\Lambda)}{N}\right)^{1/4}
\end{equation}

\noindent This is the default prescription in both \texttt{CMC} and the \texttt{MOCCA} codes, and has been shown to reproduce the mass loss rates of clusters on circular orbits \citep{Rodriguez2016c}.

To ensure that our implementation of the tidal tensor in \texttt{CMC} can have similar success, we compare the results of idealized Monte Carlo integrations to direct $N$-body models from \citet{Webb2014}.  In that work, the authors used \texttt{NBODY6} \citep{Aarseth2003} to explore the influence of non-circular galactic orbits on the mass loss rates from idealized models of star clusters (comprised of equal-mass point particles).  They consider 5 different orbits with varying eccentricities and pericenters in a galactic potential consisting of a disk, bulge, and halo with the default parameters from \texttt{NBODY6}; we reproduce this potential using a potential comprised of a $1.5\times10^{10} M_{\odot}$ Keplerian potential for the bulge/SMBH, a logarithmic potential that yields a velocity of $220$ km$/$s at 8.5 Kpc from the galactic center, and a \citet{1975PASJ...27..533M} disk potential with mass $5\times10^{10}M_{\odot}$ and scale parameters $a=4.5$ Kpc and $b=0.5$ Kpc.   The orbits are integrated using the \texttt{Gala} \citep{gala} package, which also allows us to extract the tidal tensor at each point along the orbit.  This is then fed directly into \texttt{CMC}, allowing us to reproduce the initial conditions of \citet{Webb2014}.  

In Fig.\ \ref{fig:tides}, we show the normalized mass loss rates over time for our 5 \texttt{CMC} models and the $N$-body models from \citet{Webb2014}, using both the basic radial stripping criterion and the more sophisticated energy-based criterion of \citet{giersz2008}.  Surprisingly, the radial criterion shows better agreement with the \texttt{NBODY6\_tt} models than the default energy criterion, which causes the \texttt{CMC} models to lose mass faster and disrupt sooner than their $N$-body counterparts.  We speculate that this is because the radial criterion, by depending on the apocenter of each star's orbit, requires diffusion in both energy and angular momentum for a star’s apocenter to move beyond the cluster boundary (while the energy criterion requires only the former).   However, further study will be required to confirm this.  For this work, we use the radial criterion, and strip any stars where $r_{\rm apo} > r_t$.

We reiterate that both approaches to tidal stripping we present here are highly simplified.  In practice, the escape of stars from a star cluster is a very complicated problem even in a fixed tidal field.  The time scale for stars that are technically unbound to escape from the cluster is non-trivial to calculate, and during that time, stars may linger indefinitely in unstable orbits near the Lagrange points  \citep[even though $E > \alpha \phi_t$; e.g.,][]{Henon1969,fukushige2000}, or may be scattered back to lower energies \citep[e.g.,][]{King1959,baumgardt2001}.  To account for this, recent improvements to the Monte Carlo method \citep{Giersz2013,2014MNRAS.443.3513S} have implemented more sophisticated techniques which calculate approximate timescales for the stars to escape their parent clusters, which in turn treats escaping particles as a Poisson process.  However, even with this sophisticated approach, the escape rate from Monte Carlo methods still overestimates the rate from direct $N$-body calculations \citep[e.g.,][Fig.\ 5]{Madrid2017}.  Furthermore, these effects are all compounded by the presence of a time-varying galactic potential, which changes the position of the Lagrange points over time.  It is not obvious that these improvements, while more physically motivated, would necessarily improve the comparisons in Fig.~\ref{fig:tides}.  Work to better compare the escape rate we have calculated to direct $N$-body calculations is currently underway.

\section{Tidal Shocking}
\label{apx:shocking}

While we have incorporated time-varying tidal forces based on the above implementation, we have neglected the effect of tidal shocking, where the time-dependent external potential can itself do work on the cluster itself.  These shocks both directly inject energy into the cluster, thereby increasing its radius (and susceptibility to tidal stripping), while second-order effects can serve to advance the relaxation-driven evolution of the cluster \citep{1995ApJ...438..702K}, speeding it towards an early demise.  It has been argued that the inclusion of tidal shocking is critical to explain the destruction of massive clusters with large radii \citep[e.g.,][]{Gnedin1997}.  

We showed the vital diagrams used in \citet{Gnedin1997} as evidence of the importance of tidal shocking in Fig.\ \ref{fig:vital}.  However, from that plot, it was not immediately obvious that tidal shocking was actually a necessary component for reproducing the mass and radii of present-day GCs in the MW, since only a handful of clusters were born in the region excluded by tidal shocking, and the majority of those were destroyed by tidal stripping anyway.  However, the fact that our clusters largely inhabit the correct regions predicted by the MW vital diagrams is not proof that tidal shocking is negligible.  

While implementing tidal shocking in a H\'enon-style $N$-body code presents significant challenges, we can estimate what the influence of tidal shocking would have been in our population in the following way: first, during each shock, we can compute the average change in stellar energies per unit mass at the half-mass radius, $r_{\rm hm}$, of the cluster with \citep{Prieto2008}:

\begin{equation}
\left<\Delta E_h\right> = \frac{1}{6} I_{\rm tid} r_{\rm hm}^2
\label{eqn:ehtidal}
\end{equation}

\noindent where $I_{\rm tid}$ is the tidal heating parameter, given by

\begin{equation}
I_{\rm tid} = \sum_{i, j}\left(\int \mathbf{T}^{ij}dt\right)^2\left(1+\left(\frac{\tau}{t_{\rm dyn}}\right)^2\right)^{-3/2}
\label{eqn:heating}
\end{equation}

\noindent where the sum is over all components $i$, $j$, of the tidal tensor defined in Equation \eqref{eqn:tidetensor}.  The second term in the parenthesis is the adiabatic correction, accounting for the ability of the clusters to successfully re-absorb injected energy.  We use the power-law adiabatic correction from \citep{gnedin1999}, which depends on the ratio of the length of the tidal shock $\tau$ to the dynamical time of the cluster $t_{\rm dyn}$. Once we have the average energy injected by every tidal shock, we can convert this to a typical mass loss experienced by the cluster due to its subsequent expansion.  We use the approximate relation from \cite{Gieles2006}: 

\begin{equation}
\frac{\Delta M}{M_0} \approx 0.22 \frac{\left<\Delta E_h\right>}{E_0}
\label{eqn:deltam}
\end{equation}

\noindent where $M_0$ and $E_0$ are the initial mass and energy of the cluster.

As an order-of-magnitude check, we assume that once the cumulative mass loss given by Equation \eqref{eqn:deltam} exceeds the total mass, the cluster has been disrupted.  For every cluster in our evolved catalog, we identify tidal shocks as occurring whenever any component of the tidal tensor is one standard deviation away from the median value of that component as measured over the entire cluster lifetime (Fig.\ \ref{fig:typical} shows the typical amount by which the effective tidal tensor changes).  The integral in Equation \eqref{eqn:heating} is then calculated by directly summing the component of each shock times the $\Delta t$ between the FIRE snapshots that Equation \eqref{eqn:tidetensor} was extracted from.  In practice, the vast majority of shocks only appear in a single snapshot, so we multiply each component by the spacing between snapshots ($\approx$20-30 Myr).  

After applying this procedure to our catalog, we found virtually no contribution to our final cluster population arising from our neglect of tidal shocking.  In fact, according to this procedure, only one of our 895 clusters should have been destroyed by tidal shocking, and even in that case--a particularly massive ($\sim3\times10^6M_{\odot}$) and large ($r_{\rm hm} = 26$ pc) cluster--it was destroyed by tidal shocking at the same time \texttt{CMC} recorded it as being destroyed by tidal stripping.  In other words, the strong tidal fields responsible for shocking the cluster are also responsible for the increase in mass loss through the tidal boundary.

From this, one might conclude that our implementation of tidal stripping in \texttt{CMC} naturally culls large clusters at the same rate as tidal shocking would, since they are both based in the rapid variation of the galactic tidal field, allowing us to reproduce the correct distribution of clusters in the mass-radius region as \cite{Gnedin1997}.  However, such optimism would be premature.  As previously mentioned, we have extracted our tidal tensors from snapshots of the \texttt{m12i} galaxy that are typically taken every 20-30 Myr.  But the typical dynamical time of any of our clusters during a tidal shocks is a significantly shorter ($\sim$ 1 Myr), and as a result, each tidal shock is reduced by a factor of $10^4$ from the adiabatic correction to Equation \eqref{eqn:heating}.

What this actually means is that the 20 Myr interval between snapshots is insufficient to resolve the influence of tidal shocking on the star cluster population of a single galaxy.  Work to improve the extraction of the instantaneous tidal boundaries in zoom in simulations, as well as comparisons to direct $N$-body models capable of resolving the influence of tidal shocking on the cluster survival, is currently underway.  

\label{lastpage}
\end{document}